\documentclass[12pt]{article}
\usepackage{amsmath}
\usepackage{amsfonts}
\usepackage{latexsym}
\usepackage{graphicx}
\usepackage{color}
\usepackage{xy}
\usepackage[linktocpage]{hyperref}

\usepackage{cite}

\usepackage{ifpdf}

\newcommand{\bmat}{\left(\begin{array}}
\newcommand{\emat}{\end{array}\right)}

\def\yzero{\smash{\hbox{$y\kern-4pt\raise1pt\hbox{${}^\circ$}$}}}
\def\p{\partial}
\def\a{\alpha}
\def\b{\beta}
\def\g{\gamma}
\def\d{\delta}
\def\sig{\sigma}
\def\beq{\begin{equation}}
\def\eeq{\end{equation}}
\def\beqa{\begin{eqnarray}}
\def\eeqa{\end{eqnarray}}

\def\-{\hphantom{-}}

\def\s2{\frac{1}{\sqrt2}}

\def\oh{\frac{1}{2}}

\def\IF{\relax{\rm I\kern-.18em F}}
\def\II{\relax{\rm I\kern-.18em I}}

\def\cn{{\cal N}}
\def\cl{{\cal L}}
\def\cam{{\cal M}}
\def\cp{{\cal P}}

\def\ch{{\cal H}}

\def\Dsl{\,\raise.15ex\hbox{/}\mkern-13.5mu D} 

\def\IC{\mathbb{C}}

\def\IR{\mathbb{R}}
\def\IZ{\mathbb{Z}}

\def\im{{\rm Im}\,}
\def\re{{\rm Re}\,}

\def\raw{\rightarrow}

\def\bes{\begin{subequations}}
\def\ees{\end{subequations}}
\def\cM{\mathcal{M}_6}
\def\vol{\textrm{Vol}}

\catcode`\@=11
\newdimen\@rotdimen
\newbox\@rotbox

\def\@vspec#1{\special{ps:#1}}
\def\@rotstart#1{\@vspec{gsave currentpoint currentpoint translate
   #1 neg exch neg exch translate}}
\def\@rotfinish{\@vspec{currentpoint grestore moveto}}
\def\@rotr#1{\@rotdimen=\ht#1\advance\@rotdimen by\dp#1%
   \hbox to\@rotdimen{\hskip\ht#1\vbox to\wd#1{\@rotstart{90 rotate}%
   \box#1\vss}\hss}\@rotfinish}
\def\@rotl#1{\@rotdimen=\ht#1\advance\@rotdimen by\dp#1%
   \hbox to\@rotdimen{\vbox to\wd#1{\vskip\wd#1\@rotstart{270 rotate}%
   \box#1\vss}\hss}\@rotfinish}%
\def\@rotu#1{\@rotdimen=\ht#1\advance\@rotdimen by\dp#1%
   \hbox to\wd#1{\hskip\wd#1\vbox to\@rotdimen{\vskip\@rotdimen
   \@rotstart{-1 dup scale}\box#1\vss}\hss}\@rotfinish}%
\def\@rotf#1{\hbox to\wd#1{\hskip\wd#1\@rotstart{-1 1 scale}%
   \box#1\hss}\@rotfinish}%
\def\rotate{\@ifnextchar[{\@rotate}{\@rotate[l]}}
\def\@rotate[#1]#2{\setbox\@rotbox=\hbox{#2}\@nameuse{@rot#1}\@rotbox}

\catcode`\@=12

\topmargin
-1.5cm
\textwidth
15.5cm
\textheight
23.5cm
\oddsidemargin
0.7cm
\evensidemargin
1.2cm

\begin{document}

\makeatletter
\@addtoreset{equation}{section}
\makeatother
\renewcommand{\theequation}{\thesection.\arabic{equation}}
\pagestyle{empty}
\rightline{CERN-PH-TH/2011-127}
\rightline{IFT-UAM/CSIC-11-34}
\vspace{2cm}
\begin{center}
\LARGE{\bf RR photons  \\[12mm]}
\large{Pablo G. C\'amara$^{1}$, Luis E. Ib\'a\~nez$^{2,3}$ and Fernando Marchesano$^{3}$
\\[3mm]}

\bigskip

\footnotesize{

${}^1$ PH-TH Division, CERN CH-1211 Geneva 23, Switzerland  \\[2mm]
${}^2$ Departamento de F\'{\i}sica Te\'orica,
Universidad Aut\'onoma de Madrid, 
28049 Madrid, Spain  \\[2mm]
${}^3$ Instituto de F\'{\i}sica Te\'orica UAM-CSIC, Cantoblanco, 28049 Madrid, Spain}

\bigskip

\bigskip

\bigskip

\small{\bf Abstract} \\[5mm]
\end{center}
\begin{center}
\begin{minipage}[h]{16.0cm}

Type II string compactifications to 4d generically contain massless Ramond-Ramond U(1)  gauge symmetries.
However there is no massless matter charged  under these U(1)'s, which makes a priori difficult to measure
any physical consequences of their existence. There is however a window of opportunity if these RR U(1)'s
 mix with the hypercharge U(1)$_Y$  (hence with the photon).  In this paper we study in detail different avenues
 by which U(1)$_{RR}$ bosons may mix with D-brane U(1)'s. We concentrate on Type IIA orientifolds
 and their M-theory lift, and provide geometric criteria for the existence of such mixing, which may occur either
 via standard kinetic mixing or via the mass terms induced by St\"uckelberg couplings. The latter case is particularly
 interesting, and appears whenever D-branes wrap torsional $p$-cycles in the compactification manifold.
 We also show that in the presence of torsional cycles discrete gauge symmetries and Aharanov-Bohm strings and
 particles appear in the 4d effective action, and that type IIA St\"uckelberg couplings can be understood in terms of
torsional (co)homology in M-theory. We provide examples of Type IIA Calabi-Yau orientifolds in which the required torsional
 cycles exist and kinetic mixing induced by mass mixing is present. We discuss some phenomenological consequences
 of our findings. In particular, we find that mass mixing may induce  corrections relevant for  hypercharge gauge coupling
 unification in F-theory SU(5) GUT's.

\end{minipage}
\end{center}
\newpage
\setcounter{page}{1}
\pagestyle{plain}
\renewcommand{\thefootnote}{\arabic{footnote}}
\setcounter{footnote}{0}

\vspace*{1cm}

\tableofcontents

\section{Introduction}

String theory compactifications with a semi-realistic spectrum generically lead to a number of U(1) gauge symmetries beyond the standard model hypercharge. Some of these U(1) symmetries acquire masses of the order of the string scale via the St\"uckelberg mechanism and would be  difficult to detect unless $ M_s \sim 1 \textrm{ TeV}$. They remain as global symmetries of the low energy effective Lagrangian, only broken by non-perturbative effects. The canonical example is the U(1)$_{B-L}$ symmetry which arises in many D-brane models.
 Some other U(1)'s, however, may appear in the massless spectrum or acquire very light masses (generated for instance by quantum corrections). Those can pass all the current experimental bounds (from EW precision data, searches for $\gamma - \gamma'$ oscillations, cosmological bounds, etc.) if their coupling to the Standard Model hypercharge is sufficiently small. The relevant parameter space has two quantities: the mass of the hidden photon and the kinetic mixing between the hypercharge and the hidden photon. In addition, due to the above mixing with the SM hypercharge, particles charged under the hidden U(1) acquire an effective electric (mini-)charge and can lead to further experimental signatures. Some references for U(1) mixing in the string
theory context include \cite{Dienes:1996zr}.
The possibility of having hidden U(1) gauge symmetries has also motivated interesting applications in the context of supersymmetric models. For instance, it has been suggested that hidden U(1)'s can lead to a possible mechanism for mediating SUSY breaking to the visible sector in a flavor independent way
\cite{Langacker:2007ac,Langacker:2008ip,Verlinde:2007qk,Grimm:2008ed}. Also, mixing of MSSM neutralinos with hidden U(1) gauginos can be a relevant signature at the
LHC \cite{Feldman:2006wd,Ibarra:2008kn,Arvanitaki:2009hb}.

In type II string compactifications there are two possible sources of hidden U(1) gauge symmetries: D-branes located far away from the SM D-brane sector and which do not intersect it, and U(1) gauge symmetries arising from Kaluza-Klein reduction of Ramond-Ramond closed string fields. This work intends to be a systematic study of RR U(1) gauge symmetries in Calabi-Yau compactifications and their possible mixing with D-brane gauge bosons. In particular, we find that RR gauge bosons can mix with D-brane U(1)'s through direct kinetic mixing (see also \cite{Jockers:2004yj,Grimm:2008dq,Grimm:2011dx, Kerstan:2011dy}) or through the mass matrix induced by a St\"uckelberg mechanism. The latter is generic in Calabi-Yau orientifold compactifications with torsional $p$-cycles, and can be understood in a precise way in terms of the integer homology of the Calabi-Yau.  We develop the necessary tools to describe this mixing and provide examples of type IIA CY orientifolds in which the required torsional cycles exist
and kinetic mixing is induced via St\"uckelberg mass mixing.

Mixing between Ramond-Ramond and D6-brane U(1) gauge symmetries may find interesting applications in the context of type II/F-theory SU(5) models, which we briefly describe. In particular we observe that RR U(1) gauge symmetries can provide an alternative to the standard picture that has been developed in the context of F-theory local GUT's, in which the GUT gauge symmetry is broken via a hypercharge flux along the internal dimensions \cite{Beasley:2008kw, Donagi:2008kj}. Such scenario is compatible with a massless hypercharge only if certain topological conditions are imposed on the hypercharge flux. As we discuss, such conditions are compatible with the topological conditions required for the mass mixing between the hypercharge and RR U(1)'s and so it could happen that the actual hypercharge has a contamination from RR U(1) gauge symmetries. A direct consequence of this contamination is a modification of the fine structure constant which may be crucial for achieving actual gauge coupling unification in the present setup.

The effect of mass mixing is intimately related to another interesting feature of Calabi-Yau compactifications with torsion in (co)homology, namely the appearance of RR discrete gauge symmetries. Recently Banks and Seiberg \cite{bs10} have shown that in every consistent four-dimensional quantum theory of gravity massive U(1) gauge symmetries are spontaneously broken to discrete $\mathbb{Z}_k$ gauge symmetries, and that there are Aharanov-Bohm strings and particles associated to them, with unusual charge quantization. In this sense our study reveals that this 4d picture of massive U(1)'s is consistently realized in string theory through the torsional (co)homology of the compact manifold. In fact, it is precisely this set of massive RR U(1)'s the ones that in the presence of D-branes may develop a mass mixing with open string U(1)'s, so that the massless U(1) that results from the St\"uckelberg mechanism in neither open nor closed, but a linear combination of both.

While our results are equally valid for both type IIA or type IIB compactifications, our discussion is mainly carried in the context of type IIA compactifications, since it has a more direct connection to M-theory. The M-theory picture is particularly compelling when analyzing Abelian gauge symmetries, since there both D6-brane and RR U(1) gauge symmetries arise from Kaluza-Klein reduction on the $G_2$ manifold. In this sense, our discussion shows that both sets of massive U(1)'s/discrete gauge symmetries arise from KK reduction on the torsional cohomology of the $G_2$ manifold. As a quite direct consequence of this, we observe that Freed-Witten D6-brane gauge anomalies are lifted to M-theory backgrounds where  4-form $G_4$ has a torsional cohomology class in the compactification manifold.

The paper is organized as follows. In section \ref{U1typeIIA} we describe the family of type IIA Calabi-Yau compactifications in which we will carry most of our discussion, reviewing those results in the literature which will be necessary in subsequent sections. In section \ref{kmixing} we describe the kinetic mixing that occurs between open and closed string U(1)'s, as well as its lift to M-theory. We describe the effect of torsional homology in these compactifications in section \ref{mmixing}. In particular we first discuss the relation between torsion $p$-cycles and discrete gauge symmetries and then, upon adding D-branes, to the mass mixing developed between open and closed string U(1)'s. The latter mixing is used in section \ref{pheno} in order to describe how our results may be relevant for certain scenarios, and in the particular in the F-theory setup described above. Finally, in section \ref{fluxes} we leave the realm of Calabi-Yau compactifications, and discuss certain new features that appear when we consider type IIA/M-theory compactifications with background fluxes.

We leave our final comments for section \ref{conclu}, and several technical details for the appendices. In particular in appendix \ref{apdim} we perform the dimensional reduction to 4d of a D6-brane action. Appendix \ref{typeIIB} translates the results of the main text to the mirror symmetric language of type IIB compactifications, and appendix \ref{tortopo} describes how D-branes can detect RR fields that live in the torsional cohomology of the compactification.

\section{U(1)'s in type IIA compactifications}\label{U1typeIIA}

Abelian gauge bosons  in weakly coupled type II string compactifications can originate from either open or closed strings. While the former are localized in the worldvolume of D-branes, the latter propagate along the full compactification manifold. Understanding the circumstances under which these two apparently different sectors interact with each other is the purpose of the next two sections. For concreteness, we will carry our discussion in the context of 4d $\cn = 1$ type IIA compactifications on Calabi-Yau orientifolds with intersecting D6-branes which, as shown in the literature \cite{d6mb}, constitute a rich framework for model building in string theory. Our results can however be easily translated to dual type IIB orientifold compactifications, as we show in Appendix \ref{typeIIB}.
In order to set up the stage, in this section we review those aspects of the 4d effective action of type IIA compactifications which are relevant for our purposes.

\subsection{Type IIA orientifold compactifications}

Let us consider type IIA string theory on an orientifold of $\IR^{1,3} \times \cM$, with $\cM$ a compact Calabi-Yau 3-fold. The orientifold action is given by $\Omega_p (-1)^{F_L}\sigma$, where $\Omega_p$ is the worldsheet parity reversal operator, ${F_L}$ is the space-time fermion number for the left-movers, and $\sigma$ is an internal involution of the Calabi-Yau. The involution acts on the K\"ahler 2-form $J$ and the holomorphic 3-form $\Omega$ of $\cM$ as \cite{Acharya:2002ag, Brunner:2003zm}
\begin{equation}
\sigma J=-J\ , \qquad \sigma\Omega = \overline{\Omega}\label{iiaori}
\end{equation}
The fixed locus $\Lambda$ of $\sigma$ is given by one or several 3-cycles of $\cM$, in which O6-planes are located. One can then see that each of these 3-cycles is an special Lagrangian (sLag) submanifold of $\cM$, since (\ref{iiaori}) automatically imply the sLag conditions
\begin{equation}
J|_{\Lambda}=0\ , \qquad \im \Omega|_{\Lambda}=0
\end{equation}
In order to cancel the RR charge of the O6-planes one may introduce D6-branes,\footnote{In order to cancel RR tadpoles and build 4d chiral models one may consider coisotropic D8-branes as in \cite{fim06}. It should be straightforward to generalize the results of this paper to that case.} each of them wrapping a 3-cycle $\pi_a$ within $\cM$. Consistency with $\cn=1$ supersymmetry in 4d then requires that these 3-cycles fulfill the same sLag conditions as the orientifold \cite{slag}, namely
\begin{equation}
J|_{\pi_a}=0\ , \qquad \im(\Omega)|_{\pi_a}=0 \label{d6slag}
\end{equation}
usually dubbed F-term and D-term conditions, respectively, due to how they appear in the D6-brane effective action. Cancellation of the total D6-brane charge in $\mathcal{M}_6$ can then be recast as a condition in homology \cite{Aldazabal:2000dg}
\begin{equation}
\sum_aN_a([\pi_a]+[\pi_a^*]) =4 [\Lambda]
\label{tadpole}
\end{equation}
where $[\pi_a] \in H_3(\cM, \IZ)$ is the homology class of the 3-cycle $\pi_a$, and $[\pi_a^*]=[\sigma\pi_a]$ that of the image of $\pi_a$ under the orientifold. Finally, $N_a$ stands for the number of D6-branes on top of the 3-cycle $\pi_a$.

One of the virtues of compactifications on K\"ahler manifolds resides in that there is a one-to-one correspondence between massless fields in the 4d effective theory and de Rham cohomology classes. In particular, for the closed string sector of the theory the spectrum of 4d massless fields is obtained from expanding the 10d type IIA supergravity fields in a basis of harmonic forms. To which 4d field a $p$-form corresponds to not only depends on its degree $p$, but also on its parity under the orientifold involution $\sigma$ \cite{Brunner:2003zm, Grimm:2004uq, Grimm:2004ua}. We therefore introduce a basis of cohomology representatives of definite parity under $\sigma$
\begin{center}
\begin{tabular}{ccccc}
&&\underline{$\sigma$-even}&&\underline{$\sigma$-odd}\\
2-forms && $\omega_i\ \ i=1,\ldots , h^{1,1}_+$ && $\omega_{\hat{i}}\ \ \hat{i}=1,\ldots , h^{1,1}_-$\\
3-forms && $\alpha_I\ \ I=0,\ldots , h^{1,2}$ && $\beta^I\ \ I=0,\ldots , h^{1,2}$\\
4-forms && $\tilde{\omega}^{\hat{i}}\ \ \hat{i}=1,\ldots , h^{1,1}_-$ && $\tilde{\omega}^i\ \ i=1,\ldots , h^{1,1}_+$
\end{tabular}
\end{center}
paired up and normalized such that
\begin{equation}
\int \omega_i\wedge \tilde{\omega}^j=\delta_i^j\ ,\qquad \int \omega_{\hat i}\wedge \tilde{\omega}^{\hat j}=\delta_{\hat i}^{\hat j}\ , \qquad \int \alpha_I\wedge \beta^J=\delta_I^J\label{norm}
\end{equation}

The spectrum of 4d massless fields can then be arranged into $h^{1,1}_-+h^{1,2}+1$ chiral multiplets and $h^{1,1}_+$ vector multiplets of the 4d $\cn=1$ supersymmetry preserved by the compactification \cite{Brunner:2003zm, Grimm:2004ua}. Apart from these, there are extra vector multiplets coming from the open string sector.

The moduli space of the compactification is parametrized by the scalar components of the chiral multiplets. More precisely, these are given by  $h^{1,1}_-$ K\"ahler moduli $T^{\hat{i}}$ and $h^{1,2}+1$ complex structure moduli
$N^A$, with $N^0$ the universal axio-dilaton. They result from the expansions \cite{Grimm:2004ua}
\begin{equation}
J_c\equiv B_2+iJ=T^{\hat{i}}\omega_{\hat i}\ , \qquad \Omega_c \equiv C_3+i\textrm{Re}(C\Omega)=N^I\alpha_I\label{iiamoduli}
\end{equation}
where $B_2$ is the NSNS 2-form, $C_3$ is the RR 3-form and $C$ is a compensator field defined as
\begin{equation}
C\equiv e^{-\phi_{10}\sqrt{\vol_6}}e^{K_{cs}/2}\ , \qquad K_{cs}\equiv -\log\left[-\frac{i}{8}\int\Omega\wedge\overline{\Omega}\right]
\end{equation}
with $\phi_{10}$ the 10d dilaton. The kinetic terms of the 4d chiral multiplets are then encoded in the K\"ahler potential for such moduli space, that can be expressed as \cite{Grimm:2004ua}
\begin{equation}
\frac{K}{M_{Pl}^2}=-\log\left[\frac{4}{3}\int_{\cM} J\wedge J\wedge J\right]-\log\ e^{-4\phi_4}\label{kahler}
\end{equation}
where $M_{Pl}$ is the reduced 4d Planck mass, $*_6$ stands for the Hodge star operator in $\cM$ and the 4d dilaton is given by,
\begin{equation}
e^{-2\phi_4}=2\int_{\cM}\re(C\Omega)\wedge *_6 \re(C\Omega)
\end{equation}

Particularly relevant for our purposes are the real parts of the complex structure moduli. These are invariant under shifts, and therefore behave as axions in the 4d effective theory. Their kinetic terms can be directly read from (\ref{kahler})
\begin{equation}
\mathcal{L}=\frac{1}{2}e^{2\phi_4}\mathcal{G}^{-1}_{IJ}\textrm{Re}(dN^{I})\wedge *_{4}\textrm{Re}(dN^{J})
\label{kin0}
\end{equation}
where
\begin{equation}
\mathcal{G}_{IJ}^{-1}\equiv M_{Pl}^2\int_{\cM}\alpha_I\wedge *_6\alpha_J \label{gij}
\end{equation}
is a function depending only on the complex structure moduli $N^I$. It is often convenient to express the axions in terms of 2-forms $C_2^{I}$ of $\IR^{1,3}$ that belong to the dual linear multiplets
\begin{equation}
dC_2^{I}\equiv -e^{2\phi_4}\mathcal{G}^{-1}_{IJ}*_{4}\textrm{Re}(dN^J)
\label{dual4d}
\end{equation}
and which arise from expanding the RR 5-form potential $C_5$ in $\sigma$-odd harmonic 3-forms of $\mathcal{M}_6$
\begin{equation}
C_5 \, =\, \sum_I C_2^{I} \wedge \beta^I + \dots
\label{redC5}
\end{equation}
where the dots stand for further terms giving rise to 4d gauge bosons, see eq.(\ref{f6exp}).
The 4d duality relation (\ref{dual4d}) then arises as a direct consequence of the 10d duality relation $\hat{F}_4=*_{10}\hat{F}_6$, where $\hat{F}_p=dC_{p-1}-C_{p-3}\wedge dB_2$.

\subsection{Open string U(1)'s}
\label{opensec}

In weakly coupled type II orientifolds non-Abelian gauge groups and chiral fermions charged under them arise from open strings.  As a consequence, in semi-realistic 4d compactifications the Standard Model gauge group and matter content are located in this sector.\footnote{This is no longer necessarily true at strong coupling, where the distinction between open and closed string degrees of freedom becomes rather artificial.} In particular, for type IIA intersecting D6-brane models chiral fermions are localized at the D6-brane intersections, and the corresponding gauge groups at the 3-cycles $\pi_a$, $a =1 ,\dots, K$ wrapped by the D6-branes. A single D6-brane on $\pi_a$ will contain a U(1)$_a$ gauge theory in its worldvolume, while $N_a$ coincident D6-branes wrapping $\pi_a$ will give rise to an $SU(N_a) \times U(1)_a$ gauge group.\footnote{If $\pi_a$ is invariant under the orientifold action the gauge group may instead be $SO(N_a)$ or $USp(N_a)$. Although these D6-brane can be easily incorporated into our discussion, we will not consider them in the following, as they do not give rise to U(1) factors of the gauge group.} In the following we will focus on the U(1) factors of such open string gauge group.

The gauge coupling constants of such U(1) factors are obtained at the disc level by dimensionally reducing the D6-brane  DBI action (see e.g. \cite{Grimm:2011dx, Kerstan:2011dy} and Appendix \ref{apdim})
\begin{equation}
g_a^{-2}=\textrm{Re}(f_a)\ , \qquad f_a=-iN_a\int_{\pi_a}\Omega_c
\label{gkf0}
\end{equation}
Whereas at disc level the overall gauge kinetic function is diagonal, quantum corrections may induce kinetic mixing between different D6-brane gauge factors (see e.g. \cite{Gmeiner:2009fb}).

In addition to the matter multiplets at the D6-brane intersections, there are $h^1(\pi_a)$ massless chiral multiplets transforming in the adjoint representation of $U(N_a)$ for the $a$-th stack of D6-branes. Their scalar components are given by a combination of the Wilson line moduli $\theta^j_{a}$ and the geometric deformations $\phi_{a}^i$ of the 3-cycle $\pi_a$ which preserve the sLag conditions (\ref{d6slag}), namely we have that
\begin{equation}
\Phi^{j}_{a}=\theta^{j}_{a}+\lambda_i^j\phi^i_{a}
\label{openmoduli}
\end{equation}
where $\theta^{j}_{a}$ are the components of an arbitrary Wilson line harmonic 1-form
\begin{equation}
\theta_{a}\, =\, \theta^{j}_{a} \zeta_j , \qquad \frac{\zeta_j}{2\pi} \in \ch^1(\pi_a, \IZ)\label{wilson}
\end{equation}
and $\phi^i_{a}$ are the components of a normal vector preserving the sLag condition \cite{McLean}
\begin{equation}
\phi_{a}\, =\, \phi^{i}_{a} X_i , \qquad X_i \in N(\pi_a) \ | \ \cl_{X_i} J = \cl_{X_i} \im \Omega = 0 \label{geod6}
\end{equation}
with $\cl_{X_i}$ the Lie derivative along $X_i$. Finally, $\lambda_i^j \in \IC$ is a matrix relating the two basis $\{\zeta_j\}$ and $\{X_i\}$, and can be defined as
\begin{equation}
\iota_{X_i} J_c|_{\pi_a}\, =\, \lambda_i^j\zeta_j\label{lambda}
\end{equation}
where $\iota_{X_i} J = (X_i^m J_{mn}) dx^n$ is a harmonic 1-form on the D6-brane worldvolume \cite{McLean}.

While in principle each stack of $N_a$ D6-branes wrapping a 3-cycle $\pi_a$ gives rise to a U(1)$_a$ factor, not all of these gauge symmetries survive at low energies. Indeed, several linear combinations of U(1)'s, and in particular those which are anomalous, become massive by an St\"uckelberg mechanism \cite{Ibanez:1998qp, Poppitz:1998dj,Aldazabal:2000dg}, with masses of the order of the string scale.
In order to describe those U(1)'s which remain massless let us introduce the set of numbers
\begin{equation}
c^I_a\, =\, -\int_{\pi_a} \beta^I, \qquad d_{Ia}\, =\, \int_{\pi_a} \alpha_I\label{poinc1}
\end{equation}
which define the Poincar\'e duals to the 3-cycles $\pi_a$
\begin{equation}
\hat \pi_a = c^I_a\alpha_I+d_{Ia}\beta^I \in H^3(\cM,\mathbb{R})\label{poinc2}
\end{equation}
Notice that $c^I_a$ is proportional to the coupling of the 2-forms $C_2^I$ to a D6-brane wrapping $\pi_a$, and that this coupling is the one triggering the St\"uckelberg mechanism. Indeed, dimensional reduction of the D6-brane action (c.f. Appendix \ref{apdim}) reveals that some combinations of shift symmetries in the 4d effective theory are gauged in presence of D6-branes, and so (\ref{kin0}) gets modified to
\begin{equation}
\mathcal{L}_{\rm Stk}=\frac12 e^{2\phi_4}\mathcal{G}_{IJ}^{-1}\textrm{Re}(DN^I)\wedge *_4\textrm{Re}(DN^J)\ , \qquad DN^I=dN^I+c^I_aN_aA^a\label{stuck}
\end{equation}
with $A^a$ the gauge potential for U(1)$_a$. The linear combinations of U(1) gauge symmetries which become massive are therefore
\begin{equation}
Q^I=\sum_ac_a^I N_a Q^a\label{umass}
\end{equation}
where $Q^a$ denotes the diagonal U(1) generator of the $a$-th stack of D6-branes.
The number of axions $N^I$ which are eaten in order to produce massive U(1)'s is then given by the rank of the matrix $c_a^I$.

In order to get a better picture of which U(1)'s remain massless, let us briefly detour from our discussion and consider the case where the type IIA compactification is simply given by $\IR^{1,3} \times \cM$, without any orientifold. In that case, dimensional reduction of the closed string sector yields a 4d $\cn=2$ spectrum, and in particular we now have $1+h^{2,1}$ $\cn=2$ hypermultiplets, each containing two axions instead of one. Similarly, we have doubled the number of dual 2-forms, which arise from the reduction of $C_5$ without any particular orientifold parity
\begin{equation}
C_5 \, =\, \sum_I C_2^{I} \wedge \beta^I + \sum_J C_{2J} \wedge \alpha_J + \dots
\label{redC52}
\end{equation}
As a result, the number of axions that can be eaten by the D6-brane U(1)'s is doubled with respect to the orientifold case, and we have that those open string U(1)'s that become massive are
\begin{equation}
Q^I=\sum_ac_a^I N_a Q^a \qquad {\rm and} \qquad Q_J=\sum_ad_{Ja} N_a Q^a
\label{umass2}
\end{equation}

With this information it is quite straightforward to provide a description of which U(1) bosons become massive and which ones do not. For this first notice that $\vec{\g}_a = (\vec{c}_a, \vec{d}_a)$ is nothing but a vector in $H^3(\cM, \IR) \simeq \IR^{b_3}$, with $b_3 = 2 + 2h^{1,2}$ the number of independent harmonic 3-forms of $\cM$. A stack of $N_a$ D6-branes wrapping the 3-cycle $\pi_a$ is then represented by the vector $N_a\vec{\g}_a$, and the whole set of vectors $\{ N_a\vec{\g}_a \}_{a=1}^K$ arising from the $K$ different stacks spans a vector subspace $V = \langle \{ \vec{\g}_a \} \rangle \simeq \IR^r$ of $H^3(\cM,\IR)$. The dimension $r$ of such subspace will be the number of eaten axions and massive open string U(1)'s, while $K-r$ will be the number of D6-brane U(1)'s that remain massless.

In the Poincar\'e dual language of 3-cycles this amounts to say that the number $r$ of massive U(1)'s correspond to the number of 3-cycles within $\{\pi_a\}$ which are linearly independent in homology, more precisely as elements of $H_3(\cM, \IR)$. The U(1)'s that remain massless are those whose coefficients $c_a^I$, $d_{Ja}$ vanish identically, which means that they are wrapping a trivial 3-cycle in $H_3(\cM, \IR)$. This is impossible for a single stack of D6-branes wrapping a sLag 3-cycle, but it can be achieved by taking linear combinations of 3-cycles. Indeed, a simple example of the latter would be to consider two coincident D6-branes wrapping $\pi_a$ which are separated via the adjoint Higgsing $SU(2) \times U(1)_a \raw U(1)_{a_1} \times U(1)_{a_2}$. The two 3-cycles $\pi_{a_1}$ and $\pi_{a_2}$ only differ by the values of their moduli $\Phi^j$ and so are equivalent in homology $[\pi_{a_1}] = [\pi_{a_2}]$. This means that their U(1) gauge bosons have exactly the same couplings to the 2-forms, $\vec{\g}_{a_1} = \vec{\g}_{a_2} = \vec{\g}_a$. Hence, the combination U(1)$_{a_1} - U(1)_{a_2}$ orthogonal to U(1)$_a = U(1)_{a_1} + U(1)_{a_2}$ does not couple to any axion, and it remains as a gauge symmetry of the low energy theory. Note that this massless U(1) combination corresponds to the formal difference of 3-cycles $\pi_{a_1} - \pi_{a_2}$, which is indeed trivial in homology.

In general, a massless U(1) will be given by a linear combination of the form
\begin{equation}
Q^b\, =\, \sum_a n^b_a Q^a \qquad  {\rm such\ that} \qquad \vec{\g}_b=  \sum_a n^b_a N_a \vec{\g}_a = 0
\end{equation}
and, as each vector $\vec{\g}_a$ corresponds to a 3-cycle $\pi_a$ we have that $\vec{\g}_b$ corresponds to a formal linear combination of 3-cycles
\begin{equation}
\pi_b\, =\, \sum_a n^b_a N_a \pi_a \qquad  {\rm such\ that} \qquad [\pi_b] = 0
\label{combi3}
\end{equation}
 Hence, we can identify massless U(1)'s with linear combinations of D6-branes that correspond to (sums of) 3-cycles $\pi_b$ trivial in homology. By definition, this means that there exists a 4-chain $\Sigma_4$ whose boundary is given by $\p \Sigma_4 = \pi_b$, and so it connects all the 3-cycles that participate in the massless U(1). As we discuss in section \ref{kmixing}, this fact will be crucial for computing kinetic mixing between open and closed string U(1)'s.

Let us now go back to the orientifold compactification, where the picture is quite similar. The main difference there is that the open string U(1)'s only couple to the coefficients $c_b^I$, and not to $d_{Ja}$. As a result, the number of massive U(1)'s is given by the dimension of $\langle \{ \vec{c}_a \} \rangle \subset \IR^{h^{2,1} + 1} \simeq H_{-}^3 (\cM, \IR)$. In addition, massless U(1)'s will be given by linear combinations of generators of the form
\begin{equation}
Q^b\, =\, \sum_a n^b_a Q^a \qquad  {\rm such\ that} \qquad \vec{c}_b=  \sum_a n^b_a N_a \vec{c}_a = 0
\label{combiQ-}
\end{equation}
and so its associated combination of 3-cycles $\pi_b$ built as in (\ref{combi3}) does not need to be trivial in full 3-cycle homology $H_3 (\cM, \IR)$, but only in the subspace $H_3^{-} (\cM, \IR)$ of odd 3-cycles. Nevertheless, since we now have the orientifold images $\pi_a^*$ of these 3-cycles we can construct the linear combination
\begin{equation}
\pi_b^-\, =\, \sum_a n^b_a N_a (\pi_a - \pi_a^*)
\label{combi3-}
\end{equation}
which by (\ref{combiQ-}) will be a trivial 3-cycle, in the sense that $[\pi_b^-] =0$ in $H_3(\cM, \IR)$. This again guarantees that we can build a 4-chain $\Sigma_4$ such that $\p \Sigma_4 = \pi_b^-$.

Note that in this discussion we have mainly dealt with the de Rham cohomology group $H^3(\cM, \IR)$ and its homology dual $H_3(\cM, \IR)$, rather than the more fundamental homology group $H_3(\cM, \IZ)$ that classifies topologically different 3-cycles. The difference between $H_3(\cM, \IZ)$ and  $H_3(\cM, \IR)$ does however only arise when $\cM$ contains $\IZ_N$ torsional 3-cycles, a possibility that we have implicitly ignored up to now. In fact, as we will see in section \ref{mmixing} the discussion above has to be slightly modified in the presence of torsional 3-cycles. In that case the spectrum of  massless and massive open string U(1)'s cannot be understood without considering the U(1) gauge symmetries that arise from the closed string sector, which we now turn to describe.

\subsection{Closed string U(1)'s}\label{closedU1}

Besides the gauge symmetries localized at the worldvolume of D-branes, there are generically extra U(1) gauge symmetries arising from the closed string sector.\footnote{In particular the presence of massless closed string gauge bosons in the 4d spectrum is ubiquitous in compactifications with extended supersymmetry.} For type IIA Calabi-Yau orientifold compactifications, massless closed string U(1) gauge bosons result from dimensionally reducing the RR 3-form $C_3$ on harmonic 2-forms of $\cM$ which are even under the orientifold involution
\begin{equation}
C_3\, =\, \sum_I \textrm{Re}(N^I) \alpha_I + \sum_i {A}^i\wedge\omega_i\label{f4exp}
\end{equation}
where we have included the axions $\textrm{Re}(N^I)$ discussed above.
The corresponding 4d dual magnetic degrees of freedom arise from expanding the RR 5-form in hodge dual harmonic 4-forms\footnote{There are also 3-forms in the 4d theory which result from dimensionally reducing $C_5$ on $\sigma$-odd 2-forms. In this work we do not consider them as they are not relevant for our purposes.\label{no3forms}}
\begin{equation}
C_5\, =\, \sum_{I} C_2^{I}\wedge \beta^I + \sum_{i} {V}^i\wedge\tilde{\omega}^i\label{f6exp}
\end{equation}
Thus, overall there is a $U(1)^{h^{1,1}_+}$ gauge symmetry in the 4d effective theory originating from the closed string sector of the compactification.

The gauge kinetic function for these RR U(1)'s can be obtained from dimensional reduction of the relevant kinetic term and Chern-Simons coupling in the 10d type IIA supergravity action, resulting in \cite{Grimm:2004ua}
\begin{equation}
f_{ij}=-i\mathcal{K}_{ij\hat{k}}T^{\hat k}\label{gaugerr}
\end{equation}
where triple intersection numbers $\mathcal{K}_{ij\hat{k}}$ are defined as,
\begin{equation}
\mathcal{K}_{ij\hat{k}}=\int_{\cM}\omega_i\wedge\omega_j\wedge\omega_{\hat{k}}
\end{equation}
Hence, contrary to what happens for open string U(1) gauge symmetries, kinetic mixing between different RR U(1) factors can occur already at the disc level.

In general, the only objects of the 4d effective theory which are charged under RR U(1) gauge symmetries are very massive D-particles made up from bound states of D2 and D4-branes wrapping respectively even 2-cycles and odd 4-cycles in $\cM$. At very special points of the moduli space, such as orbifold points, these states can become light and the $U(1)^{h^{1,1}_+}$ gauge symmetry gets enhanced to some non-Abelian group.

\subsection{Lift to M-theory}
\label{msec}

Whereas in weakly coupled type IIA compactifications open and closed string U(1) gauge symmetries appear as rather different sectors, at strong coupling these differences are smoothed out. As the coupling increases, D6-brane excitations become delocalized in the transverse space, whereas RR bosons may feel a non-trivial potential localizing their wavefunction. At large coupling the perturbative expansion breaks down and the distinction between open and closed string degrees of freedom also does. M-theory therefore provides a natural framework for a unified treatment of D6-brane and RR U(1) gauge symmetries.

Let us consider M-theory compactified on a $G_2$-holonomy manifold $\hat{\mathcal{M}}_7$ admitting at least one perturbative type IIA Calabi-Yau orientifold limit \cite{Kachru:2001je}
\begin{equation}
\hat{\mathcal{M}}_7\ \to \ (\cM\times S^1)/\hat \sigma \label{iialimit}
\end{equation}
with $\hat\sigma=(\sigma,-1)$ an involution which acts as the orientifold involution in $\cM$ and reverses the M-theory circle. The only bosonic degrees of freedom are the M-theory 3-form $A_3$ and the metric. Fluctuations of the latter are encoded in the covariantly constant real 3-form $\Phi_3$ of $\hat{\mathcal{M}}_7$ \cite{Joyce}. The massless fields in the 4d effective theory then result from expanding $A_3$ and $\Phi_3$ in a basis of cohomology forms,\footnote{Note that the only independent non-trivial cohomology classes in $\hat{\mathcal{M}}_7$ are $H^2(\hat{\mathcal{M}}_7)$ and $H^3(\hat{\mathcal{M}}_7)$, with the other non-trivial classes related by 7d Hodge duality.}
\begin{equation}
A_3=\textrm{Re}(M^I)\phi_I+A^\alpha\wedge\omega_\alpha \qquad \Phi_3=\textrm{Im}(M^I)\phi_I \qquad
\begin{array}{l}
I=1,\ldots,b_3(\hat{\cam}_7)\\
\alpha=1,\ldots,b_2(\hat{\cam}_7)
\end{array}
\end{equation}
The massless content of the 4d effective theory is therefore given by $b_3$ chiral multiplets and $b_2$ vector multiplets of $\cn=1$ supersymmetry. The gauge group at generic points of the moduli space is $U(1)^{b_2}$, although at those points where M2 and/or M5-branes wrapping 2-cycles and 5-cycles in $\hat{\mathcal{M}}_7$ become massless, it gets enhanced to some non-Abelian group. The gauge kinetic function has been obtained in \cite{Papadopoulos:1995da} from dimensional reduction of 11d supergravity action, and it is given by
\begin{equation}
f_{\alpha\beta}=-iM^I\int_{\hat{\mathcal{M}}_7}\phi_I\wedge\omega_\alpha\wedge\omega_\beta
\label{Mmixing}
\end{equation}
In the limit (\ref{iialimit}) harmonic 2-forms and 3-forms of $\hat{\mathcal{M}}_7$ decompose as,
\begin{align}
&H^2(\hat{\mathcal{M}}_7)=H^2_+(\cM)\ \oplus \ \Gamma^1_-(\cM)\wedge \xi\label{h2m}\\
&H^3(\hat{\mathcal{M}}_7)=H^3_+(\cM)\ \oplus \ H^2_-(\cM)\wedge \xi\ \oplus\ \Gamma^2_-(\cM)\wedge \xi
\end{align}
where $\xi$ is the harmonic vector of $S^1$ and $\Gamma^p_-(\cM)$ is a set of odd $p$-forms which are not globally well-defined in $\cM$. Hence, the $b_2(\hat{\mathcal{M}}_7)$ massless gauge bosons are mapped in the perturbative IIA orientifold limit to $b_2^+(\cM)$ closed string and $b_2(\hat{\mathcal{M}}_7)-b_2^+(\cM)$ D6-brane gauge bosons.\footnote{The later can be heuristically understood from expanding the NSNS 2-form $B_2$ in elements of $\Gamma^1_-(\cM)$.} Similarly, the $b_3(\hat{\mathcal{M}}_7)$ complex scalars correspond to $b_3^+(\cM)$ complex structure moduli, $b_2^-(\cM)$ K\"ahler moduli and $b_3(\hat{\mathcal{M}}_7)-b_3^+(\cM)-b_2^-(\cM)$ D6-brane moduli in the orientifold limit. Open and closed string U(1) gauge symmetries have therefore a common origin in M-theory, as anticipated. This unified description is also particularly useful for understanding open/closed string dualities. These occur when the $G_2$ manifold admits various perturbative limits of the form (\ref{iialimit}). In that case some RR and D6-brane U(1) gauge symmetries may appear exchanged at different type IIA orientifold limits \cite{Kachru:2001je}.

\section{Kinetic mixing with RR photons \label{kmixing}}

Given the two sets of massless U(1)'s described in the previous section, that is those arising from open and closed string degrees of freedom, it is natural to ask how they are related to each other. In particular, one may wonder if there is non-trivial kinetic mixing between them. The aim of this section is to provide a simple geometric expression for the gauge kinetic function $f_{ia}$ that mixes open and closed string U(1)'s
\begin{equation}
S_{\rm{4d,mix}}=-\int_{\IR^{1,3}}\left[\textrm{Re}(f_{ia})F_{\rm RR}^{i}\wedge *_4{F}_2^{a}+\textrm{Im}(f_{ia})F_{\rm RR}^{i}\wedge {F}_2^{a}\right]
\label{mix4d}
\end{equation}
where $F_{\rm RR}^{i} = dA^i$ and ${F}_2^{a} = dA^a$ are 4d field strengths for RR and D-brane U(1)'s, respectively.

A first hint on how $f_{ia}$ should look like comes from the Chern-Simons couplings of a single D6-brane to the RR potentials $C_5$ and $C_3$, encoded in the following action
\begin{eqnarray}
\label{CSD6}
S_{CS}&  = & \int_{\IR^{1,3} \times \pi_a} P\left[\mathcal{F}_2^{a} \wedge C_5  + \oh \mathcal{F}_2^{a} \wedge \mathcal{F}_2^{a} \wedge C_3 \right] \\ \nonumber
& = &  \int_{\IR^{1,3} \times \pi_a} \left[\mathcal{F}_2^{a} \wedge  \left(C_5 + \frac12\cl_{\phi_a} C_5 + \dots\right) + \oh  \mathcal{F}_2^{a} \wedge \mathcal{F}_2^{a}  \wedge \left(C_3 + \frac12\cl_{\phi_a} C_3 + \dots\right)\right]
\end{eqnarray}
where $\mathcal{F}^{a}_2\equiv F_2^{a}+B_2$, and $P[\ldots]$ denotes the pull-back to the worldvolume of the D6-brane. In the second line we have performed a Taylor expansion on a massless deformation (\ref{geod6}) of the D6-brane 3-cycle $\pi_a$, $\cl_{\phi_a}$ being the Lie derivative along such deformation. Following the computations of Appendix \ref{apdim} (see also \cite{Grimm:2011dx,Kerstan:2011dy}) one can dimensionally reduce such action to obtain an expression of the form (\ref{mix4d}) with
\begin{equation}
f_{ia}\, =\, - i \cam_{ij}^{a} \Phi^j_{a} + \dots
\label{kinmix}
\end{equation}
where we have dropped all terms beyond linear order in the D6-brane moduli $\Phi^j_{a}$, given by (\ref{openmoduli}). Finally we have defined
\begin{equation}
\mathcal{M}^{a}_{ij} \, \equiv\,
\int_{\pi_a} \omega_i \wedge \zeta_j\, =\, \int_{\rho_j} \omega_i
\label{kinint}
\end{equation}
with $\zeta_j$ a harmonic 1-form of $\pi_a$, $\rho_j \subset \pi_a$ its Poincar\'e dual 2-cycle and $\omega_i$ the Calabi-Yau 2-form related to the RR U(1). It is easy to check that $\cam_{ij}^{a}$ is a moduli-independent topological quantity, that vanishes unless some non-trivial 2-cycle $\rho_j$ of $\pi_a$ is also non-trivial in the Calabi-Yau $\cam_6$. More precisely, for $\cam_{ij}^{a}$ to be non-zero the 2-cycle $\rho_j$ should be a non-trivial element of $H_2^+(\cam_6,\IR)$, so that the rhs of (\ref{kinint}) does not vanish.

In fact, the kinetic mixing (\ref{kinmix}) is only well-defined up to a $\Phi$-independent term, related to the choice of 3-cycle $\pi_a$ within $[\pi_a]$ taken to describe the point $\phi_{a} = 0$. This ambiguity is however only present for massive D6-brane U(1)'s, while for those U(1)'s that are not lifted by the St\"uckelberg mechanism $f_{ia}$ is fully well-defined.\footnote{In order to fix this ambiguity for massive D6-brane U(1)'s one may resort to define a reference 3-cycle $\pi_a^0$ in the same homology class $[\pi_a]$, as in \cite{Grimm:2011dx,Kerstan:2011dy}. For the massless U(1)'s of interest for this paper such choice of reference 3-cycle is not needed.} Indeed, this is easily seen in the case of the adjoint Higgsing SU(2)$\times$U(1)$_a \raw$U(1)$_{a_1}\times$U(1)$_{a_2}$ discussed in the previous section, in which the massless combination U(1)$_{a_1}-$U(1)$_{a_2}$ is the one to be considered. Recall that the two 3-cycles $\pi_{a_1}$ and $\pi_{a_2}$ only differ by the vev of their moduli $\Phi_{a_i}$, and by consistency the kinetic mixing of RR fields with U(1)$_{a_1}-$U(1)$_{a_2}$ should vanish for $\Phi_{a_1} = \Phi_{a_2}$. We must then have
\begin{equation}
f_{i(a_1-a_2)}\, =\, - i \cam_{ij}^{a} (\Phi^j_{a_1} - \Phi^j_{a_2}) + \dots
\label{kinmixU2}
\end{equation}
without any $\Phi$-independent contribution. One may also see that in general this local expression translates into the more geometrical one\footnote{Indeed, both (\ref{kinmixU2}) and (\ref{kinmixU2chain}) have the same dependence with respect to the open string moduli $\Phi^j_{a_i}$, and both vanish for $\Phi_{a_1} = \Phi_{a_2}$. For further details see \cite{Hitchinrw,Grimm:2011dx,Kerstan:2011dy}.}
\begin{equation}
f_{i(a_1-a_2)}\, =\, - i \int_{\Sigma_4^{a_1-a_2}} (J_c + F_2^{a_1-a_2}) \wedge \omega_i
\label{kinmixU2chain}
\end{equation}
where $\Sigma_4$ is a 4-chain such that $\p \Sigma_4^{a_1-a_2} = \pi_{a_1} - \pi_{a_2}$, and we are identifying
\begin{equation}
 \int_{\Sigma_4^{a_1 - a_2}} F_2^{a_1-a_2} \wedge \omega_i \, =\, \int_{\p \Sigma_4^{a_1-a_2}} A^{a_1-a_2} \wedge \omega_i
 \equiv \int_{\pi_{a_1}} A^{a_1} \wedge \omega_i -  \int_{\pi_{a_2}} A^{a_2} \wedge \omega_i \, .
\end{equation}

It is now easy to generalize the expression (\ref{kinmixU2chain}) to any massless D6-brane U(1). Recall from the previous section that such U(1) can be characterized by a linear combination of 3-cycles $\pi_b = n^b_aN_a\pi_a$ trivial in homology, so that there exists a 4-chain $\Sigma_4^b$ such that $\partial \Sigma_4^b = \pi_b$. It is then natural to expect a kinetic mixing of the form
\begin{equation}
f_{ib}\, =\, - i \int_{\Sigma_4^{b}} (J_c + F_2^{b}) \wedge \omega_i
\label{kinmixchain}
\end{equation}
where again the integral over $F_2^{b} \wedge \omega_i$ should be understood as a surface integral
\begin{equation}
\int_{\Sigma_4^{b}} F_2^{b} \wedge \omega_i\, =\, \int_{\p \Sigma_4^b} A^{b} \wedge \omega_i\, .
\end{equation}

As before, this expression has the same $\Phi$-dependence as the linear combination $f_{ia}N_an_a^b$, with $f_{ia}$ given by (\ref{kinmix}), that one would obtain by expanding the CS action (\ref{CSD6}).
However, in (\ref{kinmixchain}) the $\Phi$-independent contribution to the kinetic mixing is fixed, up to a subtle point that we now describe. Given a boundary $\pi_b$, the 4-chain $\Sigma_4^b$ such that $\p \Sigma_4^b= \pi_b$ is defined only up to a 4-cycle $\pi_4$, since by definition $\p \Sigma_4^b = \p ( \Sigma_4^b + \pi_4)$. Each smooth 4-chain of the form $ \Sigma_4^b + \pi_4$ will then be equally valid to enter into the expression for the kinetic mixing and, if $\pi_4$ is non-trivial in the homology of $\cam_6$, then the $\Phi$-independent contribution to (\ref{kinmixchain}) will depend on the homology class $[\pi_4]$. More precisely, the kinetic mixing computed over $\Sigma_4^b$ or over $\Sigma_4^{b\, \prime} = \Sigma_4^b + \pi_4^j$, with $\pi_4^j$ the Poincar\'e dual to the 2-form $\omega_j$, will differ by $f_{ij} = -i \int_{\pi_4^j} J_c \wedge \omega_i$ where $f_{ij}$ is the mixing (\ref{gaugerr}) between two RR U(1)'s. Hence, it would seem that given an open string  massless U(1)$_b$ and its associated boundary $\pi_b$, the expression (\ref{kinmixchain}) gives a discrete set of possibilities for the kinetic mixing $f_{ib}$.

In practice, however, one is able to distinguish between all these choices from the physical context, so that no real ambiguity arises. Let us for instance consider the case where, by performing a loop in the open string moduli space, the initial 4-chain $\Sigma_4^b$ is deformed to $\Sigma_4^{b\, \prime} = \Sigma_4^b + \pi_4^j$. The kinetic mixing between open and closed string U(1)'s should then vary accordingly. That is
\begin{equation}
\Sigma_4^b\ \raw\ \Sigma_4^b + n \pi_4^j\quad \quad {\rm implies} \quad \quad f_{ib} \ \raw \ f_{ib} + n f_{ij} \quad \quad  n \in \IZ
\label{mono}
\end{equation}
with $n$ the number of loops that we have performed.
Such kind of behavior is well-known in $\cn=1$ string compactifications, where the closed string moduli space is fibered over the open string moduli space, and so performing certain loops on the D-brane moduli space is equivalent to shift the values of the closed string variables \cite{wolf, Camara:2009uv}. In the case at hand, performing loops is equivalent to redefine our U(1) sector. Namely,
\begin{equation}
\Sigma_4^b\ \raw\ \Sigma_4^b + n \pi_4^j\quad \quad {\rm is \ equivalent\ to} \quad \quad U(1)_{b} \ \raw U(1)_b + n U(1)_j \quad \quad  n \in \IZ
\label{monu}
\end{equation}
and so we deduce that the 4-chains $\Sigma_4^b$ and $\Sigma_4^b + n \pi_4^j$ correspond to two different U(1)'s, hence the discrepancy in their kinetic mixing with U(1)$_i$.

While the above discussion may seem slightly speculative, one may put it in firmer grounds by understanding the expression (\ref{kinmixchain}) from the viewpoint of its M-theory lift. Indeed, upon fibering the M-theory circle on the 4-chain $\Sigma_4^b$ it is easy to see that we should obtain a 5-cycle $\Lambda_5^\beta \subset \hat{\cam}_7$ related by Poincar\'e duality to some harmonic 2-form $\omega_\beta$ of the kind described in subsection \ref{msec}, and that corresponds to a massless U(1)$_\beta$. Hence, upon lifting our D6-brane configuration to M-theory we have to perform the replacements
\begin{equation}
\begin{array}{rcl}
U(1)_b & \raw & U(1)_\beta\\
\Sigma_4^b & \raw & \Lambda_5^\beta\\
J_c +F_2 & \raw & M^I \phi_I
\end{array}
\end{equation}
and so we obtain
\begin{equation}
f_{ib}\, \raw\, - i \int_{\Lambda_5^\beta} M^I \phi_I \wedge \omega_i \, =\, - i M^I \int_{\hat{\cam}_7} \phi_I \wedge \omega_i \wedge \omega_\beta \, =\, f_{i\beta}
\end{equation}
reproducing eq.(\ref{Mmixing}). Had we instead fibered the M-theory circle over the 4-chain $\Sigma_4^b + n \pi_4^j$, we would have ended up with a different 5-cycle $\Lambda_5^\g$ whose dual 2-form $\omega_\g$ is different from $\omega_\beta$. More precisely, it is easy to see that we should have $[\omega_\g] = [\omega_\beta] + n [\omega_j]$, from which the relation (\ref{monu}) follows.

Before closing this section, let us point out that the expression for the kinetic mixing (\ref{kinmixU2chain}) is quite similar to the one obtained for the open string superpotential of a D6-brane. Indeed, following \cite{martucci06} we have that
\begin{equation}
W_{\rm D6} \, =\, -\frac{i}{2} \int_{\Sigma_4^a} (J_c + F_2^{a}) \wedge (J_c + F_2^{a})
\label{supoD6}
\end{equation}
where $\Sigma_4^a$ is a 4-chain such that $\p \Sigma_4^a = \pi_a - \pi_a^0$, with $\pi_a^0$ a reference 3-cycle. Compared to the D6-RR gauge kinetic mixing (\ref{kinmixU2chain}), the D6-brane superpotential (\ref{supoD6}) is basically obtained from  performing  the replacement $\omega_i \raw T^{\hat{k}} \omega_{\hat{k}}$. Following our above discussion, we then see that a D6-brane may develop a non-trivial superpotential of the form (\ref{supoD6}) only if some of the 2-cycles $\rho_j$ of $\pi_a$ are non-trivial in the Calabi-Yau $\cam_6$ and, more precisely, if they are non-trivial elements of $H_2^-(\cam_6,\IR)$.

The similarities between $W_{\rm D6_a}$ and $f_{ia}$ are perhaps not that surprising since, from the unorientifolded $\cn=2$ perspective these two quantities are essentially the same one. In the same sense that (\ref{supoD6}) is known to be corrected by worldsheet instantons, we would expect that the kinetic mixing between D6-branes and RR photons is corrected as well. Computing such worldsheet corrections is however beyond the scope of the present paper.

\section{Mass mixing with RR photons \label{mmixing}}

In our description above, each RR photon arises from an RR potential whose internal profile is an harmonic wavefunction of the compactification manifold $\cam_6$. In this section we would like to argue that these are not the only RR U(1)'s of interest for phenomenology. There are less obvious RR symmetries, which from the 4d viewpoint can be understood as massive U(1)'s Higgsed down to $\IZ_k$ gauge symmetry by a St\"uckelberg mechanism, as in \cite{bs10}. In the following  we would like to argue that in Calabi-Yau compactifications such RR U(1)'s appear whenever the topology $\cam_6$ allows for torsional $p$-cycles and $p$-forms, by simply analyzing the 4d strings and particles that are charged under such discrete gauge symmetries. For simplicity, we first perform such analysis in the absence of orientifolds of D-branes. Remarkably, we find that when we include D-branes into the picture a mass mixing arises between certain open string U(1)'s and RR torsional U(1)'s, the massless U(1) being a linear combination of the two. We also analyze this effect from the viewpoint of M-theory, concluding that the discrete gauge symmetries of a 4d vacuum can be understood in terms of the torsional (co)homology groups of the M-theory compactification manifold $\hat{\cam}_7$. Finally, we provide an explicit example of a compactification where such mass mixing occurs, and which illustrates different mass mixing scenarios whose phenomenology will be analyzed in section \ref{pheno}.

\subsection{Torsion and discrete gauge symmetries}\label{gaugesym}

All along the above discussion, a key role has been played by the topology of the compactification manifold $\cam_6$. In particular, we have been able to derive rather general features of the 4d low energy effective action thanks to the fact that each object of the compactification corresponds to a topological class of $\cam_6$. Indeed, each massless mode of the closed string sector, including RR U(1)'s, corresponds to a harmonic $p$-form of $\cam_6$, and so to an element of the de Rham cohomology group $H^p(\cam_6, \IR)$. On the other hand, each D6-brane wrapping a sLag 3-cycle $\pi_3 \subset \cam_6$ corresponds to a non-trivial element of the homology group $H_3(\cam_6, \IR)$, while the non-trivial 2-cycles of $\pi_3$ may also be non-trivial elements of $H_2(\cam_6, \IR)$. The well-known relations between $H^p(\cam_6, \IR)$ and $H_p(\cam_6, \IR)$, namely the integrals of closed $p$-forms over $p$-cycles, allows then to compute the couplings between open and closed string sectors, and from there all the analysis follows.

Given this fact, one may wonder if that is all the topological information of $\cam_6$ that is relevant for the 4d effective action. After all, a $p$-cycle $\pi_p \subset \cam_6$ not only defines an element of $H_p(\cam_6, \IR)$, but rather one of
  the more fundamental group $H_p(\cam_6, \IZ)$. In general, $H_p(\cam_6, \IZ)$ contains more information than $H_p(\cam_6, \IR)$, the difference being the torsion homology groups Tor $H_p(\cam_6, \IZ)$, which are generated by $p$-cycles of $\cam_6$ with a $\IZ_k$ structure. As discussed below, a D-brane wrapping one of these torsion cycles cannot be detected by an element of $H^p(\cam_6, \IR)$ and so it is invisible to the closed string massless spectrum. It may however be detected by the {\em massive} closed string spectrum, and in particular by massive sectors of the theory related to a topological class of $\cam_6$. In the following, we would like to argue that this is indeed the case, and that in our setup the torsion groups of $\cam_6$ are related to massive RR U(1)'s Higgsed down to $\IZ_k$ gauge symmetries, as in the analysis of \cite{bs10}.

In general, the homology group $H_r(\cam_D, \IZ)$ of a $D$-dimensional K\"ahler manifold $\cam_D$ consists of a free part, given by $b_r$ copies of $\mathbb{Z}$, and a torsional part, given by a set of finite $\mathbb{Z}_k$ groups,
\begin{equation}
H_r(\cam_D, \IZ) \, =\, \underbrace{\mathbb{Z}\oplus\ldots\oplus\mathbb{Z}}_{b_r}\, \oplus\, \mathbb{Z}_{k_1} \oplus\ldots\oplus\mathbb{Z}_{k_n}
\label{homor}
\end{equation}
Here $b_r \equiv {\rm dim\, } H_r(\cam_D, \IR)$ stands for the $r^{\rm th}$ Betti number of $\cam_D$, which also counts the number of harmonic $r$-forms of $\cam_D$. The correspondence between elements of $\IZ^{b_r} \subset H_r(\cam_D, \IZ)$ and harmonic $r$-forms can be made via de Rham's and Hodge's theorems, and amounts to the fact that given a basis of $r$-cycles $\{\pi_r^j\}$ generating the lattice $\IZ^{b_r}$, one can construct a basis of harmonic $r$-forms $\{\omega_r^i\}$ such that $\int_{\pi_r^j} \omega_r^i = \delta_{ij}$.

The elements of Tor $H_r(\cam_D, \IZ) = \mathbb{Z}_{k_1} \oplus\ldots\oplus\mathbb{Z}_{k_n}$ are much harder to describe via differential geometry. A generator of $\IZ_{k}$ consist of a non-trivial $r$-cycle $\pi_r^{\rm tor}$ in the homology of $\cam_D$, but wrapping $k$ times $\pi_r^{\rm tor}$ corresponds to a trivial $r$-cycle. Otherwise said, $\pi_r^{\rm tor}$ is not the boundary of any $(r+1)$-chain on $\cam_D$, but we can always  construct a chain $\Sigma_{r+1} \subset \cam_D$ such that $\p \Sigma_{r+1} = k \pi_r^{\rm tor}$. This implies that the integral of any closed $r$-form $\omega_r$ over $\pi_r^{\rm tor}$ vanishes identically, since $\int_{\pi_r^{\rm tor}} \omega_r = k^{-1} \int_{\Sigma_{r+1}} d \omega_r = 0$. As a result, D-branes wrapped on torsional cycles of a Calabi-Yau $\cam_6$ cannot be detected by the 4d massless closed string modes, since the internal wavefunctions of the latter are described by harmonic $p$-forms. In addition, D-branes wrapping torsional 2, 3 and 4-cycles are necessarily non-BPS since their central charge, respectively measured by the integral of $J$, $\Omega$ and $J^2$ over them, also vanishes.\footnote{This is not necessarily true for type II flux compactifications on SU(3)-structure manifolds, where the forms $\Omega$ and $J$ are no longer necessarily closed \cite{ckt05,D6torsion}.\label{torfoot}}

While non-BPS, D-branes wrapping torsional $p$-cycles of $\cam_6$ are stable objects of the 4d effective theory, since they have discrete conserved charges. Let us consider type IIA string theory compactified on a manifold $\cam_6$ with torsional 3-cycles, and more precisely such that Tor $H_3(\cam_6, \IZ) = \IZ_k$. The relations between torsional groups discussed in the next subsection imply that Tor $H_2(\cam_6, \IZ) = \IZ_k$ as well. Hence, together with a $k$-torsional 3-cycle $\pi_3^{\rm tor}$ we will always have a $k$-torsional 2-cycle $\pi_2^{\rm tor}$ within $\cam_6$. Let us now wrap a D2-brane around $\pi_2^{\rm tor}$, seen in 4d as a massive particle, and a D4-brane around $\pi_3^{\rm tor}$, seen in 4d as a massive string. Both 4d objects are non-BPS but nevertheless stable, at least mod $k$. That is, it is possible that $k$ D-strings combine and disappear, but this can only happen in groups of $k$, and not for less than $k$ D-strings. Note that this property has also been observed from a 4d field theory viewpoint in strings dubbed as Aharanov-Bohm strings in \cite{ABstrings} and $\IZ_k$ strings in \cite{bs10}, and which are associated to a U(1) gauge symmetry broken down to $\IZ_k$ via a St\"uckelberg mechanism. In fact, the main property of these strings is that certain particles, also stable mod $k$, can detect a non-trivial $\IZ_k$ holonomy when circling around the string. As we will now show, that property is precisely reproduced by those 4d particles and strings that arise from wrapping D-branes on torsion cycles of $\cam_6$.

Indeed, let us again consider a D4-brane on $\IR^{1,1} \times \pi_3^{\rm tor}$ and a D2-brane wrapped on $\pi_2^{\rm tor}$ and performing a closed loop $\g\subset \IR^{1,3}$ around the 4d D-string. The phase picked up by our D-particle upon performing such loop reads
\begin{equation}
{\rm hol}(\g) \, =\, {\rm exp\, } \left(2\pi i \int_{\g \times \pi_2^{\rm tor}} C_3 \right) \, =\, {\rm exp\, } \left(2\pi i \int_{D \times \pi_2^{\rm tor}} F_4 \right)
\label{holo4}
\end{equation}
where $F_4 = dC_3$ is the RR field strength sourced by the D4-brane. In particular, we have that $dF_4 = \delta_5$, with $\delta_5$ a $\delta$-like 5-form concentrated around $\IR^{1,1} \times \pi_3^{\rm tor}$ and with components transverse to it. Finally, $D \subset \IR^{1,3}$ is given by a disk such that $\p D = \g$ and it intersects the 4d D-string once.

As the holonomy (\ref{holo4}) is an observable 4d quantity, it should not depend on the precise embedding of $\pi_2^{\rm tor}$. In particular, (\ref{holo4}) should not vary if we perform a continuous deformation of the 2-cycle $\pi_2^{\rm tor}$ or if we pick a different representative $\pi_2^{{\rm tor}\, \prime}$ within the homology class $[\pi_2^{\rm tor}] \in H_2(\cam_6, \IZ)$. Indeed, a D2-brane wrapped on any representative of $[\pi_2^{\rm tor}]$ is supposed to represent the same kind of D-particle in 4d, and so the holonomy (\ref{holo4}) for any of them should be the same. One can check this by considering another D2-brane wrapping $\pi_2^{{\rm tor}\, \prime}$ and performing the same 4d loop $\g$. Let us denote the phase picked by this D-particle by ${\rm hol }'(\g)$. Since $[\pi_2^{{\rm tor}\, \prime}] = [\pi_2^{{\rm tor}}]$, we can construct a 3-chain $\Sigma_3 \subset \cam_6$ such that $\p \Sigma_3 = \pi_2^{{\rm tor}\, \prime} - \pi_2^{{\rm tor}}$. We then have that ${\rm hol}'(\g) = e^{2\pi i n} {\rm hol}(\g)$, with
\begin{equation}
n\, =\, \int_{D \times \pi_2^{{\rm tor}  \prime}} F_4 - \int_{D \times \pi_2^{{\rm tor}}} F_4\, =\, \int_{D \times \Sigma_3} \delta_5
\label{linkingn}
\end{equation}
where we have applied Stockes' theorem. It is easy to see that the rhs of (\ref{linkingn}) is an integer, more precisely a product of signed intersections: $n = \# (\IR^{1,1} \cap D) \cdot \# (\pi_3^{\rm tor} \cap \Sigma_3)$. Hence, we deduce that ${\rm hol}'(\g) = {\rm hol}(\g)$ as expected from four-dimensional grounds. It does then make sense to denote the holonomy (\ref{holo4}) as ${\rm hol}(\g, [\pi_2^{\rm tor}])$.

Let us now consider the case where the D-particle above performs $k$ times the loop $\g$. Since $k\g \times \pi_2^{\rm tor}$ is the same integration domain as $\g \times k \pi_2^{\rm tor}$ we have that
\begin{equation}
\left[{\rm hol}(\g, [\pi_2^{\rm tor}])\right]^k \, \equiv\, {\rm hol}(k \g, [\pi_2^{\rm tor}])\, =\, {\rm hol}(\g, [k \pi_2^{\rm tor}])\, =\, 1
\label{holo4b}
\end{equation}
where we have used the fact that $[k\pi_2^{\rm tor}]$ is trivial in the homology of $\cam_6$ and so, by the discussion above, its holonomy should be trivial. Hence, we deduce that ${\rm hol}(\g, [\pi_2^{\rm tor}])$ should be a $k^{\rm th}$ root of unity, just like for the Aharanov-Bohm strings of \cite{ABstrings,bs10}.

In fact, we can be more precise about ${\rm hol}(\g, [\pi_2^{\rm tor}])$. Notice that
\begin{equation}
\frac{1}{2\pi i}{\rm log\, } \left[{\rm hol}(\g, [\pi_2^{\rm tor}])\right] \, \stackrel{{\rm mod\, } 1}{=}\, \frac{1}{k} \int_{D \times k\pi_2^{\rm tor}} F_4\, =\, \frac{1}{k} \int_{D \times \Sigma_3} \delta_5\,= \, \frac{p}{k}
\label{holo4c}
\end{equation}
where $\Sigma_3$ is a 3-chain such that $\p \Sigma_3 = k\pi_2^{\rm tor}$. Again, $p \in \IZ$ since it can be defined as the product of transverse intersections $\#(\IR^{1,1} \cap D)\cdot\#(\pi_3^{\rm tor} \cap\Sigma_3)$. By construction $\#(\IR^{1,1} \cap D) = 1$, while $\#(\pi_3^{\rm tor} \cap\Sigma_3)$ is (mod 1) the exact definition of the torsion linking form $L([\pi_2^{\rm tor}], [\pi_3^{\rm tor}])$: a topological invariant used to classify manifolds with torsion, and which is the equivalent of the intersection product for non-torsional cycles \cite{bt24,spanier}.

Recall that the intersection product $I([\pi_r],[\pi_{D-r}]) = [\pi_r] \cdot [\pi_{D-r}]$ is a bilinear form between a $r$ and a $(D-r)$-cycle of $\cam_D$, which only depends on the homology class of each cycle. Similarly, the torsion linking form $L([\pi_r^{\rm tor}],[\pi_{D-r-1}^{\rm tor}])$ is a bilinear form between torsional cycles of $\cam_D$ that only depends on their homology classes, and that is symmetric for $D = {\rm even}$. In our setup, such quantity not only computes the holonomy of a torsional D-particle around a torsional D-string, but also the holonomy of a torsional D-string around a torsional D-particle.

Indeed, let us consider a D4-brane wrapping $\pi_3^{\rm tor}$ and whose 4d worldsheet sweeps a two-sphere $S^2 \subset \IR^{1,3}$ that surrounds our torsional D-particle. Similarly to (\ref{holo4c}) we obtain that the holonomy for such D-string is given by
\begin{equation}
\frac{1}{2\pi i}{\rm log\, } \left[{\rm hol}(S^2, [\pi_3^{\rm tor}])\right] \, \stackrel{{\rm mod\, } 1}{=}\, \frac{1}{k} \int_{B \times k\pi_3^{\rm tor}} F_6\, =\, \frac{1}{k} \int_{B \times \Sigma_4} \delta_7\,\stackrel{{\rm mod\, } 1}{\equiv} \, L([\pi_2^{\rm tor}],[\pi_{3}^{\rm tor}])
\label{holo4d}
\end{equation}
where $B \subset \IR^{1,3}$ is a 3-ball such that $\p B = S^2$, $\Sigma_4 \subset \cam_6$ is a 4-chain with $\p \Sigma_4 = k \pi_3^{\rm tor}$,
and $F_6$ is the RR field strength sourced by the D2-brane, so that $dF_6 = \delta_7$ is a $\delta$-like 7-form on $\IR \times \pi_2^{\rm tor}$.

To sum up we have shown that, in compactification manifolds $\cam_6$ with torsional cycles, Aharanov-Bohm strings and particles appear in the 4d effective theory. The fractional holonomies that such strings and particles induce on each other is controlled by a topological invariant of $\cam_6$, namely the torsion linking number $L([\pi_r^{\rm tor}],[\pi_{6-r-1}^{\rm tor}])$. As shown in \cite{bs10}, such kind of Aharanov-Bohm strings are the smoking gun for a set of discrete gauge symmetries in 4d field theories, which arise from a massive U(1) gauge symmetry higgsed down to $\IZ_k$. As is easy to infer from our discussion, one should have a different kind of Aharanov-Bohm string for each $\IZ_{k_i}$ factor in (\ref{homor}), and so we would expect to also have a massive U(1) for each of these factors. This will be our working assumption in the following and, as we will see, several non-trivial consequences can be derived from it.

\subsection{Massive RR U(1)'s from torsion}
\label{massivesec}

Let us now explore the implications of having a massive U(1) for each generator of Tor $H_3(\cam_6, \IZ)$, which is where the Aharanov-Bohm D-strings were constructed from. Notice that our discussion above was carried in the context of type IIA string theory compactified on a Calabi-Yau 3-fold $\cam_6$, without the need of any orientifold projection or $\IR^{1,3}$-filling D6-branes. In the following we will continue to assume such class of 4d $\cn=2$ compactifications, leaving the effect of the orientifold projection for the end of this subsection.

If Aharanov-Bohm strings and particles arise from wrapping D$p$-branes on elements of Tor $H_r(\cam_6, \IZ)$, then massive U(1)'s should arise from reducing RR $p$-forms in elements of Tor $H^r(\cam_6, \IZ)$. That is, one should expand the RR potentials $C_p$ in the torsional analogues of the harmonic forms of section \ref{closedU1}, which should moreover be eigenvectors of the Laplacian $\nabla^2 = dd^\dag + d^\dag d$. Constructing such torsional analogues of harmonic forms is quite similar to finding an appropriate basis of $p$-forms to perform dimensional reduction on $SU(3)$-structure manifolds \cite{glw05,ckt05,Benmachiche:2006df,minasian}, since both problems deal with $p$-forms that are invisible to de Rham cohomology and correspond to the internal profile of massive 4d modes.

From the viewpoint of de Rham cohomology $H^r (\cam_D, \IR)$, a torsional $r$-form $\a_r^{\rm tor}$ of a manifold $\cam_D$ is trivial. Given an $r$-form $\a_r^{\rm tor}$ that represents a torsional element $[\a_r^{\rm tor}] \in$ Tor $H^r(\cam_D, \IZ) = \IZ_k$ we should have $\int_{\pi_r} \a_r^{\rm tor} = 0$ for any $r$-cycle $\pi_r$ of $\cam_D$, for the same reason that integrals of closed forms over torsional cycles vanish. Hence, such form can be written as
\begin{equation}
k \a_r^{\rm tor}\, = \, d \omega_{r-1}^{\rm tor}
\label{torform}
\end{equation}
with $\omega_{r-1}^{\rm tor}$ a globally well-defined $(r-1)$-form, and $k \in \IZ$ such that $k \a_r^{\rm tor}$ is trivial also in Tor $H^r(\cam_D, \IZ)$. Since $\omega_{r-1}^{\rm tor}$ is globally well-defined, we can expand an RR potential $C_p$ on it.\footnote{In fact, since we are dealing with RR potentials, we should think of $\a_r^{\rm tor}$ as a gerbe. Then it is no longer true that $\omega_{r-1}^{\rm tor}$ is globally well-defined but exp$(2\pi i \int_{\pi_{r-1}} \omega_{r-1}^{\rm tor})$ must be so for any cycle $\pi_{r-1}$, which is enough for our purposes. See Appendix \ref{tortopo} for further details.} Indeed, we will argue below that both $\a_r^{\rm tor}$ and $\omega_{r-1}^{\rm tor}$ are related to an isolated set of massive modes of the compactification and so, in a spirit similar to \cite{minasian}, we will demand that the set of representatives $\{\omega_{r-1}^{\rm tor}\}$ should be closed under the action of the Laplacian, as in eq.(\ref{eigensp}). For concreteness, we will denote by $\widehat{\textrm{Tor}}\ H^{r-1}$ the set $\{\omega_{r-1}^{\rm tor}\}$ of non-closed forms which describe such 4d massive modes.

Let us now relate this set of forms to the torsional cycles of a compactification. For this one needs to make use of Poincar\'e duality \cite{Munkres30}
\begin{equation}
H_r (\cam_D, \IZ) \simeq H^{D - r} (\cam_D, \IZ)
\label{poincared}
\end{equation}
as well as of the universal coefficient theorem \cite{bt24}
\begin{equation}
{\rm Tor}\, H_r (\cam_D, \IZ) \simeq {\rm Tor}\, H^{r+1} (\cam_D, \IZ)
\label{unitheorem}
\end{equation}
For a six-dimensional manifold $\cam_6$, these two results imply that the only two finite groups that describe torsional classes in $\cam_6$ are
\begin{equation}
{\rm Tor}\, H_3 (\cam_6, \IZ) \simeq {\rm Tor}\, H_2 (\cam_6, \IZ) \simeq {\rm Tor}\, H^4 (\cam_6, \IZ) \simeq {\rm Tor}\, H^3 (\cam_6, \IZ)
\label{together}
\end{equation}
and
\begin{equation}
{\rm Tor}\, H_1 (\cam_6, \IZ) \simeq {\rm Tor}\, H_4 (\cam_6, \IZ) \simeq {\rm Tor}\, H^2 (\cam_6, \IZ) \simeq {\rm Tor}\, H^5 (\cam_6, \IZ)
\label{together2}
\end{equation}
We will be mainly interested in (\ref{together}), since ${\rm Tor}\, H_3 (\cam_6, \IZ)$ classifies Aharanov-Bohm (AB) strings built from D4-branes, and ${\rm Tor}\, H_2 (\cam_6, \IZ)$ dual 4d particles from wrapped D2-branes.

Given a torsion homology group
\begin{equation}
{\rm Tor}\, H_3 (\cam_6, \IZ) \, =\, \mathbb{Z}_{k_1} \oplus\ldots\oplus\mathbb{Z}_{k_n} \, =\, {\rm Tor}\, H_2 (\cam_6, \IZ)
\label{torH3}
\end{equation}
then by our previous discussion we have $n$ different kinds of 4d AB-strings and particles. In addition we will also have $4n$ forms in which the RR potentials $C_3$ and $C_5$ can be reduced. In order to describe (\ref{torH3}) such forms will satisfy the relations
\begin{equation}\label{formbasis}
d \omega^{\rm tor}_\a \, =\, k_\a{}^\b \a_\b^{\rm tor} \quad \quad \quad  \quad \quad \quad d \beta^{{\rm tor}, \b}  \, =\, - k^\b{}_\a \tilde{\omega}^{{\rm tor}, \a}
\end{equation}
where $k_\a{}^\b \in \IZ$, $\a,\b = 1, \dots, n$ is an invertible symmetric matrix,  and
\begin{align*}
[\a_\a^{\rm tor}] &\in {\rm Tor}\, H^3 (\cam_6, \IZ) & [\tilde{\omega}^{{\rm tor}, \a}] &\in {\rm Tor}\, H^4 (\cam_6, \IZ)\\
\omega_\a^{\rm tor} &\in \widehat{\textrm{Tor}}\ H^{2} &  \beta^{{\rm tor}, \a} &\in \widehat{\textrm{Tor}}\ H^{3} \nonumber
\end{align*}
The numbers $k_\a$ in (\ref{torH3}) will constrain the choice of $k_\a{}^\b$, having $k_\a = k_\a{}^\a$ if $k$ is diagonal. For a matrix $k$ with off-diagonal entries, $k_\a$ is the smallest integer such that $k_\a (k^{-1})^\a{}_\b \in \IZ,\, \forall \b$. As discussed in appendix \ref{tortopo} in this formalism the torsion linking form is given by
\begin{equation}
L_{\a}{}^{\b} = L([\pi_{2,\a}^{{\rm tor}}],[\pi_{3}^{{\rm tor},\b}]) \, =\, (k^{-1})_\a{}^\b
\label{tlnform}
\end{equation}
and the integrals of these forms satisfy
\begin{equation}
\int_{\cam_6} \alpha_{\rho}^{\rm tor} \wedge \b^{{\rm tor}, \sigma}\, =\,   \int_{\cam_6} \omega_{\rho}^{\rm tor} \wedge \tilde{\omega}^{{\rm tor}, \sigma}\, =\, \delta^\sigma_\rho
\label{normtor}
\end{equation}
being the analogue of (\ref{norm}) for torsional cohomology.

Clearly, this set of forms are the torsional analogues of the forms $\a_I$, $\omega_i$, $\tilde{\omega}^i$ and $\beta^I$ that were introduced in section \ref{U1typeIIA} in order to dimensionally reduce $C_3$ and $C_5$. Performing the same kind of expansion in the present basis
\begin{eqnarray}
\label{torexpC3}
C_3 &  = & \sum_\a \textrm{Re}(N^\a) \alpha^{\rm tor}_\a +  {A}^\a\wedge\omega^{\rm tor}_\a\\
C_5 & = & \sum_{\a} C_2^{\a}\wedge \beta^{{\rm tor}, \a} +  {V}^\a\wedge\tilde{\omega}^{{\rm tor}, \a}
\label{torexpC5}
\end{eqnarray}
we obtain $n$ pairs of electric and magnetic 4d RR U(1) gauge bosons $(A^\a, V^\a)$, as well as a set of $n$ axions $\textrm{Re}(N^\a)$ and 2-forms $C_2^\a$. These 4d modes are massive, since they correspond to a massive U(1)$^n$ gauge symmetry broken down to the discrete subgroup (\ref{torH3}). In particular, $A^\a$ are the electric gauge bosons which couple to $\IZ_{k_\a}$ particles, while $C_2^\a$ are the 2-forms coupled to the dual AB strings. Note that $\omega^{\rm tor}_\a$ and $\beta^{{\rm tor}, \a}$ are non-closed forms and so, unlike harmonic forms, they can have non-zero integrals over torsional cycles $\pi_2^{\rm tor}$ and $\pi_3^{\rm tor}$, respectively.

Since $\{\omega^{\rm tor}_\a\}$ and $\{\beta^{{\rm tor}, \a}\}$ should produce a well-defined massive 4d sector, we should impose that they are eigenvectors of the Laplacian of $\cam_6$ or, more generally, that they generate a vector space closed under the action of $\nabla^2 = dd^\dag +d^\dag d$. That is, we require that
\begin{equation}
\nabla^2 \omega^{\rm tor}_\a \, =\, - M_{Pl}^2\, {\bf M}_\a{}^\b  \omega^{\rm tor}_\b\quad \quad \quad
\nabla^2 \beta^{{\rm tor}, \a} \, =\, - M_{Pl}^2\, {\bf \tilde{M}}^{\a}{}_\b  \beta^{{\rm tor}, \b}
\label{eigensp}
\end{equation}
with ${\bf M}$ and ${\bf \tilde{M}}$ constant matrices. Then, because $[\nabla^2, d] = 0$, we also have that
\begin{equation}
\nabla^2 \a^{\rm tor}_\a \, =\, - M_{Pl}^2\, (k^{-1}\cdot{\bf M}\cdot k)_\a{}^\b  \a^{\rm tor}_\b\quad \quad \quad
\nabla^2 \tilde{\omega}^{{\rm tor}, \a} \, =\, - M_{Pl}^2\, (k^{-1}\cdot{\bf \tilde{M}}\cdot k)^{\a}{}_\b  \tilde{\omega}^{{\rm tor}, \b}
\label{eigensp2}
\end{equation}
These two mass matrices are actually related to each other, since plugging (\ref{eigensp}) and (\ref{eigensp2}) into (\ref{normtor}) we obtain that ${\bf M} = k\cdot{\bf \tilde{M}}\cdot k^{-1}$. Finally, it is useful to define the quantities
\begin{equation}
\check{f}_{\alpha\beta}\equiv\int_{\mathcal{M}_6}\omega_\alpha^{\rm tor}\wedge*_6\, \omega_\beta^{\rm tor}\ ,\qquad \check{\mathcal{G}}_{\alpha\beta}^{-1}\equiv M_{Pl}^2\int_{\mathcal{M}_6}\alpha_\alpha^{\rm tor}\wedge*_6\,\alpha_\beta^{\rm tor}
\label{fg}
\end{equation}
which satisfy the relation
\begin{equation}
\check{\mathcal{G}}_{\alpha\beta}^{-1}\, =\, (k^{-1}\cdot {\bf M}\cdot \check{f}\cdot k^{-1})_{\a\b}
\label{relfg}
\end{equation}

Let us now show that all these geometric relations provide a consistent effective field theory, and in particular the 4d field theory Lagrangian describing discrete gauge symmetries put forward in \cite{bs10}. From (\ref{torexpC3}) and (\ref{formbasis}) we have
\begin{equation}
dC_3\, =\, [\textrm{Re}(dN^\b)+k^\b{}_{\alpha}{A}^\alpha]\wedge \alpha_\b^{\rm tor} + d{A}^\alpha\wedge\omega_\alpha^{\rm tor}
\end{equation}
Plugging this expression into the $C_3$ 10d kinetic term $\int F_4 \wedge *_{10} F_4$, and integrating over $\cam_6$ we obtain the 4d Lagrangian density
\begin{equation}
\mathcal{L}^{\rm tor}_{\rm Stk}=\frac12 e^{2\phi_4}\check{\mathcal{G}}_{\alpha\beta}^{-1}\textrm{Re}(DN^\alpha)\wedge *_4\textrm{Re}(DN^\beta)\ , \qquad DN^\b=dN^\b+k^\b{}_{\alpha}{A}^\alpha
\label{stucktor}
\end{equation}
which indeed corresponds to a St\"uckelberg Lagrangian for $n$ RR massive U(1)'s, as in \cite{bs10}. Note that the rather abstract relations described in the context of torsional cohomology acquire an elegant physical interpretation in the context of massive RR U(1)'s. In particular, we observe that relations (\ref{together}) ensure an equal number of electric and magnetic degrees of freedom, whereas eqs.(\ref{formbasis}) provide a one-to-one correspondence between massive axions and massive vector bosons. Finally, the universal coefficient theorem, eq.(\ref{unitheorem}), sets a correspondence between charges of 4d particles and U(1) gauge symmetries.

From (\ref{stucktor}) one can read the mass matrix for the gauge bosons with canonically normalized kinetic terms, which is as expected given by $M_{Pl}^2 {\bf M}$. The quantities $\check{\mathcal{G}}_{\alpha\beta}$ and $\check{f}_{\alpha\beta}$ defined in (\ref{fg}) are the torsional analogues of (\ref{gij}) and (\ref{gaugerr}), respectively. The mass of a RR U(1) gauge boson is thus controlled by the ratio between some combination of complex structure moduli and some combination of K\"ahler moduli, a rough estimation being
\begin{equation}
m^2_{RR}\simeq \frac{M_{Pl}^2}{\textrm{Vol}_{\textrm{3-cycle}}^2\textrm{Vol}_{\textrm{2-cycle}}}
\end{equation}
where $\textrm{Vol}_{\textrm{3-cycle}}$ and $\textrm{Vol}_{\textrm{2-cycle}}$ are the typical volumes of torsional 3- and 2-cycles, measured in string units. In particular, for regions of the moduli space where the volume of the 2-cycle becomes large,  the RR U(1) vector boson can become light as compared to massive D6-brane gauge bosons.

While the above discussion is carried in the context of type IIA 4d $\cn=2$ Calabi-Yau compactifications, one can easily adapt the above results to include the presence of an orientifold projection. Indeed, recall from section \ref{U1typeIIA} that due to the orientifold parity of $C_3$ and $C_5$, massless RR U(1) gauge bosons are associated to $\sigma$-even harmonic 2-forms $\omega_i$ and $\sigma$-odd harmonic 4-forms $\tilde{\omega}^i$, classified by the groups $H^2_+(\cam_6, \IR)$ and $H^4_-(\cam_6, \IR)$. Similarly, in orientifold compactifications $\IZ_k$ discrete gauge symmetries are classified by the torsion groups
\begin{equation}
{\rm Tor}\, H_3^- (\cam_6, \IZ) \simeq {\rm Tor}\, H_2^+ (\cam_6, \IZ) \simeq {\rm Tor}\, H^4_- (\cam_6, \IZ) \simeq {\rm Tor}\, H^3_+ (\cam_6, \IZ)
\label{togetherori}
\end{equation}
rather than by (\ref{together}). In fact, when we consider the whole set of closed string degrees of freedom that may give rise to a 4d massive U(1) symmetry via reduction on torsional $p$-forms, much more possibilities appear. We have summarized in Table \ref{tabla1} the 10d origin of the electric degrees of freedom of massive closed string U(1) vector bosons for type IIA Calabi-Yau orientifold compactifications. We also give the 10d origin of the particles which are charged electrically under these U(1)'s, and of the axions which mediate the St\"uckelberg mechanism giving masses to the vector bosons. Similarly, in Table \ref{tabla2} we present the dual magnetic degrees of freedom and 2-forms.

\begin{table}[!ht]
\begin{center}
\begin{tabular}{|c|c||c|c||c|c|}
\hline
U(1)$_{elec.}$ & group & charged particles & cycle & axions & group \\
\hline \hline
$g^m{}_\mu$&$\widehat{\textrm{Tor}}\ H^{1}_+$& $P$& $\textrm{Tor}\ H_{1}^{+}$ & $g_{ij}$&$\textrm{Tor}\ H^{2}_+$ \\
\hline
$B^m{}_\mu$&$\widehat{\textrm{Tor}}\ H^{1}_-$& $F1$ & $\textrm{Tor}\ H_{1}^{-}$ & $B_{ij}$&$\textrm{Tor}\ H^{2}_-$ \\
\hline
$C_{\mu}{}^{mn}$&$\widehat{\textrm{Tor}}\ H^{2}_+$& $D2$ & $\textrm{Tor}\ H_{2}^{+}$ & $C_{ijk}$&$\textrm{Tor}\ H^{3}_+$ \\
\hline
$C_{\mu}{}^{mnop}$& $\widehat{\textrm{Tor}}\ H^{4}_-$ & $D4$ & $\textrm{Tor}\ H_{4}^{-}$ &$C_{ijklm}$& $\textrm{Tor}\ H^{5}_-$ \\
\hline
\end{tabular}
\caption{\small{Complete set of massive closed string gauge symmetries and charged states in weakly coupled type IIA Calabi-Yau orientifold compactifications. $P$ denotes the gravity wave and $F1$ the fundamental string. We present also the axions which mediate the St\"uckelberg mechanism giving masses to the corresponding vector boson.}\label{tabla1}}
\end{center}
\end{table}

\begin{table}[!ht]
\begin{center}
\begin{tabular}{|c|c||c|c||c|c|}
\hline
U(1)$_{mag.}$ & group & charged strings & cycle & $C_2^{I}$ & group \\
\hline \hline
$KK_{\mu}{}^{mnopq}$&$\textrm{Tor}\ H^{5}_-$ & $KK$ & $\textrm{Tor}\ H_{4}^{-}$  &$KK_{\mu\nu}{}^{ijkl}$& $\widehat{\textrm{Tor}}\ H^{4}_-$ \\
\hline
$B_{\mu}{}^{mnopq}$& $\textrm{Tor}\ H^{5}_+$ & $NS5$ & $\textrm{Tor}\ H_{4}^{+}$ &$B_{\mu\nu}{}^{ijkl}$& $\widehat{\textrm{Tor}}\ H^{4}_+$ \\
\hline
$C_{\mu}{}^{mnop}$& $\textrm{Tor}\ H^{4}_-$ & $D4$ & $\textrm{Tor}\ H_{3}^{-}$ &$C_{\mu\nu}{}^{ijk}$& $\widehat{\textrm{Tor}}\ H^{3}_-$ \\
\hline
$C_{\mu}{}^{mn}$&$\textrm{Tor}\ H^{2}_+$& $D2$ & $\textrm{Tor}\ H_{1}^{+}$ & $C_{\mu\nu}{}^{i}$&$\widehat{\textrm{Tor}}\ H^{1}_+$ \\
\hline
\end{tabular}
\caption{\small{Dual U(1) magnetic degrees of freedom and 2-forms mediating the St\"uckelberg mechanism. $KK$ denotes the Kaluza-Klein monopole.}\label{tabla2}}
\end{center}
\end{table}

Needless to say, for each massive U(1) the identities (\ref{poincared}) and (\ref{unitheorem}) (or rather their orientifold version) insure that the degrees of freedom arising from torsional groups arrange into complete $\cn=1$ massive vector multiplets.\footnote{Beside axions and gauge bosons, these multiplets contain scalars that control a FI-term. For the case of vector multiplets that arise from expanding $C_3$ as in (\ref{torexpC3}), such scalars parametrize massive deformations of the metric that spoil the Calabi-Yau condition. In order to write down the corresponding FI-terms we need to expand $\Omega$ in elements of $\widehat{\textrm{Tor}}\ H^{3}_-$, obtaining
\begin{equation*}
\frac{\xi_\alpha}{g_\alpha^2}\simeq \int_{\cM} \im(d\Omega) \wedge \alpha_\alpha^{\rm tor}\ , \qquad \alpha_\alpha^{\rm tor}\in\textrm{Tor}\ H^{3}_+
\end{equation*}
which vanishes because of the Calabi-Yau condition $d\Omega = 0$.}
The total number of massive closed string vector multiplets in a type IIA Calabi-Yau orientifold compactification is therefore
\begin{equation}
\# \ \textrm{torsional U(1)'s} \ = \ \textrm{dim}\left(2\textrm{Tor}\ H^{2}_+\oplus\textrm{Tor}\ H^{2}_-\oplus\textrm{Tor}\ H^{3}_+\right)
\end{equation}
Notice that some of these U(1) symmetries may actually correspond to massive graviphotons. In that case even in the presence of the orientifold, 4d $\mathcal{N}\geq 2$ supersymmetry is approximately recovered at points near the boundary of the moduli space where these vector states become light. Compactifications of this type were intensively studied for instance in \cite{Vafa:1995gm, Kiritsis:1997ca}, and the particular example of section \ref{example} belongs to this class.

From this point of view $\cn=2$ and $\cn=1$ CY$_3$ orientifold compactifications do not seem so different, since in order to describe $\cn=1$ massive U(1) sectors we just need to perform an orientifold projection of the $\cn=2$ spectrum. The latter turns out to be a naive statement, in particular for those compactifications that contain D-branes. Indeed, just like torsional Aharanov-Bohm D-strings, space-time filling D-branes wrapping torsional cycles can detect torsional U(1) symmetries. Hence, in the presence of such open string sectors which U(1) symmetries are massless and which ones are massive needs to be reconsidered, as we now proceed to describe.

\subsection{The St\"uckelberg mechanism revisited}
\label{stuckrev}

Once that we consider type II orientifold compactifications we should also consider space-time filling D-branes. An obvious question is therefore whether such D-branes feel the presence of torsion in homology. In particular, in type IIA CY$_3$ orientifold compactifications space-time filling D6-branes wrap 3-cycles of the compactification manifold $\cam_6$, for which we assume a torsion group Tor $H_3$ of the form (\ref{torH3}). Of course, if aiming for a 4d $\cn=1$ compactification one would never wrap a D6-brane in a purely torsional 3-cycle since (for $\cam_6$ a Calabi-Yau, c.f. footnote \ref{torfoot}) it would be automatically non-BPS. However, recall from section \ref{U1typeIIA} that open string U(1) gauge symmetries are not associated to a particular 3-cycle, but rather to a formal sum of them.
More precisely, we saw there that each massless open string U(1) should be related to a linear combination of 3-cycles $\pi_b^-$ which is trivial in $H_3(\cam_6, \IR)$ or, otherwise said, the integral of any harmonic 3-form of $\cam_6$ vanishes over $\pi_b^-$.
But from our discussion above it is easy to see that this does not imply that $\pi_b^-$ in (\ref{combi3-}) is trivial in the more fundamental group $H_3(\cam_6, \IZ)$: $[\pi_b^-]$ could still be a non-trivial element of ${\rm Tor\, } H_3(\cam_6, \IZ)$.

In the following we would like to argue that if $\pi_b^-$ is non-trivial in torsional homology (more precisely if $[\pi_b^-]$ is non-trivial in ${\rm Tor\, } H_3^-(\cam_6, \IZ)$) then the corresponding open string U(1) will not be free of $\int _{\IR^{1,3}}C_2 \wedge F$ couplings that mediate the St\"uckelberg mechanism. Instead, a St\"uckelberg coupling will be generated with the 2-forms $C_2^\a$ in the expansion (\ref{torexpC5}) of the RR potential $C_5$. As a result, $[\pi_b^-]$ should be a trivial 3-cycle in $H_3(\cam_6, \IZ)$ for an open string U(1)$_b$ to be massless. If it is only trivial in $H_3(\cam_6, \IR)$ but not in ${\rm Tor\, } H_3^-(\cam_6, \IZ)$ then a mass mixing term will be generated with torsional RR U(1)'s, and the massless U(1) will be given by a linear combination U(1)$_b + \sum_\a n_\a U(1)_\a$, where U(1)$_\a$ are the RR U(1)'s.

In order to argue for such class of St\"uckelberg couplings let us consider a D4-brane wrapping a 3-cycle $\pi_3^{\rm tor}$ homologous to $\pi_b^-$. This setup is precisely the one considered in section \ref{gaugesym}, up to the orientifold projection whose effect amounts to consider the torsion groups (\ref{togetherori}) instead of (\ref{together}). As these D-strings are the 4d $\IZ_k$ strings of \cite{bs10}, their 4d worldsheet $\Sigma_2$ contains couplings of the form
\begin{equation}
-\sum_{\b} c_{b}^{\b} \int_{\Sigma_2} C_2^\b
\label{Dstringcoupling}
\end{equation}
where the 2-form $C_2^\b$ is dual to the axion $\textrm{Re}(N^\b)$, specified by the torsion classes $[\pi_{3}^{{\rm tor},\b}] \in {\rm Tor}\, H_3^- (\cam_6, \IZ)$ and $[\pi_{2,\b}^{\rm tor}] \in {\rm Tor}\, H_2^+ (\cam_6, \IZ)$ respectively (c.f. Tables \ref{tabla1} and \ref{tabla2}).  The (mod $k_\b$) integer coefficients $c_{b}^{\beta}$ can be obtained from the expansion
\begin{equation}
[\pi_b^-]\, =\, \sum_\b c_{b}^{\b} [\pi_3^{{\rm tor},\b}]
\label{decomtor}
\end{equation}
so that, in terms of the linking form $L$, we get
\begin{equation}
c_{b}^{\b} \, =\, \sum_\a k_\b{}^\a\, L( [\pi_{2,\a}^{\rm tor}],[\pi_b^-]) \, =\, \sum_\a k_\b{}^\a L_{\a}{}^{b}
\label{decomlink}
\end{equation}

Since a D4-string wrapped on $\pi_3^{\rm tor}$ can be seen as a vortex defect of the U(1)$_b$ gauge symmetry upon D6-brane annihilation or recombination \cite{Dstrings}, it follows that the open string gauge symmetry U(1)$_b$ has the 4d couplings
\begin{equation}
-\sum_\b c_b^\b \int_{\IR^{1,3}} C_2^\b \wedge F_2^{b}
\label{D6coupling}
\end{equation}
where $F_2^{b} = dA^b$ is the field strength for the U(1)$_b$ gauge boson. Otherwise said, as (\ref{Dstringcoupling}) arises from dimensional reduction of the CS coupling $\int_{\rm D4} C_5$ of a D4-brane, eq.(\ref{D6coupling}) should equally arise from dimensional reduction of the coupling $\int_{\rm D6} C_5 \wedge F$ of a D6-brane in the same topological sector $[\pi_b^-]$. We provide a more direct derivation of this result in Appendix \ref{tortopo}.

Given the couplings (\ref{D6coupling}), it is clear that the St\"uckelberg mechanism has to be reconsidered if open and closed string U(1)'s are both present. In particular, the 4d Lagrangian (\ref{stucktor}) has to be modified, since now the open string gauge bosons $A^a$ also couple to the massive RR axions $\re (N^\b)$. Putting all pieces together we arrive to a full St\"uckelberg Lagrangian of the form
\begin{equation}
\begin{array}{c}\vspace*{.2cm}
\mathcal{L}^{\rm tor}_{\rm Stk}\, =\, \frac12 e^{2\phi_4}\left[\mathcal{G}^{-1}_{IJ}\textrm{Re}(DN^{I})\wedge *_{4}\textrm{Re}(DN^{J}) + \check{\mathcal{G}}_{\alpha\beta}^{-1}\textrm{Re}(DN^\alpha)\wedge *_4\textrm{Re}(DN^\beta)\right] \\
DN^I=dN^I+\sum_a c^I_aN_aA^a \qquad DN^\b=dN^\b+k^\b{}_{\a} {A}^\alpha + \sum_a  c_a^{\b} N_a A^a
\end{array}
\label{stucktotal}
\end{equation}
where $I = 1, \dots, h^{1,2} +1$ label the RR axions of section \ref{U1typeIIA}, and $\b = 1, \dots, \textrm{dim}\left(\textrm{Tor}\ H^{3}_+\right)$ the massive axions of this section. Finally, the index $a$ runs over each stack of $N_a$ D6-branes wrapped on a sLag 3-cycle $\pi_a$ and carrying a gauge group U($N_a$). If any of the coefficients $c_a^{\b}$ is non-zero (that is, if $[\pi_a]$ has a component  in the torsional homology group ${\rm Tor\, } H^{3}_+(\cM,\mathbb{Z})$), then there is some mixing between open and closed string U(1)'s in the mass matrix, and massless gauge symmetries are a combination of both types of U(1)'s. It is easy to see that the linear combinations of RR and D6-brane U(1)'s which become massive due to this St\"uckelberg mechanism are
\begin{eqnarray}
\label{massive}
Q^I & = & \sum_a c_a^{I}N_aQ^a \\
Q^{\b} & = & \sum_\a k^\b{}_\a Q_{RR}^{\a}+\sum_ac_a^\b N_aQ^a
\label{massivetor}
\end{eqnarray}
where $Q_{RR}^\b$ is the generator of the torsional RR U(1)$_\a$
associated to $[\pi_{2, \a}^{\rm tor}]$.

The set of RR and D6-brane U(1) gauge symmetries which remain massless admits an elegant interpretation in terms of integer homology classes, generalizing the results for open string U(1) gauge symmetries of section \ref{opensec}. We have just argued that to each RR U(1) generator entering in (\ref{massivetor}) we can associate a torsional 2-cycle class
$[\pi_{2,\a}^{\rm tor}]$, as well as a dual torsional 3-cycle class $k^\a{}_{\g}[\pi_{3}^{{\rm tor},\g}]$.
Hence, each linear combination of D6-brane and torsional RR U(1) generators is mapped to an element of $ H_3^- (\cam_6, \IZ)$
\begin{equation}
Q_0=\sum_a n_a Q^a+\sum_{\a}\check{n}_{\a}Q_{RR}^{\a} \ \longrightarrow \ \pi_0= \sum_a\frac{N_an_a}{2}[\pi_a-\pi_a^*]+\sum_{\a,\g}\check{n}_{\a} k^\a{}_{\g}[\pi^{{\rm tor},\g}_{3}]
\end{equation}
for $n_a, \, \sum_\a \check{n}_\a k^\a{}_\g \in \IZ$.  Extending the reasoning of section \ref{opensec} to this case, we observe that massless combinations of RR and D6-brane U(1) gauge symmetries correspond to linear combinations for which $[\pi_0]$ is trivial in the integer homology of $\cM$
\begin{equation}
\sum_a\frac{N_an_a}{2}([\pi_a]-[\pi_a^*])+\sum_{\a,\g}\check{n}_{\a}k^\a{}_{\g}[\pi^{{\rm tor},\g}_{3}]=0\label{combifinal}
\end{equation}
We can illustrate this expression with a simple toy model. For that, consider the case of two D6-branes wrapping 3-cycles $\pi_a$ and $\pi_b$. As we discussed in section \ref{opensec}, if $\pi_a$ and $\pi_b$ are in the same homology class, $[\pi_a]=[\pi_b]$ (and $[\pi_{a,b}^*]\neq[\pi_{a,b}]$), the linear combination U(1)$_a-$U(1)$_b$ remains in the massless spectrum, whereas the orthogonal combination, U(1)$_a$+U(1)$_b$, acquires a mass by means of the St\"uckelberg mechanism. We can now consider a slightly different situation on which the two 3-cycles wrapped by the D6-branes differ by a $\sigma$-odd torsional 3-cycle, $[\pi_b]-[\pi_a]=[\pi^{\rm tor}_3]$. According to eq.(\ref{massivetor}), some of the axions  which couple to the branes $a$ and $b$ by means of St\"uckelberg couplings, couple also to the RR U(1) gauge boson. The linear combination which remains massless in this case is 2[U(1)$_a-$U(1)$_b$]+U(1)$_{\rm RR}$, whereas the two orthogonal combinations, U(1)$_a-$U(1)$_b-$4U(1)$_{\rm RR}$ and U(1)$_a$+U(1)$_{b}$, are massive.

\subsection{M-theory and discrete gauge symmetries}\label{Mtorsion}

We have seen in section \ref{msec} that \emph{massless} D6-brane and RR U(1) gauge symmetries share a common origin in M-theory compactified on a $G_2$ manifold $\hat{\mathcal{M}}_7$, namely, they both come from dimensional reduction of the M-theory 3-form in elements of $H_2(\hat{\mathcal{M}}_7,\mathbb{R})$. From that perspective, it is not surprising that D6-brane and RR U(1) gauge symmetries appear in eq.(\ref{combifinal}) on the same footing. Indeed, one may easily show  that \emph{massive} D6-brane and RR U(1) gauge symmetries also have a common lift to M-theory. For that, one has to consider the more fundamental group $H_{2}(\hat{\mathcal{M}}_7,\mathbb{Z})$, instead of $H_2(\hat{\mathcal{M}}_7,\mathbb{R})$. Electrically charged 4d particles arise from M2-branes wrapping $k_\a$-torsional 2-cycles $\hat{\pi}_{2,\alpha}^{\rm tor}\in\textrm{Tor}\ H_{2}(\hat{\mathcal{M}}_7,\mathbb{Z})$ whereas 4d Aharanov-Bohm strings are M5-branes wrapping dual $k_\a$-torsional 4-cycles $\hat{\pi}_{4}^{{\rm tor},\a}\in\textrm{Tor}\ H_{4}(\hat{\mathcal{M}}_7,\mathbb{Z})$ (recall that for a 7d manifold $\textrm{Tor}\ H_{2}(\hat{\mathcal{M}}_7,\mathbb{Z})\simeq \textrm{Tor}\ H_{4}(\hat{\mathcal{M}}_7,\mathbb{Z})$). The linking form in $\hat{\mathcal{M}}_7$ then relates the classes $[\hat{\pi}_{2,\alpha}^{\rm tor}]$ and $[\hat{\pi}_{4}^{{\rm tor},\a}]$ unambiguously. Hence, following a similar reasoning that the one in section \ref{gaugesym}, it is natural to associate to each element of $\textrm{Tor}\ H_{2}(\hat{\mathcal{M}}_7,\mathbb{Z})$ a 4d U(1) gauge symmetry broken down to a $\mathbb{Z}_{k_\a}$ subgroup.  In the perturbative type IIA Calabi-Yau orientifold limit (\ref{iialimit}), these U(1)'s reduce to the massive D6-brane and RR U(1) gauge symmetries discussed in the previous section. The general picture described in \cite{bs10} (see also \cite{hs10}) for 4d quantum theories of gravity is therefore realized in M-theory through torsion.

Following our discussion in section \ref{massivesec}, we can introduce a set of torsional forms, $\phi_\alpha^{\rm tor}\in \textrm{Tor}\ H^{3}(\hat{\mathcal{M}}_7,\mathbb{Z})$ and $\omega_\b^{\rm tor}\in \widehat{\textrm{Tor}}\ H^{2}(\hat{\mathcal{M}}_7,\mathbb{Z})$, such that
\begin{equation}
\hat{k}_{\a}{}^\b\phi_\b^{\rm tor}=d\omega_\a^{\rm tor}\label{g2fib}
\end{equation}
with $\hat{k}_{\a}{}^\b\in\mathbb{Z}$. These are the torsional analogues of the harmonic forms $\phi_I$ and $\omega_\a$ that we made use of to dimensionally reduce the M-theory 3-form $A_3$. Performing the same kind of expansion in this basis we get
\begin{equation}
dA_3=\left(\textrm{Re}(dM^\alpha)+\hat{k}^\a{}_{\b} A^{\b}\right)\wedge \phi_{\a}^{\rm tor}+dA^{\b}\wedge \omega_\b^{\rm tor}
\end{equation}
and therefore the 4d effective Lagrangian contains St\"uckelberg couplings which arise from dimensional reduction of the 11d $A_3$ kinetic term (see also \cite{Grimm:2010ez}). In the perturbative type IIA limit (\ref{iialimit}), the M-theory St\"uckelberg mechanism reduces to eq.(\ref{stucktotal}).

The fact that massive D6-brane and RR U(1)'s are both related to torsional 2-cycles of the $G_2$ manifold in M-theory  has some interesting consequences. Indeed, consider a type IIA orientifold compactification on a given $\textrm{CY}_3$ $\cM$. There are typically many possible consistent configurations of D6-branes which cancel the global charge of the O6-planes. All of them are connected through brane recombination processes. The number of massless and massive D6-brane U(1)'s depends on the particular configuration of D6-branes at angles. Thus, according to the above discussion there should be a family of $G_2$ manifolds associated to the above compactification, where each manifold corresponds to a different configuration of D6-branes in $\cM$. We can build such a family starting from the case on which all D6-branes are parallel to the O-planes. Let $\hat{\mathcal{M}}_7^{||}$ be the corresponding $G_2$ manifold, with Betti numbers $(b_2,b_3)$. Different configurations of D6-branes at angles can be then obtained by fibering the (co)homology of $\hat{\mathcal{M}}_7^{||}$ accordingly to (\ref{g2fib}). The new $G_2$ manifolds constructed in this way have Betti numbers $(b_2-n,b_3-n)$, with $n=\textrm{rank}(\hat{k})$, and $n$ more torsional 2-cycles than $\hat{\mathcal{M}}_7^{||}$ has. The matrix $\hat{k}$ obviously cannot be arbitrary and, in particular, it has to satisfy global consistency conditions such as compactness of the resulting $G_2$ manifold.

It is also enlightening to consider in this context the open/closed string dualities that were introduced in section \ref{msec} and which result from different perturbative type IIA limits of the $G_2$ manifold. We saw there that massless D6-brane and RR U(1) gauge symmetries can be exchanged under these dualities, due to different splits (\ref{h2m}) of $H_2(\hat{\mathcal{M}}_7,\mathbb{R})$. This statement obviously still holds true for the more fundamental group $H_2(\hat{\mathcal{M}}_7,\mathbb{Z})$. Massive D6-brane and torsional RR U(1) gauge symmetries are therefore also exchanged under open/closed string dualities.  In particular, different configurations of D6-branes at angles within the same type IIA $\textrm{CY}_3$ orientifold are mapped to families of type IIA $\textrm{CY}_3$ orientifolds, which result from
 twisting the (co)homology of a torsion-free Calabi-Yau as
\begin{equation}
k \ : \ H^2(\cM,\mathbb{R})_+\ \to \ H^3(\cM,\mathbb{R})_+\ , \quad \textrm{such that} \quad d\omega_i=k_i{}^I\alpha_I\label{twist}
\end{equation}
in the same spirit than \cite{Cvetic:2007ju}.

In section \ref{fwsec} we discuss yet another consequence of massive D6-brane U(1)'s being lifted to torsional homology in M-theory, namely that D6-brane Freed-Witten anomalies in type IIA $\textrm{CY}_3$ orientifolds \cite{Camara:2005dc, Villadoro:2006ia, D6torsion} correspond to 4-form backgrounds in M-theory whose cohomology class $[G_4]$ is torsion.

\subsection{An explicit example}
\label{example}

There are many examples of Calabi-Yau orientifold compactifications which have RR U(1) gauge symmetries in their 4d spectrum. Simplest models include toroidal orbifold compactifications, see e.g., \cite{Klein:2000hf,z6,z6p,Forste:2010gw}. 
In this section we consider a type IIA orientifold of the Enriques Calabi-Yau \cite{vb,Ferrara:1995yx}. The large amount of symmetry of this manifold allows to perform very explicit computations, whereas  its moduli space is rich enough to contain massless and massive RR U(1)'s and D6-branes at angles. Thus, it is an appealing  setup where to illustrate some of the above ideas on mass mixing with RR photons explicitly.

We can think of the Enriques Calabi-Yau as the smooth manifold which results from blowing-up the singularities of a $(T^2\times K3)/g_1$ orbifold, where $g_1$ reverses the coordinates of $T^2$ and acts on the $K3$ lattice as \cite{Ferrara:1995yx},
\begin{center}
\begin{tabular}{c}
$H^2(K3,\mathbb{R})=-\Gamma_{E_8}\oplus-\Gamma_{E_8}\oplus\Gamma_{1,1}\oplus\Gamma_{1,1}\oplus\Gamma_{1,1}$\\
$\downarrow$\\
$H^2(K3/g_1,\mathbb{R})=-\Gamma_{E_8}\oplus\Gamma_{1,1}$
\end{tabular}
\end{center}
At the $T^4/\mathbb{Z}_2$ orbifold point of $K3$, the Enriques Calabi-Yau therefore becomes a $T^6/(\mathbb{Z}_2\times\mathbb{Z}_2)$ freely-acting orbifold with generators
\begin{align}
g_1\ : \ & (z^1,\ z^2,\ z^3)\ \to \ (-z^1,\ -z^2,\ z^3+\pi R_3)\label{g1}\\
g_2\ : \ & (z^1,\ z^2,\ z^3)\ \to \ (-z^1,\ z^2+\pi R_2,\ -z^3)\nonumber\\
g_3\ : \ & (z^1,\ z^2,\ z^3)\ \to \ (z^1,\ -z^2+\pi R_2,\ -z^3-\pi R_3)\nonumber
\end{align}
where $z^i=dx^i+\tau_idx^{i+3}$, $i=1,2,3$, are the three complex coordinates of $T^2\times T^2\times T^2$. For simplicity, we work at this orbifold point of the moduli space and, moreover, we set $2\pi R_i=1$. Generalization to arbitrary radii is straightforward.

The integer homology of the Enriques Calabi-Yau was first computed in \cite{Aspinwall:1995mh} by means of the Hochschild-Serre spectral sequence. We have summarized the result in Table \ref{tablacy}. Different elements are identified as follows. The free part of the homology is given by eleven 2-cycles (and their dual 4-cycles) and twenty-four 3-cycles. In the $T^6/(\mathbb{Z}_2\times\mathbb{Z}_2)$ limit of the Enriques Calabi-Yau, these correspond to the canonical three 2-cycles and eight 3-cycles of the covering space, $T^2\times T^2\times T^2$, plus eight exceptional 2-cycles and sixteen exceptional 3-cycles attached to the fixed points of (\ref{g1}). Apart from these, there are three torsional 1-cycles and one torsional 2-cycle (plus their dual torsional 4-cycles and 3-cycle, c.f. eq.(\ref{together})).

In order to gain more intuition on the torsional part of the homology, we can look at the explicit loci of the torsional cycles. For that, we take oriented segments in the covering $T^2\times T^2\times T^2$ and draw their images under the orbifold generators, eq.(\ref{g1}). We identify
\begin{align}
\eta_{1}^{\rm tor}\ &=\ x^1\in \left[0,\frac12\right),\ \ x^4,x^5,x^6\in\left\{0,\frac12\right\}, \ \ x^2=x^3=\frac14\cup \frac34\\
\eta_{2}^{\rm tor}\ &=\ x^4\in \left[0,\frac12\right),\ \ x^1,x^5,x^6\in\left\{0,\frac12\right\}, \ \ x^2=x^3=\frac14\cup \frac34\nonumber \\
\eta_{3}^{\rm tor}\ &=\ x^2\in\left[\frac14,\frac34\right)\cup x^3\in\left[\frac14,\frac34\right), \ \ x^1,x^4,x^5,x^6\in\left\{0,\frac12\right\}\nonumber
\end{align}
as the loci of the three torsional 1-cycles, and
\begin{equation}
\rho^{\rm tor}\ = \ x^5\in\left[0,1\right),\ \ x^6\in \left[0,\frac12\right),\ \ x^1,x^4\in\left\{0,\frac12\right\}, \ \ x^2=x^3=\frac14\cup \frac34
\end{equation}
as the locus of the torsional 2-cycle. For latter purposes we also give the locus of the torsional 3-cycle, obtained by means of the same procedure,
\begin{equation}
\pi^{\rm tor}\ = \ x^2\in\left[\frac14,\frac34\right)\cup x^3\in\left[\frac14,\frac34\right),\ \ x^1\in\left[0,1\right),\ \ x^4\in\left[0,\frac12\right),\ \ x^5,x^6\in\left\{0,\frac12\right\}
\end{equation}

\begin{table}
\begin{center}
\begin{tabular}{|c|c|c|c|c|c|c|}
\hline
$H_0(\cM)$&$H_1(\cM)$&$H_2(\cM)$&$H_3(\cM)$&$H_4(\cM)$&$H_5(\cM)$&$H_6(\cM)$\\
\hline
$\mathbb{Z}$&$(\mathbb{Z}_2)^3$&$(\mathbb{Z})^{11}\oplus \mathbb{Z}_2$&$(\mathbb{Z})^{24}\oplus \mathbb{Z}_2$&$(\mathbb{Z})^{11}\oplus (\mathbb{Z}_2)^3$&$0$&$\mathbb{Z}$\\
\hline
\end{tabular}
\end{center}
\caption{Integer homology of the Enriques Calabi-Yau.\label{tablacy}}
\end{table}

We now consider a type IIA orientifold of the above $T^6/(\mathbb{Z}_2\times\mathbb{Z}_2)$ orbifold, where the orientifold involution $\sigma$ reverses the coordinates $x^4$, $x^5$ and $x^6$ of $T^6$. O6-planes wrap the 3-cycles,
\begin{align}
&\Lambda_0\ = \ x^1,x^2,x^3\in\left[0,\frac12\right),\ \ x^4,x^5,x^6\in\left\{0,\frac12\right\}\\
&\Lambda_1\ = \ x^1,x^5,x^6\in\left[0,\frac12\right), \ \ x^2=x^3=\frac14\cup\frac34,\ \ x^4\in\left\{0,\frac12\right\}\nonumber
\end{align}
The reader may easily check that $\eta_1^{\rm tor}$, $\eta_3^{\rm tor}$ and $\rho^{\rm tor}$ are even under $\sigma$, whereas $\eta_2^{\rm tor}$ is odd. Hence, according to the results of previous subsections (c.f. Table \ref{tabla1}), there are 6 massive closed string vector bosons arising from the torsional part of the homology. Four of these come from dimensional reduction of the metric on $\eta_1^{\rm tor}$, the NSNS 2-form on $\eta_2^{\rm tor}$, the RR 3-form on $\pi^{\rm tor}$ and the RR 5-form on the torsional 4-cycle dual to $\eta_1^{\rm tor}$.  There is a U(1)$_L^2\times $U(1)$_R^2$ gauge symmetry spontaneously broken to $(\mathbb{Z}_2)^4$. These states are identified with the graviphoton and the 3 gauge bosons in the $S-T-U$ vector multiplets of $\cn=2$ orientifold compactifications on $T^2\times K3$. The fact that they appear in the 4d spectrum is understood by noting that part of the supersymmetry is only spontaneously broken in the Enriques CY \cite{Ferrara:1995yx}. At large volumes of the first 2-torus, $\textrm{Im}(T^{\hat 1}) >>1$, these vector multiplets become light and 4d $\mathcal{N}=2$ supersymmetry is approximately recovered. In addition, there are 2 extra massive vector bosons coming from dimensionally reducing the metric on $\eta_3^{\rm tor}$ and the RR 5-forms on the dual torsional 4-cycle.

Let us now focus on the massive RR photon associated to $\pi^{\rm tor}$, which we have identified as a massive graviphoton. Its mass is acquired by combining with a complex structure axion, namely the one which results from expanding $C_3$ on the exact 3-form related to $\rho^{\rm tor}$ by eq.(\ref{togetherori}). Hence, D6-brane U(1) gauge bosons which couple to the same complex structure axion will develop a non-trivial mixing with the RR photon  via the St\"uckelberg mechanism, as described in subsection \ref{stuckrev}.

Supersymmetric D6-branes wrap calibrated 3-cycles. Geometrically we can distinguish two different cases: \emph{bulk} D6-branes wrapping 3-cycles in the covering space, and \emph{fractional} D6-branes, wrapping 3-cycles which only close in the quotient space. Bulk D6-branes have three massless chiral multiplets transforming in the adjoint representation and therefore can move freely in the $T^6$. Fractional D6-branes, on the other hand, are stuck at fixed points of one or more generators in eq.(\ref{g1}). Whereas a precise determination would require a detailed CFT computation which is beyond the scope of this work, we assume that the gauge group of fractional D6-branes is U($N$).

In what follows we present three different configurations of D6-branes which lead to qualitatively different scenarios of mixing between RR and D6-brane U(1) gauge symmetries:

- \underline{Two stacks of bulk branes in the same homology class.} Consider for instance two bulk D6-branes with same wrapping numbers on $T^2\times T^2\times T^2$,
\begin{equation}
\pi_a,\ \pi_b\ \ : \ \ (1,0)\otimes(n^2,m^2)\otimes(n^3,-m^3)\label{bulk}
\end{equation}
According to our previous discussion, since the D6-branes wrap 3-cycles in the same homology class, $[\pi_a]=[\pi_b]$, they do not couple to the axion which gives mass to U(1)$_{\rm RR}$. The linear combination U(1)$_{G_1}\equiv \frac{1}{\sqrt{2}}($U(1)$_a+$U(1)$_b)$ becomes massive by combining with one of the complex structure moduli of the covering $T^2\times T^2\times T^2$ and the universal axion, whereas the orthogonal combination, U(1)$_{Y}\equiv \frac{1}{\sqrt{2}}($U(1)$_a-$U(1)$_b)$, remains massless. The corresponding gauge kinetic functions for the mass eigenstates read
\begin{align}
f_{YY}&=f_{G_1G_1}=-i(n^2n^3N^0+m^2m^3N^1)\\
f_{G_2G_2}&=-iT^{\hat 1}\nonumber
\end{align}
where U(1)$_{G_2}\equiv $U(1)$_{\rm RR}$. In particular there is no kinetic mixing between massless and massive linear combinations of U(1)'s.\\

- \underline{Two stacks of fractional branes which differ by $\pi^{\rm tor}$.} Consider now the D6-branes $a$ and $b$ to be fractional, so that generically $[\pi_a]\neq[\pi_b]$. We take them to coincide in the second and third 2-tori, whereas they are located at different fixed points in the first $T^2$. For simplicity we take them to wrap the 3-cycles
\begin{align}
\pi_a \ &= \ x^1\in\left[0,\frac12\right),\ \ x^2,x^3\in\left[\frac14,\frac34\right),\ \  x^4,x^5,x^6=0\\
\pi_b \ &= \ x^1\in\left[0,\frac12\right),\ \ x^2,x^3\in\left[\frac14,\frac34\right),\ \ x^4=\frac12, \ \ x^5,x^6=0
\end{align}
so that the bulk component of the branes is along the direction $(1,0)\otimes(1,0)\otimes(1,0)$. It is possible to check that the 4-chain which connects $\pi_a$ and $\pi_b$ has also $\pi^{\rm tor}$ as part of the boundary, and therefore $[\pi_b]-[\pi_a]=[\pi^{\rm tor}]$. The massless combination of U(1) gauge symmetries is U(1)$_{Y}\equiv \frac{1}{\sqrt{5}}(2$U(1)$_a-2$U(1)$_b+$U(1)$_{\rm RR})$, whereas the two orthogonal combinations, U(1)$_{G_1}\equiv \frac{1}{\sqrt{2}}($U(1)$_a+$U(1)$_b)$ and U(1)$_{G_2}\equiv \frac{1}{\sqrt{6}}($U(1)$_a-$U(1)$_b-4$U(1)$_{\rm RR})$, develop St\"uckelberg couplings. Thus, in this case the massless photon is a linear combination of D6-brane and RR U(1) gauge bosons. The corresponding gauge kinetic functions are
\begin{align}
f_{YY}&=-\frac{5i}{81}(T^{\hat 1}+8N^0)\ , \quad
f_{G_1G_1}=N^0\ , \quad
f_{G_2G_2}=-\frac{i}{27}(8T^{\hat 1}+N^0)\\
&\qquad\qquad \quad f_{YG_2}=-\frac{4i}{27}\sqrt{\frac{10}{3}}(N^0-T^{\hat 1})\nonumber
\end{align}
and there is non-trivial kinetic mixing between the massless photon and one of the massive combinations of U(1)'s.

Had we instead taken two D6-branes per stack, we would have recovered the case of various bulk branes in the same homology class, $[2\pi_b]-[2\pi_a]=[2\pi^{\rm tor}]=0$. More generically, we can consider fractional D6-branes of the above type whose bulk component is given by eq.(\ref{bulk}). In that case we may argue that  for $n^2$ and $n^3$ arbitrary integers but $m^2=m^3=0$, one has $[\pi_b]-[\pi_a]=n^2n^3[\pi^{\rm tor}]$ which is homologically non-trivial whenever $n^2n^3$ is an odd integer. Similar arguments show that for $n^2=n^3=0$ and $m^2=m^3=1$ the 3-cycles $\pi_a$ and $\pi_b$ instead differ by some exceptional 3-cycle. Hence, we conclude that if the ratios $m^2/n^2$ and $m^3/n^3$ are even integers and $n^2n^3$ is odd, then $[\pi_b]-[\pi_a]=[\pi^{\rm tor}]$.\\

\begin{figure}[!ht]
\vspace{0.5cm}
\begin{center}
\includegraphics[width=15cm]{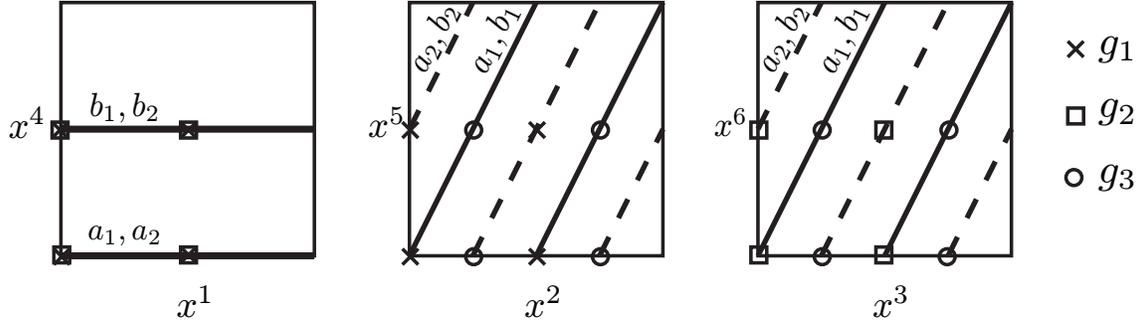}
\end{center}
\vspace{-0.3cm}
\caption{\label{figura} Configuration of 4 fractional D6-branes leading to two mutually hidden sectors which communicate via RR photons.}
\end{figure}
%

- \underline{Two mutually hidden brane sectors which communicate via RR photons.} Finally we can consider two copies of the previous configuration of fractional D6-branes. We locate each pair of branes, $\{a_1,b_1\}$ and $\{a_2,b_2\}$, at different fixed points in the second and/or third 2-torus. An explicit example is depicted in Figure \ref{figura}. The above pairs are completely isolated from each other, since they carry different twisted charge (as they wrap different exceptional 3-cycles). They couple however to the same RR U(1) gauge boson, since they carry the same torsional charge. Thus, the two pairs $\{a_1,b_1\}$ and $\{a_2,b_2\}$ communicate only via the RR photon. The two massless combinations of U(1) gauge bosons are,
\begin{equation}
U(1)_{Y_k}\equiv\frac{1}{\sqrt{5}}(2U(1)_{a_k}-2U(1)_{b_k}+U(1)_{\rm RR})\ , \quad k=1,2
\end{equation}
whereas massive U(1) symmetries are,
\begin{align}
U(1)_{G_k}&\equiv \frac{1}{\sqrt{2}}(U(1)_{a_k}+U(1)_{b_k})\ , \quad k=1,2\\
U(1)_{G_3}&\equiv \frac{1}{\sqrt{8}}(U(1)_{a_1}-U(1)_{b_1}+U(1)_{a_2}-U(1)_{b_2}-4U(1)_{\rm RR})\nonumber
\end{align}
The reader may easily check that there is kinetic mixing between the massless $U(1)_{Y_k}$ gauge bosons and the massive $U(1)_{G_3}$ boson
\begin{equation}
f_{Y_1G_3}=-\frac{i}{10\sqrt{10}}(9f_1-f_2-8T^{\hat 1})\ , \qquad f_{Y_2G_3}=-\frac{i}{10\sqrt{10}}(9f_2-f_1-8T^{\hat 1})
\end{equation}
with $f_k\equiv f_{a_k}=f_{b_k}$, $k=1,2$ the gauge kinetic functions of the D6-branes $\{a_k,b_k\}$, whose explicit expression we omit for briefness. Moreover, the two massless U(1) gauge bosons also mix through the following component of the gauge kinetic function,
\begin{equation}
f_{Y_1Y_2}=-\frac{i}{80}(8T^{\hat 1}-9f_{1}-9f_{2})
\end{equation}
Hence, in this toy example the presence of a massive RR U(1) gauge boson induces kinetic mixing between the two D6-brane sectors $\{a_1,b_1\}$ and $\{a_2,b_2\}$, which otherwise would be completely hidden from each other at low energies.

\section{Some phenomenological implications}
\label{pheno}

We have seen in the previous section that under certain conditions (namely, in the presence of torsional cycles)  there may appear {\it mass mixing}
between RR and D-brane U(1) gauge symmetries. In particular, massless eigenstates may be linear combinations of D-brane and RR gauge bosons.
It is natural to ask whether such a mixing may have some effect of phenomenological interest.  At first sight it seems that no
effect should appear at all since there are no perturbative light fields which could couple to the RR U(1)'s. Hence, if the
SM hypercharge contained some RR contamination we would be unable to tell it.  There are however situations in which this
mass mixing may turn  out to be phenomenologically interesting. For instance, the rigid D6-brane configurations presented at the end of last section are explicit realizations of the U(1) mediation mechanism proposed in \cite{Langacker:2007ac, Verlinde:2007qk} (see also \cite{Grimm:2008ed}).  Moreover,
in   section \ref{kmixing} we described {\it kinetic  mixing}  between RR and D-brane U(1)'s and
in  the previous section we have also seen another mechanism for the generation of kinetic mixing between  visible and  hidden sector
massless U(1)'s.  These sources of  kinetic mixing have  potential phenomenological applications to the mixing of the hypercharge
U(1)$_Y$ (and hence the photon) with hidden U(1)'s, as studied e.g. in
refs. \cite{Dienes:1996zr,Feldman:2006wd,Feldman:2007wj,Ibarra:2008kn,Arvanitaki:2009hb}.

In this section we discuss yet another interesting effect of RR U(1) gauge bosons, this time in the context of  SU(5) unification within type IIB orientifolds (or their F-theory extension).
In these constructions the SU(5) degrees of freedom live on a 7-brane which wraps a 4-cycle $S$, whereas matter fields are localized at the intersection
with other U(1) 7-branes (leading to matter curves in the F-theory language).
In some of these constructions the SU(5) symmetry is broken down to the SM one by turning on a non-zero flux along the hypercharge generator,
$\overline{F}_Y\not= 0$.  Generically such fluxes give rise to St\"uckelberg masses for the hypercharge gauge boson,
through the couplings
\beq
\int_{\mathbb{R}^{1,3}\times S}C_4 \wedge F_Y\wedge \overline{F}_Y \ \rightarrow \ \int_{\mathbb{R}^{1,3}} C_2^Y\wedge F_Y
\eeq
with
\beq
C_2^Y\ \equiv \int_S C_4\wedge \overline{F}_Y  \ =\ \int_{\rho^Y} C_4
\eeq
where $\rho^Y$ denotes the Poincar\'e dual of $\overline{F}_Y$ in $S$.
This is unacceptable since U(1)$_Y$ disappears from the massless spectrum. One way to solve this problem
is to assume that $\rho^Y$ is trivial in the homology of the full Calabi-Yau, although non-trivial in $S$ \cite{Buican:2006sn}. In this case the
dangerous $C_2^Y\wedge F_Y$ coupling disappears and the problem goes away.  This is the {\it standard }
solution within F-theory model building \cite{Beasley:2008kw, Donagi:2008kj}.

In view of our results in the previous section
(or rather their type IIB version discussed  in Appendix \ref{typeIIB}), there is however a particularly compelling alternative.  Indeed, let us assume that there is a  RR U(1) gauge boson
$V_{RR}$ which results from the expansion of the RR 4-form in torsional forms, $C_4=A_{\rm RR}\wedge \alpha^{\rm tor}+V_{\rm RR}\wedge \beta^{\rm tor}+\ldots$. The gauge boson is massive and the U(1)$_{\rm RR}$ symmetry is spontaneously broken to a discrete $\mathbb{Z}_{k_{\rm RR}}$ gauge symmetry due to a $C_2\wedge dV_{\rm RR}$
St\"uckelberg coupling, as may be seen from eq.(\ref{dc4}). If the hypercharge flux is also along the associated torsional cycle, $F^{Y}_2=\overline{F}_Y\omega^{\rm tor}$,
 then the same 4d 2-form $C_2$ couples both to
U(1)$_{RR}$ and U(1)$_Y$ and there is a St\"uckelberg mass term of the form
\beq
\mathcal{L}\supset -\frac12\left(\textrm{Re}(dT) + k_{\rm RR}A_{\rm RR}+ \frac{5k_Y}{3}A_Y\right)^2
\eeq
where $\textrm{Re}(T)$ is the 4d axion dual of $C_2$ and we have included the SU(5) normalization factor
for the hypercharge.
In terms of gauge bosons ${\tilde A}_{\rm RR}\equiv A_{\rm RR}/g_{\rm RR}$ and ${\tilde A}_Y\equiv A_Y/g_Y$ with canonical kinetic terms, there
is a massless ($A_1$) and a massive ($A_X$) linear combination of U(1) gauge symmetries
\beq
A_1\ = \textrm{cos}(\theta) {\tilde A}_Y \ -\ \textrm{sin}(\theta) {\tilde A}_{\rm RR} \ \ ; \ \  A_X\ = \textrm{sin}(\theta) {\tilde A}_Y \ +\  \textrm{cos}(\theta) {\tilde A}_{\rm RR}
\label{autoestados}
\eeq
where
\beq
\textrm{sin}(\theta) \ \equiv\ \frac {g_Yk_Y}{\sqrt{g_{\rm RR}^2k_{\rm RR}^2+g_Y^2k_Y^2}} \ .
\eeq
Explicit expressions for the gauge coupling constants $g_{\rm RR}^2$ and $g_Y^2$ can be obtained from the gauge kinetic functions (\ref{fr7}) and (\ref{fd7}) respectively.
Note that for $g_Y^2\ll g_{\rm RR}^2$ the massless eigenstate mostly corresponds to the brane hypercharge U(1)$_Y$ generator, whereas
in the opposite case it is the U(1)$_{\rm RR}$ factor the dominant component. The massless boson, $A_1$, couples to
the D7-brane matter fields with coupling constant $g_Y\textrm{cos}(\theta)$. The inverse fine structure constant $\alpha_1$ of the
massless  U(1) is therefore given by
\beq
\frac {1}{\alpha_1} \ =\ \frac {3}{5\alpha_G} \ +\ \frac {k_Y^2}{k_{\rm RR}^2\alpha_{\rm RR}}
\label{correccionu1}
\eeq
with $\alpha_{\rm RR}=g_{\rm RR}^2/4\pi$ and $\alpha_G$ the $SU(5)$ fine structure constant. This implies the existence
of a correction to the standard unification of hypercharge given by the last term in this expression. Since the $SU(5)$
unification boundary conditions work quite well, with a precision of a few percent,  this correction should not be much larger than
$\sim O(1)$. This implies that
\beq
\alpha_{\rm RR} \sim \frac {k_Y^2}{k_{\rm RR}^2} \ .
\eeq

\begin{figure}[t]
\begin{center}
\includegraphics[width=11cm]{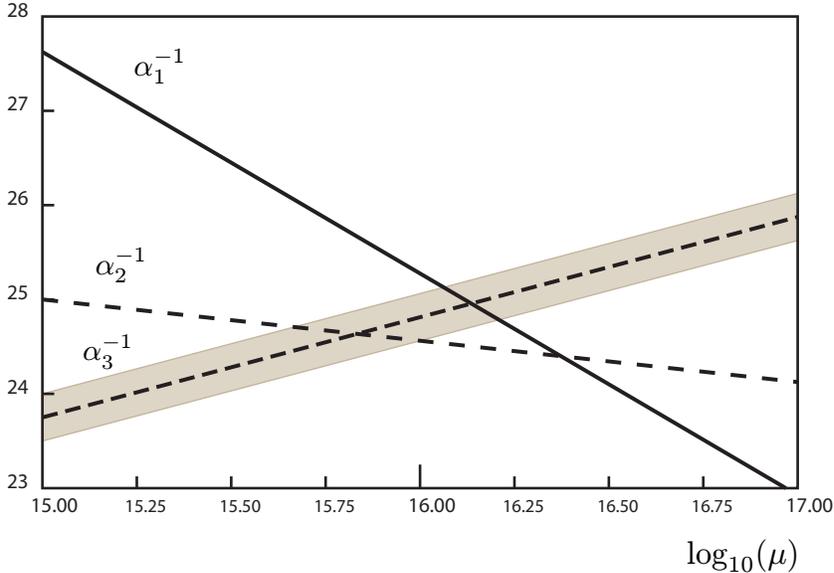}
\end{center}
\caption{\label{gcu} Two-loop running of the MSSM gauge coupling constants in the region
 around $10^{16}$ GeV \cite{Ibe:2003ys}. The shaded region represents the uncertainty in the measurement of the QCD gauge coupling constant.}
\end{figure}
%

This solution to the St\"uckelberg mass problem of the hypercharge flux can actually be though as a different avatar of a similar idea proposed
 for heterotic compactifications in Ref.\cite{tw}. In that case the extra U(1) gauge symmetry was coming from the second $E_8$ factor of the $E_8\times E_8$ heterotic
gauge group. A strong coupling regime for this second $E_8$ was assumed.
 In our case, however, the structure is simpler since the extra U(1) is a RR field with no perturbative couplings to any massless field
 and assuming that the U(1) is strongly coupled is rather natural.

The above correction could in fact be of phenomenological interest to describe a known small discrepancy
 in gauge coupling unification. Figure \ref{gcu} shows the two-loop running of the MSSM gauge couplings in the region
 around $10^{16}$ GeV adapted from \cite{Ibe:2003ys}.  The fact that there is not exact unification may be interpreted
by saying that the line $1/\alpha_1$ is one unit higher than it should. This is precisely the kind of correction
 provided by eq.(\ref{correccionu1}) for $\alpha_{\rm RR} \sim {k_Y^2}/{k_{\rm RR}^2}$.

Of course this should be taken with some care since additional threshold effects may be also present, leading to extra
 contributions to the gauge couplings.
 In particular, additional  corrections may come from the
 $\overline{F}_2\wedge \overline{F}_2$ term in eq.(\ref{fd7}). For the MSSM gauge kinetic functions these corrections read \cite{Donagi:2008ca, Blumenhagen:2008aw, Conlon:2009qa}
 \beqa
 f_{SU(3)}\ &=& T\ -\frac{1}{2}\tau \int_S{\overline F}_a\wedge {\overline F}_a \label{lastres}\\
 f_{SU(2)}\ & = & T - \frac{1}{2}\tau \int_S \left({\overline F}_a\wedge {\overline F}_a+{\overline F}_Y\wedge{\overline F}_Y+2{\overline F}_a\wedge{\overline F}_Y\right) \nonumber\\
\frac {3}{5} f_{U(1)}\ & = & T - \frac{1}{2}\tau \int_S \left({\overline F}_a\wedge {\overline F}_a+\frac {3}{5}({\overline F}_Y\wedge {\overline F}_Y+2{\overline F}_a\wedge {\overline F}_Y)\right) \ .
\nonumber
\eeqa
where $\tau$ is the complex dilaton and  ${\overline F}_a$ are fluxes along the U(1) contained in
the U(5) gauge group of the D7-branes (see  \cite{Blumenhagen:2008aw}).
These corrections by themselves would imply an ordering of the size of the fine structure constants
at the string scale given by
\beq
\frac {1}{\alpha_3} \ < \frac {1}{\alpha_1} \  < \frac {1}{\alpha_2} \ .
\eeq
As remarked in \cite{Blumenhagen:2008aw}, this ordering seems incompatible with that appearing in the unification region (see Figure \ref{gcu}), so that it was suggested in \cite{Blumenhagen:2008aw} that threshold  corrections from the Higgs triplets in SU(5) combined
with those from eq.(\ref{lastres}) could adjust the results for the couplings.  In our scheme such Higgs triplet threshold
corrections would be unnecessary.

\section{Adding background fluxes}\label{fluxes}

Closed string background fluxes are a prominent mechanism for generating non-trivial scalar potentials for the moduli of the compactification \cite{Grana:2005jc}. In type IIA orientifold compactifications, solutions to the equations of motion in presence of non-vanishing RR flux require the internal space to be a half-flat manifold \cite{Gurrieri:2002wz}, instead of Calabi-Yau. Alternatively, it is possible to keep the Calabi-Yau condition for the internal manifold\footnote{Neglecting backreaction of the fluxes and localized sources.} if NSNS 3-form fluxes and a non-zero VEV for the Romans parameter are also considered \cite{DeWolfe:2005uu}.

Having $\cn=1$ supersymmetry in 4d requires the compactification to preserve an $SU(3)$ structure \cite{Chiossi:2002tw, Gauntlett:2001ur}. The latter can be still completely characterized in terms of an $SU(3)$ invariant non-degenerate 2-form $J$ and a holomorphic 3-form $\Omega$ but, in contrast to the SU(3) holonomy case, these are not necessarily closed forms, $dJ\neq 0$, $d\Omega\neq 0$. In particular, for half-flat manifolds $dJ$ and $d\Omega$ satisfy the conditions,
\begin{equation}
J\wedge dJ=0\ , \qquad \textrm{Im}(d\Omega)=0
\end{equation}
Hence, families of half-flat orientifolds can be built by  twisting the $\sigma$-odd cohomology of a Calabi-Yau orientifold as
\begin{equation}
f \ : \ H^2(\cM,\mathbb{R})_-\ \to \ H^3(\cM,\mathbb{R})_-\ , \quad \textrm{such that} \quad d\omega_{\hat i}=f_{\hat i I}\beta^I\label{torflat}
\end{equation}
generalizing the construction that we presented at the end of section \ref{Mtorsion}. In the following we discuss two main features that appear in this type of SU(3)-structure manifolds: the appearance of F-terms and their interplay with D-terms and the fact that D-brane gauge kinetic functions may depend on open string moduli.

\subsection{F-terms and Freed-Witten anomalies}
\label{fwsec}

The equations of motion for a type IIA SU(3) structure orientifold compactification with fluxes can be conveniently expressed (in the limit of diluted RR fluxes) as the vanishing of the F-terms of the following 4d effective superpotential \cite{Grimm:2004ua, Villadoro:2005cu},
\begin{equation}
W=\int_{\cM}\left[\Omega_c\wedge (H_{\rm NS}+idJ)\ + \ e^{J_c}\wedge F_{\rm RR}\right]\label{fluxsuper}
\end{equation}
Here $F_{\rm RR}$ denotes the formal sum of RR field-strengths, $F_{\rm RR}=F_0+F_2+F_4+F_6$, whereas $H_{\rm NS}$ is the NSNS 3-form. Note that this superpotential may a priori depend on all moduli of the compactification. In particular the vev's of K\"ahler moduli governing the gauge kinetic function of RR U(1) gauge symmetries (and therefore their mass, for massive RR U(1)'s) can be fixed in this way.\footnote{Apart from superpotential (\ref{fluxsuper}), the torsion in eq.(\ref{torflat}) induces also a superpotential in the worldvolume of D6-branes for the open-string moduli (c.f. eq.(\ref{supoD6})) \cite{D6torsion}. Thus, the amount of kinetic mixing between RR and D6-brane U(1) symmetries can be also stabilized in half-flat orientifold compactifications.}

We have summarized in Table \ref{fdtable} the higher dimensional origin of 4d F-terms and D-terms in general type IIA SU(3) structure orientifold compactifications. We have seen already that, neglecting torsional 1-cycles, D-terms in the 4d theory are associated to massive RR U(1) vector multiplets coming from $\sigma$-even torsional 2-cycles ($\textrm{Tor }H_{2}^+(\cM,\mathbb{Z})$) and to massive D6-brane U(1) vector multiplets. All of these have a common origin in the torsional 2-cycles of the $G_2$ manifold in M-theory ($\textrm{Tor }H_{2}(\hat{\mathcal{M}}_7,\mathbb{Z})$). Similarly, from eq.(\ref{fluxsuper}) we observe that F-terms are associated to background fluxes of the NSNS and RR forms of type IIA supergravity and to $\sigma$-odd torsional 2-cycles ($\textrm{Tor }H_{2}^-(\cM,\mathbb{Z})$). These have an M-theory origin on background fluxes of the M-theory 4-form and the torsional 3-cycles of the $G_2$ manifold ($\textrm{Tor }H_{3}(\hat{\mathcal{M}}_7,\mathbb{Z})$) encoded in the non-closure of the $G_2$ invariant 3-form, $d\Phi_3\neq 0$.

\begin{table}
\begin{center}
\begin{tabular}{|c||c|c|}
\hline
& F-terms & D-terms \\
\hline \hline
type IIA & $F_{\rm RR},\ H_{\rm NS},\ \textrm{Tor }H_{2}^-\simeq \textrm{Tor }H_{3}^+$&$\textrm{Tor }H_{2}^+\simeq \textrm{Tor }H_{3}^-, \ \textrm{D6-branes}$\\
\hline
M-theory & $G_4,\ \textrm{Tor }H_{3}$ & $\textrm{Tor }H_{2}\simeq \textrm{Tor }H_{4}$ \\
\hline
\end{tabular}
\end{center}
\caption{Higher dimensional origin of F-terms and D-terms of the 4d effective theory in a general type IIA $SU(3)$ structure orientifold compactification. $G_4$ denotes the M-theory 4-form field-strength. We have not considered torsional 1-cycles.\label{fdtable}}
\end{table}

The interplay between F-terms and D-terms in the 4d effective theory is subtle. Shift symmetries of axions which participate in some St\"uckelberg mechanism should not be spoiled by quadratic or higher order couplings induced by superpotential (\ref{fluxsuper}). As it was shown in \cite{Camara:2005dc}, for massive D6-brane U(1) gauge symmetries this leads to a set of constraints which turn out to be equivalent to the cancelation of Freed-Witten (FW) anomalies \cite{Freed:1999vc, Maldacena:2001xj, Cascales:2003zp} in the worldvolume of D6-branes. Indeed, from eq.(\ref{stuck}) we observe that the RR 3-form transforms under a D6-brane U(1)$_a$ gauge transformation as,
\begin{equation}
A^a\to A^a+d\chi \quad \Rightarrow \quad \delta_aC_3=-c^I_aN_a\alpha_I\chi
\end{equation}
Requiring this to be a symmetry of the superpotential (\ref{fluxsuper}) leads to the generalized FW condition \cite{Camara:2005dc, Villadoro:2006ia},
\begin{equation}
\delta_aW=0\quad \Rightarrow \quad \int_{\pi_a}(H_{\rm NS}+idJ)=0 \quad \forall J\label{fw}
\end{equation}
Moreover, it was noticed in \cite{Cascales:2003zp} that this condition can be relaxed if D4-branes stretching between D6-branes and their orientifold images are also present in the compactification.

In the context of the more general St\"uckelberg Lagrangian (\ref{stucktotal}), we have seen that $C_3$ can also transform under RR U(1) gauge transformations,
\begin{equation}
A^\alpha\to A^\alpha+d\chi \quad \Rightarrow \quad \delta_\alpha C_3=-k_{\alpha}{}^\b\alpha_{\b}^{\rm tor}\chi
\end{equation}
Following the same reasoning than before, we obtain the following additional consistency condition,
\begin{equation}
\delta_\alpha W=0\quad \Rightarrow \quad \int_{\pi_3^{{\rm tor},\alpha}}(H_{\rm NS}+idJ)=0 \quad \forall J\label{fwtor}
\end{equation}
for any $\pi_3^{{\rm tor},\a}\in \textrm{Tor }H_{3}^-(\cM,\mathbb{Z})$. Let us look in more detail to this condition. First of all, it requires that the net $H_{NS}$ flux threading any $\sigma$-odd torsional 3-cycle vanishes. If there were a non-zero flux of $H_{NS}$, then $dH_{NS}\neq 0$, and the Bianchi identity for $H_{NS}$ would not be satisfied. By this argument we therefore also expect that (\ref{fwtor}) can be relaxed in the presence of NS5-branes wrapping dual torsional 2-cycles belonging to $\textrm{Tor }H_{2}^+(\cM,\mathbb{Z})$.

Similarly, the constraint (\ref{fwtor}) for $dJ$ admits also a natural interpretation. We can express it equivalently as,
\begin{equation}
\textrm{Tor}\ H^3_{-}(\cM,\mathbb{Z})\cap \widehat{\textrm{Tor}}\ H^3_{-}(\cM,\mathbb{Z})=0
\end{equation}
which, from the point of view of bijections (\ref{twist}) and (\ref{torflat}), simply accounts for the nilpotency of the exterior derivative, $d^2=0\ \Rightarrow \ f_{\hat i I}k_j{}^I=0$ \cite{Cvetic:2007ju}.

The conditions (\ref{fw}) and (\ref{fwtor}) can be discussed in a unified way from the point of view of their M-theory lift. Indeed, they both reduce to the M-theory constraint
\begin{equation}
\int_{\pi^{{\rm tor},\a}_4}(G_4+d\Phi_3)=0 \qquad \forall \Phi_3 \label{fwm}
\end{equation}
for every $\pi^{{\rm tor},\a}_4\in \textrm{Tor }H_4(\hat{\mathcal{M}}_7,\mathbb{Z})$. This condition could have been directly derived by requiring the M-theory superpotential \cite{Beasley:2002db} to be invariant under U(1) gauge transformations of massive torsional U(1) symmetries. By similar arguments, eq.(\ref{fwm}) can be relaxed if M5-branes wrapping dual torsional 2-cycles in $\textrm{Tor }H_2(\hat{\mathcal{M}}_7,\mathbb{Z})$ are present.

\subsection{Adjoint-dependent gauge kinetic functions}

We have seen in section \ref{kmixing} that the kinetic mixing $f_{ia}$ between open and closed string U(1)'s is a non-trivial holomorphic function of the open string moduli $\Phi^j_{a}$ that describe the embedding of the D6-brane 3-cycle $\pi_a$. The only requirement for this to be the case is that the 2-cycle $\rho_j \subset \pi_a$ associated to $\Phi^j_a$ is a non-trivial element of $H_2^+(\cam_6, \IR)$. The D6-brane gauge kinetic function $f_{a}$ has on the other hand a constant value all over the open string moduli space, simply because
\begin{equation}
f_{a}\, =\, \int_{\pi_a} \Omega_c\, =\, \int_{\pi_a} \left(C_3 + i e^{4A -\phi_{10}} \re(\Omega)\right)
\label{gkfCY}
\end{equation}
and for CY$_3$ orientifolds $d\Omega_c = 0$, at least in the constant warp factor limit $dA = 0$.

For flux compactifications on half-flat manifolds, however, this does no longer need to be true, since in general $\re d\Omega \neq 0$. Indeed, let us consider type IIA compactifications to 4d $\cn=1$ Minkowski vacua. Supersymmetry imposes the following conditions on the background \cite{Kaste:2002xs}
\begin{eqnarray}
\label{fluxrel1}
d(3A- \phi_{10})\, =\, H_{\rm NS} + i dJ\, =\,  0 & \quad & F_0\, =\, F_4\, =\, F_6 \, =\, 0  \\
d(e^{2A-\phi_{10}}\im \Omega)\, =\, 0 & \quad & d(e^{4A-\phi_{10}} \re \Omega)\, =\,  - e^{4A} *_6 F_2
\label{fluxrel2}
\end{eqnarray}
where $F_2 = dC_1$ is the RR 2-form field strength (not to be confused with the D6-brane gauge field strength $F_2^{a}$). Even if $\im \Omega_c$ is non-closed, for a D6-brane wrapped on a sLag 3-cycle $\pi_a$ eq.(\ref{gkf0}) is still true. Hence, we see that $f_{a}$ depends on the particular embedding of $\pi_a$, and therefore on the open string moduli $\Phi_{a}^j$. Part of this dependence is due to the fact that the warp factor is non-constant, and it arises even in the absence of any twist (\ref{torflat}), by simply taking into account the backreaction of the D6-branes. We will not be interested in this warp factor dependence of $f_{a}$, which following \cite{threshold} can be interpreted as a threshold correction to the gauge kinetic function, but rather on a $\Phi$-moduli dependence that remains even in the limit of constant warp factor.

Indeed, in the limit of constant warp factor we have that $F_2$ is a primitive (1,1)-form, and so
\begin{equation}
d(e^{4A-\phi_{10}} \re \Omega)\, =\, e^{4A} J \wedge F_2
\end{equation}
On the other hand, the Chern-Simons part of the D6-brane action contains a coupling of the form
\begin{equation}
\label{CSD6b}
S_{CS}\,  =\,  \oh \int_{\IR^{1,3}} F_2^{a} \wedge F_2^{a} \int_{\pi_a} \mathcal{F}_2^{a} \wedge C_1
\end{equation}
Combining both terms and taking a Lie derivative of the DBI + CS actions we obtain that the gauge kinetic function depends on the open string moduli $\Phi^j_a$ as
\begin{equation}
f_{a}\, =\, f_{a}|_{\Phi^j_{a}=0} \, -\, i \cp_{j}^{a} \Phi^j_{a} + \dots
\label{gkfdep}
\end{equation}
where
\begin{equation}
\mathcal{P}^{a}_{j} \, \equiv\,
\int_{\pi_a} F_2 \wedge \zeta_j\, =\, \int_{\rho_j} F_2
\label{Fint}
\end{equation}
Hence, if the pull-back of the RR field strength $F_2$ is topologically non-trivial over a 2-cycle $\rho_j$ within a D6-brane 3-cycle $\pi_a$, then the gauge kinetic function $f_{a}$ will depend non-trivially on the corresponding open string modulus $\Phi_a^j$.

This result is quite similar to the one obtained for the kinetic mixing, eq.(\ref{kinmix}). Indeed, if we compare (\ref{gkfdep}) with the expression for the kinetic mixing (\ref{kinmix}), we just need to replace $\omega_i \raw F_2$. The 2-form $F_2$ is however quite different from $\omega_i$. Indeed, from eqs.(\ref{fluxrel1}) and (\ref{fluxrel2}) we observe that $F_2$ is a non-closed $\sigma$-odd primitive (1,1)-form.
Moreover, as shown in \cite{D6torsion}, $[dF_2]/N$ is Poincar\'e dual to some torsional 3-cycle $[\Lambda_{F}]$ wrapped by some O6-plane, a fact that relaxes the RR tadpole conditions and allows certain D6-branes to be BPS while wrapping purely torsional 3-cycles. Hence, using the language of section \ref{mmixing} we conclude that $F_2 \in \widehat{\textrm{Tor}\, }H^{2}_-(\cam_6, \IZ)$, and therefore (\ref{Fint}) is nothing but the torsion linking number of $[\Lambda_F]$ and $[\rho_j]$. That is, in order for $f_{a}$ to depend on some open string modulus $\Phi_a^j$, the associated 2-cycle $\rho_j$ should have a non-trivial component on the torsion homology group ${\textrm{Tor}\, }H_2^-(\cam_6, \IZ)$.

\section{Conclusions}\label{conclu}

In this paper we have analyzed an important aspect of 4d type II compactifications and their M/F-theory relatives, namely the structure of Abelian gauge symmetries that survive at low energies. We have in particular considered those Abelian symmetries that in one way or another couple to the Standard Model (SM) degrees of freedom of any realistic compactification of this kind. Naively, these amount to the D-brane U(1)'s that remain massless after the St\"uckelberg couplings of \cite{Aldazabal:2000dg} have been taken into account. We have however seen that Abelian symmetries arising from the closed string RR sector of the theory can also play a non-trivial role in describing the visible sector of a realistic compactification.

One simple way this can happen is via the kinetic mixing of the SM hypercharge and a massless RR U(1) gauge symmetry. Such kind of kinetic mixing between open and closed string U(1)'s have been previously discussed in the D-brane literature, and are in general quite difficult to compute. Here we have provided a global geometric description of such mixing, which may help computing this U(1)$_Y-U(1)_{RR}$ kinetic mixing in specific type II models. In particular, in type IIA intersecting D6-brane models an open string U(1) is given by a formal sum of 3-cycles in the compactification manifold $\cam_6$, namely those 3-cycles wrapped by the D6-branes, together with a 4-chain $\Sigma_4$ that connects them. The open-closed kinetic mixing is then expressed as an integral over this 4-chain $\Sigma_4$, see eq.(\ref{kinmixchain}). Note that previous expressions in the literature rely on the existence of open string moduli $\Phi^j$ for the D6-branes, and basically provide the dependence of the kinetic mixing $f_{ib}$ on them. These $\Phi^j$ are however massless adjoint fields which are unwanted in a realistic model, and so in practice one needs an expression like (\ref{kinmixchain}) that provides the kinetic mixing even in the absence of any open string modulus.

Kinetic mixing is however not the most direct interplay between RR and open string Abelian symmetries. One can see this by first realizing that RR U(1)'s are not the only class of Abelian gauge symmetries that arise from the RR sector of a compactification. In general one will also have discrete $\IZ_k$ gauge symmetries which, as shown in \cite{bs10} are actually a massive U(1) gauge symmetry broken down to $\IZ_k$ via an St\"uckelberg mechanism. As argued in \cite{bs10} these $\IZ_k$ gauge symmetries should be accompanied by Aharanov-Bohm strings and particles charged under them, and we have seen that for type II/M-theory compactifications this is the case if the compactification manifold $\cam$ contains a very specific topological feature: a non-trivial torsion homology group ${\rm Tor\, } H_*(\cam, \IZ)$. Torsion homology groups are generic in type II/M-theory compactification manifolds, but oftentimes ignored because they are invisible to usual methods of dimensional reduction. In particular, for Calabi-Yau compactifications torsional groups in (co)homology are not associated to any massless sector of the theory. From our findings we see that they are however related to a very special massive sector: a RR U(1) gauge symmetry with a topological, built-in St\"uckelberg coupling.

The above result would perhaps not be very relevant for phenomenology was it not for the fact that D-brane U(1) can also participate in such built-in St\"uckelberg mechanism. Indeed, a careful analysis shows that, e.g., D6-brane wrapping torsional 3-cycles couple to the 4d 2-forms that mediate this mechanism. Hence, in order to  know if a D6-brane U(1) is massless, we should know if its associated 3-cycle contains a torsional piece or not. If it does, then the built-in St\"uckelberg mechanism induces a mass mixing between this D6-brane and several torsional RR U(1)'s, and the resulting massless U(1) will be a linear combination of all of them. Hence, for many D-brane models the naive spectrum of massless open-string U(1)'s is not so. Several of them are actually contaminated by RR torsional U(1)'s.

We have provided an explicit type IIA example in which such mass mixing occurs, and which illustrates several scenarios of phenomenological interest. In fact, even if our discussion has mainly taken place in the context of type IIA compactifications, we have found that the most direct application of our results takes place in the context of type IIB/F-theory GUT models. Indeed, most GUT F-theory constructions are based on relating the hypercharge U(1)$_Y$ to a 2-cycle $\rho^Y$ trivial in $H_2(\cam, \IR)$. This however leaves the possibility for $\rho^Y$ to be non-trivial in ${\rm Tor\, }H_2(\cam, \IZ)$. If that were the case then the open string U(1)$_Y$ would not be massless, but rather U(1)$_Y^\prime =$ U(1)$_Y + $U(1)$_{RR}$. In particular, this would mean that the fine structure constant $\a_1$ for such models should be recomputed, with a non-trivial contribution coming from $\a_{RR}$. Interestingly, we find that this contribution substantially alleviates the gauge coupling unification problems pointed out in \cite{Blumenhagen:2008aw}. It would be remarkable if the key for gauge coupling unification in F-theory relied in the torsional homology of the compact manifold.

On a more formal side, along our discussion of U(1)'s in type IIA models we have found that a key role is played by the 2-cycles $\rho^j$ within the 3-cycles $\pi_3$ wrapped by the D6-branes. Recall that for a D6-brane wrapped on a BPS 3-cycle $\pi_3$ the open string adjoint moduli $\Phi^j$ are in one-to-one correspondence with the non-trivial 2-cycles $\rho^j$ of $\pi_3$. In general, it is not known whether such 2-cycles are trivial in the ambient space $\cam_6$ or not. We have however found that the interesting physics happens whenever they are non-trivial, in the sense that then $\Phi^j$ enters into some effective theory quantity. We have summarized these results in Table \ref{conctable}. It would be very interesting to explore if, via some effective field theory argument, one can obtain a general result on when the 2-cycles of a special Lagrangian are non-trivial in the compactification manifold.

\begin{table}
\begin{center}
\begin{tabular}{|c||c|}
\hline
$\rho^j \subset \pi_a$ is non-trivial on & $\Phi_a^j$ appears on \\
\hline \hline
$H_2^+(\cam_6, \IZ)$ & $f_{ia}$ \ (\ref{kinmixchain}) \\
\hline
$H_2^-(\cam_6, \IR)$ & $W_{{\rm D6}_a}$ \ (\ref{supoD6}) \\
\hline
${\rm Tor\, }H_2^-(\cam_6, \IZ)$ & $f_a$ \ (\ref{gkfdep}) \\
\hline
\end{tabular}
\end{center}
\caption{Relation between the topology of the non-trivial 2-cycles $\rho^j$ of a D6-brane 3-cycle $\pi_a$ and the quantities of the low energy effective action in which it appears. We have included the equations that describes this quantity in the main text. The last line is only true for the flux compactifications of section \ref{fluxes}.
\label{conctable}}
\end{table}

\vspace*{2cm}

\begin{center}{\bf Acknowledgments}\\\end{center}

We thank E.~Dudas, I.~Garc\'{\i}a-Etxebarria, E.~Palti, R.~Savelli, G.~Shiu, A.~Uranga and J.~Walcher for useful discussions. P.G.C. thanks IFT UAM/CSIC and CPhT Ecole Polytechnique and F.M. thanks the CERN TH group and HKUST-IAS for hospitality during the completion of this paper.
This work has been partially supported by the grants FPA 2009-09017, FPA 2009-07908, Consolider-CPAN (CSD2007-00042) from the MICINN, HEPHACOS-S2009/ESP1473 from the C.A. de Madrid and the contract ``UNILHC" PITN-GA-2009-237920 of the European Commission.
F.M. is supported by the MICINN Ram\'on y Cajal programme through the grant RYC-2009-05096.

\newpage

\appendix

\section{D6-brane dimensional reduction}
\label{apdim}

In this appendix we dimensionally reduce the terms of the D6-brane DBI-CS action that are relevant for the purposes of this work (see also \cite{Grimm:2011dx, Kerstan:2011dy} for the reduction of these and other terms in the action). In particular we are interested in computing St\"uckelberg couplings and mixed terms between RR and D6-brane U(1) factors in the gauge kinetic function. These arise from the piece of the action which contains the RR 3-form and 5-form,
\begin{align}
S_{CS}^{(a)}&=\mu_6\int_{\mathbb{R}^{1,3}\times \pi_a}P\left[C_5\wedge \mathcal{F}^a_2+\frac12C_3\wedge \mathcal{F}^a_2\wedge \mathcal{F}^a_2\right]\label{csa}\\
&= \mu_6\int_{\mathbb{R}^{1,3}\times \pi_a}\left(1+\frac12\mathcal{L}_{\phi_{a}}+\ldots\right)\left[C_5\wedge \mathcal{F}^a_2+\frac12C_3\wedge \mathcal{F}^a_2\wedge \mathcal{F}^a_2\right]\nonumber
\end{align}
where
\begin{equation}
\mathcal{F}^a_2\equiv F_2^a+B_2
\end{equation}
In this expression $\mu_6$ is the D6-brane charge and $P[\ldots]$ denotes the pull-back to the worldvolume of the D6-brane. We have performed a normal coordinate expansion to linear order in the geometric deformations (\ref{geod6}).

We follow the usual procedure for dimensional reduction. That is, we expand $C_3$ and $C_5$ in the basis of forms, as in eqs.(\ref{f4exp}) and (\ref{f6exp}). In addition we have argued in section \ref{massivesec} that it is possible to introduce an extra set of torsional forms in order to also account for the torsional cycles of the Calabi-Yau (c.f. eqs.(\ref{torexpC3})-(\ref{torexpC5})). The complete field strength expansions read (see footnote \ref{no3forms}),
\begin{align}
F_4&=\textrm{Re}(dN^{I})\wedge \alpha_{I} + dA^i\wedge\omega_i+\left(\textrm{Re}(dN^{\alpha})+k^\alpha{}_{\beta} A^{\b}\right)\wedge \alpha_{\alpha}^{\rm tor}+dA^{\alpha}\wedge\omega_\alpha^{\rm tor}\label{f4ap}\\
F_6&=dV^{i}\wedge \tilde{\omega}^{i}+dC_2^{I}\wedge\beta^{I}+\left(dV^{\alpha}-k^\a{}_{\b} C_2^{\b}\right)\wedge \tilde{\omega}^{{\rm tor},\alpha}+dC_2^{\alpha}\wedge\beta^{{\rm tor},\alpha}\label{f6ap}
\end{align}
Plugging these expressions into (\ref{csa}) and integrating by parts we obtain
\begin{multline}
S_{CS}^{(a)}=
\mu_6\int_{\mathbb{R}^{1,3}}\left[-\left(c_a^\beta C_2^\beta+c_a^IC_2^I\right)\wedge F_2^{a} +\frac12 d_{Ia}\textrm{Re}(dN^{I})\wedge A^{a}\wedge F_2^{a}\right.\\
\left.+\frac12 \left(\mathcal{R}_{i,j}^{a}\phi^j_{a}dV^i+ \mathcal{M}_{ij}^{a}\theta^j_{a}dA^i + \mathcal{S}_{i,j,\hat k}^{a}\phi^j_{a}\left(\textrm{Re}(T^{\hat k})dA^i -A^i\wedge\textrm{Re}(dT^{\hat k})\right) \right)\wedge F_2^{a}+\ldots\right]\label{csfree}\end{multline}
where $B_2=\textrm{Re}(T^{\hat k})\omega_{\hat k}$, Wilson line moduli $\theta^j_{a}$ were defined in eq.(\ref{wilson}) and the topological invariants $c_a^I$, $d_{Ia}$ and $c_a^\beta$ in eqs.(\ref{poinc1}) and (\ref{decomlink}) (see also Appendix \ref{tortopo}). Moreover, we have introduced the integrals,
\begin{equation}
\mathcal{M}_{ij}^{a}=\int_{\pi_a}\omega_i\wedge\zeta_j\ , \qquad
\mathcal{R}_{i,j}^{a}=\int_{\pi_a}\iota_{X_j}\tilde{\omega}^i\ , \qquad
\mathcal{S}_{i,j,\hat k}^{a}=\int_{\pi_a}\iota_{X_j}\omega_i\wedge\omega_{\hat k}\ , \qquad
\end{equation}
where inclusion of the integrand to the 3-cycle $\pi_a$ should be understood in all these expressions.

Notice that both electric and magnetic degrees of freedom appear explicitly in eq.(\ref{csfree}). The reason is that CS actions are given in a democratic formulation, so that all RR forms appear explicitly in the action. In order to express (\ref{csfree}) in terms of only electric degrees of freedom, we note that
\begin{multline}
\mathcal{R}_{i,j}^{a}dV^i-\mathcal{S}^{a}_{i,j,\hat k}A^i\wedge\textrm{Re}(dT^{\hat k})=-\mathcal{R}_{i,j}^{a}\mathcal{K}_{il\hat k}\textrm{Im}(T^{\hat k})*_4dA^{l}=\\
=-\mathcal{S}_{l,j,\hat k}^{a}\textrm{Im}(T^{\hat k})*_4dA^{l}=-\mathcal{M}_{lk}^{a}\textrm{Im}(\lambda_{j}^k)*_4A^{l}
\end{multline}
In this expression the first equality is obtained from applying the 10d relation $\hat{F}_4=*_{10}\hat{F}_6$, with $\hat{F}_p=dC_{p-1}-C_{p-3}\wedge dB_2$, in eqs.(\ref{f4ap}) and (\ref{f6ap}), whereas  for the second equality we have made use of $\mathcal{K}_{ij\hat k}\tilde{\omega}^i=\omega_j\wedge\omega_{\hat k}$. Finally, we have made use of eq.(\ref{lambda}) in order to express the result in terms of $\lambda_{j}^k$. Moreover, one may also check that
\begin{equation}
\mathcal{S}_{i,j,\hat k}^{a}\textrm{Re}(T^{\hat k})dA^i=\mathcal{M}_{ik}^{a}\textrm{Re}(\lambda_{j}^k)dA^i
\end{equation}
Putting all pieces together we finally obtain,
\begin{multline}
S_{CS}^{(a)}=
\mu_6\int_{\mathbb{R}^{1,3}}\left[-\left(c_a^\beta C_2^\beta+c_a^IC_2^I\right)\wedge F_2^{a} +\frac12 d_{Ia}\textrm{Re}(dN^{I})\wedge A^{a}\wedge F_2^{a}\right.\\
\left.+\frac12\mathcal{M}_{ik}^{a}\left(\textrm{Re}(\Phi^k_{a})dA^i\wedge F_2^{a}-\textrm{Im}(\Phi_{a}^k)*_4dA^{i}\wedge F_2^{a}\right)+\ldots\right]\label{csfinal}\end{multline}
where we have expressed the result in terms of the complex open string moduli $\Phi_a^k$, defined in eq.(\ref{openmoduli}). The first term in the integrand is the St\"uckelberg coupling giving mass to some linear combination of U(1) gauge bosons that we discussed in section \ref{stuckrev}. Indeed, adding the kinetic term for the 2-forms (which can be obtained by dimensionally reducing the 10d $F_6$ kinetic term) and integrating out $C_2^I$ (see for instance \cite{Ghilencea:2002da}), leads to eq.(\ref{stucktotal}). The second term in (\ref{csfinal}) corresponds to the coupling of complex structure axions to D6-brane U(1) gauge bosons.  It combines with the kinetic term for D6-brane U(1) gauge bosons (obtained by dimensionally reducing the DBI action \cite{Grimm:2011dx, Kerstan:2011dy}) to give the tree-level gauge kinetic function of  D6-brane U(1) gauge bosons,
\begin{equation}
f_{a}=-i\int_{\pi_a}\Omega_c
\end{equation}
The remaining terms in the integrand of eq.(\ref{csfree}) correspond to the kinetic mixing between RR and D6-brane U(1) gauge symmetries discussed. These can be expressed in terms of a mixed gauge kinetic function,
\begin{equation}
f_{ia}=-i\Phi^k_{a}\int_{\pi_a}\omega_i\wedge\zeta_k+\ldots
\end{equation}
which is well-defined up to a $\Phi$-independent term, as discussed in section \ref{kmixing}.

\section{Type IIB compactifications}
\label{typeIIB}

For the most part of this work we have discussed RR U(1) gauge symmetries arising in type IIA Calabi-Yau orientifold compactifications. Similar considerations, however, apply to type IIB Calabi-Yau orientifold compactifications and their F-theory relatives. In this Appendix we rephrase the main results of this paper in the language of type IIB Calabi-Yau orientifolds. Since both types of compactifications are related by mirror symmetry and most of the ingredients are topological, the discussion follows closely the one in the main part of the paper. This alternative exposition, however, is better adapted to some of the phenomenological applications with D7-branes which we describe in section \ref{pheno}.

We consider type IIB Calabi-Yau orientifold compactifications with D3 and/or D7-branes. The orientifold action is given by $\Omega_p(-1)^{F_L}\sigma$, and the involution $\sigma$ satisfies \cite{Acharya:2002ag, Brunner:2003zm},
\begin{equation}
\sigma J=-J\ , \qquad \sigma\Omega=-\Omega
\end{equation}
Fixed loci of $\sigma$ are points and/or complex 4-cycles in $\cM$, and lead to O3 and O7-planes respectively. In order to cancel the RR-charge of the O-planes one may therefore introduce D3-branes and/or magnetized D7-branes wrapping complex 4-cycles in $\cM$.

Since the roles of $h^{1,1}(\cM)$ and $h^{1,2}(\cM)$ are exchanged under mirror symmetry, the closed string spectrum of 4d massless fields now consists of $h^{1,1}+h^{1,2}_-+1$ chiral multiplets and $h^{1,2}_+$ vector multiplets of the 4d $\cn=1$ supersymmetry \cite{Brunner:2003zm, Grimm:2004uq}. The moduli space, spanned by the scalar components of the chiral multiplets, consists of $h^{1,1}$ K\"ahler moduli, $h^{1,2}_-$ complex structure moduli, and a complex axiodilaton, $\tau=C_0+ie^{-\phi_{10}}$. To simplify the discussion, we set $h_-^{1,1}=0$ in what follows, without loss of generality of our results. With that assumption, all K\"ahler moduli of the compactification come from the expansion \cite{Grimm:2004uq}.
\begin{equation}
\mathcal{J}_c\equiv C_4-\frac{i}{2}e^{-\phi_{10}}J\wedge J=-T^i\tilde\omega_i\ , \label{kahleriib}
\end{equation}
with $\tilde\omega_i$ a basis of $\sigma$-even 4-forms.

Chiral matter in type IIB orientifold compactifications typically arise from D3 and/or magnetized D7-brane intersections. We are particularly interested in the case of D7-branes, as they play a prominent role in F-theory GUT model building \cite{Heckman:2010bq}. At generic points of the moduli space, each  stack of $N_a$ D7-branes with equal magnetization carries a $U(N_a)$ gauge theory in its worldvolume. The 4d gauge kinetic function is given by \cite{Jockers:2005zy}
\begin{equation}
f_a=-iN_a\int_{S_a}\left[\mathcal{J}_c+ \tau \textrm{Tr}(\mathcal{F}_2\wedge \mathcal{F}_2)\right]\label{fd7}
\end{equation}
where $S_a$ is the complex 4-cycle wrapped by the stack of D7-branes. There are complex scalar fields transforming in the adjoint representation of the gauge group. These span the open string moduli space of the D7-brane \cite{Jockers:2004yj} and are given by $h^{1,0}(S_a)$ complex Wilson line moduli, $a^i_{a}$, and $h^{2,0}(S_a)$ geometric moduli, $\Phi^k_{a}$.

Magnetized D7-branes generically develop St\"uckelberg couplings in their 4d effective action, so that their diagonal U(1) gauge boson becomes massive, SU$(N_a)\times $U(1)$_a\to$ SU$(N_a)$. This can be explicitly seen by dimensionally reducing the following piece of the D7-brane Chern-Simons action \cite{Jockers:2005zy}
\begin{equation}
S_{CS}=\int_{\mathbb{R}^{1,3}\times S_a}P[C_4\wedge \mathcal{F}_2^{a}\wedge\mathcal{F}_2^{a}]=\mu_7\int_{\mathbb{R}^{1,3}}C_2^i\wedge F_2^{a}\int_{S_a}\omega_i\wedge \overline{F}_2^{a}+\ldots
\label{stuckd7}
\end{equation}
where $\overline{F}_2^{a}$ denotes the background of $F_2^{a}$ in $S_a$, $\omega_i$ is a basis of 2-forms even under $\sigma$ and $C_2^i$ are the 4d 2-forms dual to the K\"ahler axions $\textrm{Re}(T^i)$. This St\"uckelberg coupling is mirror symmetric to the one described in section \ref{opensec} for D6-branes. As occurs in that case, the discussion can be rephrased in terms of homology classes, however, for D7-branes the relevant homology group is $H_2^+(\cM,\mathbb{Z})$ instead of $H_3^-(\cM,\mathbb{Z})$. Indeed, if $\rho^F_a$ denotes the Poincar\'e dual of $\overline{F}_2^a$ in $S_a$, we can express (\ref{stuckd7}) as
\begin{equation}
S_{CS}=\int_{\mathbb{R}^{1,3}}C_2^i\wedge F_2^{a} \int_{\rho^F_a}\omega_i
\end{equation}
Massless U(1) gauge bosons thus correspond to combinations for which $[\rho^F_a]$ is trivial in $H_2^+(\cM,\mathbb{Z})$, so that there is a 3-chain $\Sigma_3\subset \cM$ whose boundary is $\partial\Sigma_3=\rho^F_a \subset S_a$.

Besides the gauge symmetries coming from the open string sector, there are $h^{1,2}_+(\cM,\mathbb{R})$ massless RR U(1) gauge bosons in the 4d spectrum. These result from dimensionally reducing $C_4$ in a symplectic basis of even 3-forms, $(\alpha_I,\beta^I),\ I=0,\ldots,h^{1,2}_+$. The complete expansion of $C_4$ is thus given by
\begin{equation}
C_4=\sum_I(A^I\wedge\alpha_I+V^I\wedge\beta^I)+\sum_i \left(C_2^i\wedge \omega_i-\textrm{Re}(T^i)\tilde{\omega}^i\right)
\end{equation}
where electric and magnetic vectors, $A^I$ and $V^I$, are related by the 10d self-duality condition $\hat F_5=*_{10}\hat F_5$, with $\hat F_5=dC_4+\frac12 C_2\wedge dB_2-\frac12 B_2\wedge dC_2$. The 4d gauge kinetic function of these RR U(1)'s can be obtained from dimensional reduction of the $\hat F_5$ kinetic term in the 10d type IIB supergravity action. The final result is given by \cite{Grimm:2004uq}
\begin{equation}
f_{IJ}=-i\left.\frac{\partial^2 \mathcal{F}}{\partial\hat \tau_I\partial\hat\tau_J}\right|_{\hat\tau_K=0}\ ,\label{fr7}
\end{equation}
where $\mathcal{F}$ is the $\cn=2$ prepotential of the Calabi-Yau 3-fold, which is a holomorphic function of the $\cn=1$ complex structure moduli $\tau_I$ and of the additional $\cn=2$ complex structure deformations $\hat\tau_K$. The latter ones are projected out by the orientifold, so that $f_{IJ}$ is a holomorphic function depending only on the $\cn=1$ complex structure moduli  \cite{Grimm:2004uq}.

Let us now turn to the discussion of kinetic mixing between RR and D7-brane U(1) gauge symmetries. Kinetic mixing between both types of U(1)'s can be triggered by geometric deformations of the D7-branes. Indeed, expanding the pull-back in the rhs  of eq.(\ref{stuckd7}) to linear order in the geometric deformations of the D7-brane, one obtains
\begin{multline}
S_{CS}^{kin,(1)}=\int_{\mathbb{R}^{1,3}\times S_a}\iota_{\Phi_{a}} F_5\wedge \mathcal{F}_2^{a}\wedge\mathcal{F}_2^{a}+\ldots\\
=\int_{\mathbb{R}^{1,3}}F_2^{a}\wedge dA^I\int_{S_a}\iota_{\Phi_{a}}\alpha_I\wedge \overline{F}_2^{a}+\int_{\mathbb{R}^{1,3}}F_2^{a}\wedge dV^I\int_{S_a}\iota_{\Phi_{a}}\beta^I\wedge \overline{F}_2^{a}+\ldots
\end{multline}
Eliminating the magnetic vectors $dV^I$ by means of the 10d self-duality condition of $\hat F_5$, this leads to the following 4d mixed gauge kinetic function
\begin{equation}
f_{Ia}=-i\int_{S_a}\iota_{\Phi_{a}}\gamma_I\wedge \overline{F}_2^{a}+\ldots=-i\int_{\rho^F_a}\iota_{\Phi_{a}}\gamma_I+\ldots\ , \qquad \gamma_I\equiv \alpha_I + if_{IJ}\beta^J
\end{equation}
As occurs with the analogous expression for D6-branes, eq.(\ref{kinmix}), this derivation has a $\Phi$-independent ambiguity which can be explicitly fixed for massless D7-brane U(1) gauge bosons. In that particular case, following the same reasoning than in section \ref{kmixing}, we can express the 4d mixed gauge kinetic function (up to shifts of the open string moduli) as an integral over the 3-chain $\Sigma_3$ related to the massless combination of D7-brane U(1)'s,
\begin{equation}
f_{Ia}=-i\int_{\Sigma_3}\gamma_I
\end{equation}

Apart from the gauge kinetic mixing triggered by the geometric deformations of the D7-branes, it is also possible to have kinetic mixing between D7-brane and RR U(1) gauge symmetries triggered by Wilson line deformations \cite{Jockers:2004yj}, in models where these are present. Indeed, integrating by parts the r.h.s. of eq.(\ref{stuckd7}) and proceeding as before we get \cite{Jockers:2004yj},
\begin{equation}
S_{CS}^{kin,(2)}=-\int_{\mathbb{R}^{1,3}\times S_a}F_5\wedge F_2^{a}\wedge A^a \quad \rightarrow \quad f_{Ia}=-i\int_{\rho^F_a} a_{a}\wedge \gamma_I
\end{equation}

Finally, D7-brane and RR U(1) gauge symmetries can also mix through the mass matrix induced by the St\"uckelberg mechanism. This is only possible if both types of gauge bosons couple to a common set of 4d 2-forms. As it was thoroughly discussed in section \ref{mmixing}, massive closed string U(1) vector bosons arise from torsional cycles of the Calabi-Yau. We have summarized in Tables \ref{tab5} and \ref{tab6} the 10d origin of the electric and magnetic degrees of freedom of massive closed string U(1) symmetries in type IIB Calabi-Yau orientifold compactifications. These tables are the mirror symmetric counterparts of Tables \ref{tabla1} and \ref{tabla2}.

\begin{table}[!ht]
\begin{center}
\begin{tabular}{|c|c||c|c||c|c|}
\hline
U(1)$_{elec.}$ & group & charged particles & cycle & axions & group \\
\hline \hline
$g^m{}_\mu$&$\widehat{\textrm{Tor}}\ H^{1}_+$& $P$& $\textrm{Tor}\ H_{1}^+$ & $g_{ij}$&$\textrm{Tor}\ H^{2}_+$ \\
\hline
$B^m{}_\mu$&$\widehat{\textrm{Tor}}\ H^{1}_-$& $F1$ & $\textrm{Tor}\ H_{1}^-$ & $B_{ij}$&$\textrm{Tor}\ H^{2}_-$ \\
\hline
$C_{\mu}{}^{m}$&$\widehat{\textrm{Tor}}\ H^{1}_-$& $D1$ & $\textrm{Tor}\ H_{1}^-$ & $C_{ij}$&$\textrm{Tor}\ H^{2}_-$ \\
\hline
$C_{\mu}{}^{mno}$& $\widehat{\textrm{Tor}}\ H^{3}_+$ & $D3$ & $\textrm{Tor}\ H_{3}^+$ &$C_{ijkl}$& $\textrm{Tor}\ H^{4}_+$ \\
\hline
\end{tabular}
\caption{\small{Complete set of massive closed string gauge symmetries and charged states in weakly coupled type IIB Calabi-Yau orientifold compactifications. $P$ denotes the gravity wave and $F1$ the fundamental string. We present also the axions which mediate the St\"uckelberg mechanism giving masses to the corresponding vector boson.}\label{tab5}}
\end{center}
\end{table}

\begin{table}[!ht]
\begin{center}
\begin{tabular}{|c|c||c|c||c|c|}
\hline
U(1)$_{mag.}$ & group & charged strings & cycle & $C_2^{I}$ & group \\
\hline \hline
$KK_{\mu}{}^{mnopq}$&$\textrm{Tor}\ H^{5}_+$ & $KK$ & $\textrm{Tor}\ H_{4}^+$  &$KK_{\mu\nu}{}^{ijkl}$& $\widehat{\textrm{Tor}}\ H^{4}_+$ \\
\hline
$B_{\mu}{}^{mnopq}$& $\textrm{Tor}\ H^{5}_-$ & $NS5$ & $\textrm{Tor}\ H_{4}^-$ &$B_{\mu\nu}{}^{ijkl}$& $\widehat{\textrm{Tor}}\ H^{4}_-$ \\
\hline
$C_{\mu}{}^{mnopq}$& $\textrm{Tor}\ H^{5}_-$ & $D5$ & $\textrm{Tor}\ H_{4}^-$ &$C_{\mu\nu}{}^{ijkl}$& $\widehat{\textrm{Tor}}\ H^{4}_-$ \\
\hline
$C_{\mu}{}^{mno}$&$\textrm{Tor}\ H^{3}_+$& $D3$ & $\textrm{Tor}\ H_{2}^+$ & $C_{\mu\nu}{}^{ij}$&$\widehat{\textrm{Tor}}\ H^{2}_+$ \\
\hline
\end{tabular}
\caption{\small{Dual U(1) magnetic degrees of freedom and 2-forms mediating the St\"uckelberg mechanism. $KK$ denotes the Kaluza-Klein monopole.}\label{tab6}}
\end{center}
\end{table}

Massive RR U(1) gauge bosons come from reduction of $C_2$ on $\textrm{Tor}\ H_{1}^-$ and $C_4$ on $\textrm{Tor}\ H_{3}^+$. We are particularly interested on massive RR U(1) symmetries which arise from the expansion of $C_4$. The reason is that those are the ones which can couple to the same type of axions than magnetized D7-branes do, namely to K\"ahler axions. In order to show this explicitly, we can introduce torsional forms $\omega_\alpha^{\rm tor}\in \widehat{\textrm{Tor}}\ H^{2}_+$ and $\alpha_\alpha^{\rm tor}\in \textrm{Tor}\ H^{3}_+$, with $d\omega_\alpha^{\rm tor}=k_{\alpha}{}^\b\alpha_\b^{\rm tor}$, accordingly to the procedure described in section \ref{massivesec}. We then have
\begin{multline}
dC_4\ =\ \sum_\alpha\left[\left(dA^\alpha+k^\a{}_{\b}C_2^\b\right)\wedge \alpha_\alpha^{\rm tor}\ +\ \left(\textrm{Re}(dT^{\alpha})-k_{\alpha}{}^\b V^{\b}\right)\wedge \tilde{\omega}^{{\rm tor},\alpha}\right.\\
\left.+\ dC_2^\alpha\wedge \omega_\alpha^{\rm tor}\ +\ dV^\alpha\wedge \beta^{{\rm tor},\alpha}\right]\ +\ \ldots
\label{dc4}
\end{multline}
where $\tilde{\omega}^{{\rm tor},\alpha}\in \textrm{Tor}\ H^{4}_+$ and $\beta^{{\rm tor},\alpha}\in \widehat{\textrm{Tor}}\ H^{3}_+$ are the dual forms to $\omega^{\rm tor}_\alpha$ and $\alpha_\alpha^{\rm tor}$ through eq.(\ref{normtor}). Dimensionally reducing the kinetic term of $\hat F_5$ in the 10d type IIB supergravity action we therefore obtain a 4d St\"uckelberg Lagrangian analogous to eq.(\ref{stucktor}) \cite{Grimm:2008ed}.

From eq.(\ref{stuckd7}) we observe that for a stack of magnetized D7-branes to develop a St\"uckelberg coupling to the same 2-form $C_2^\alpha$, the 4-cycle wrapped by the D7-branes must contain the torsional 2-cycle associated to the massive RR U(1) gauge symmetry. Moreover, the Poincar\'e dual of the magnetization should have a non-vanishing component along it,  $\rho^{F}_a\in \textrm{Tor}\ H_{2}^+(\cM,\mathbb{Z})$. In that case we can express the St\"uckelberg coupling in the worldvolume of the D7-branes as,
\begin{equation}
S_{CS}=\int_{\mathbb{R}^{1,3}}C_2^\alpha\wedge F_2^{a} \int_{\rho^{F,{\rm tor}}_a}\omega_\alpha^{\rm tor}+\ldots
\end{equation}
Note that $\omega_\alpha^{\rm tor}$ and $\rho^{F,{\rm tor}}_a$ are torsional on $\cM$, but not necessarily
on the 4-cycle $S_a$. Indeed, defining the 3-chain $\Sigma_3^{\rm tor}$ such that $\partial\Sigma_3^{\rm tor}=k\rho^{F,{\rm tor}}_a$, with $k$ the rank of the torsion, one often finds that $\rho^{F,{\rm tor}}_a\subset S_a$ but $\Sigma_3\not\subset S_a$.

The discussion of which combination of RR and D7-brane U(1) gauge symmetries remain massless then closely follows the one for D6-branes. As we have argued, we can associate an element of $\textrm{Tor}\ H_{2}^+(\cM,\mathbb{Z})$ to each RR U(1) gauge symmetry developing a St\"uckelberg coupling. Hence, given a homology class $[S_a]\in H^+_4(\cM)$, massless combinations of U(1) gauge symmetries are in one to one correspondence with homologically trivial combinations of elements in $H_{2}^+(\cM,\mathbb{Z})$ with non-zero pull-back to $[S_a]$.

\section{D-branes and torsion invariants}\label{tortopo}

One of the most important results regarding torsion in (co)homology is the Universal Coefficient Theorem \cite{bt24}. Rather than (\ref{unitheorem}), the canonical version of this theorem is
\begin{equation}
{\rm Tor\, } H^{r}(\cam_D, \IZ)\, \simeq \, {\rm Hom} \left({\rm Tor\, }H_{r-1}(\cam_D, \IZ), \mathbb{Q}/ \IZ\right)
\label{uctrue}
\end{equation}
That is, each class of torsional $r$-forms $[\tilde{\omega}^{\rm tor}]$ should be understood as a function that maps torsional cycles $\pi_{r-1}^{\rm tor}$ to phases
\begin{equation}
\pi_{r-1}^{\rm tor}\ \mapsto\ \textrm{exp}\left({2\pi i \varphi(\pi_{r-1}^{\rm tor})}\right)
\label{mapuct}
\end{equation}
such that $\varphi(\pi_{r-1}^{\rm tor})$ is the same for each cycle on the same homology class $[\pi_{r-1}^{\rm tor}] \in {\rm Tor\, }H_{r-1}(\cam_D, \IZ)$, and $\varphi([\pi_{r-1}^{\rm tor}]) + \varphi([{\pi_{r-1}^{\rm tor}}']) = \varphi([\pi_{r-1}^{\rm tor}+{\pi_{r-1}^{\rm tor}}'])$. This gives a one-to-one correspondence between the possible choices for $\varphi$ and the elements of ${\rm Tor\, }H_{r-1}(\cam_D, \IZ)$, from which (\ref{unitheorem}) follows.

In terms of this more fundamental definition, it is easy to see why in the main text we have identified certain $p$-forms with elements of ${\rm Tor\, } H^*(\cam_6, \IZ)$. For instance, if we take a torsional 2-cycle $\pi_{2, \a}^{\rm tor}$ of $\cam_6$ we can construct a bump 4-form $\d_4^\a = \d_4(\pi_{2,\a}^{\rm tor})$ that has components transverse to $\pi_{2, \a}^{\rm tor}$ and a $\d$-like support on it. In order to associate $[\d_4^\a]$ with an element of ${\rm Tor\, } H^4(\cam_6, \IZ)$ we should provide a map of the form (\ref{mapuct}) for the set of torsional 3-cycles of $\cam_6$. But we can do this by simply taking a 3-form $F_3^\a$ such that $dF_3^\a = \d_4^\a$ and integrating it over each torsional 3-cycle $\pi_3^{\rm tor}$. Indeed we have that
\begin{equation}
\varphi^\a(\pi_3^{{\rm tor},\b})\,
\equiv\, \int_{\pi_3^{{\rm tor},\b}} F_3^\a\, =\, \int_{\pi_3^{{\rm tor},\b\, \prime}} F_3^\a + \int_{\Sigma_4} \d_4^\a
\label{dmap}
\end{equation}
where we have taken another torsional 3-cycle $\pi_3^{{\rm tor},\b\, \prime}$ such that $[\pi_3^{{\rm tor},\b\, \prime}] = [\pi_3^{{\rm tor},\b}]$ and a 4-chain $\Sigma_4$ such that $\p \Sigma_4 = \pi_3^{{\rm tor},\b} - \pi_3^{{\rm tor},\b\, \prime}$. Notice that (\ref{dmap}) is independent of the choice of $F_3^\a$ that we take, so in the following we will replace $F_3^\a \raw d^{-1}(\d_4^\a)$. Moreover, since the integral of $\d_4^\a$ over this 4-chain is necessarily an integer number, it follows that the map
\begin{equation}
\pi_{3}^{{\rm tor}, \b}\ \mapsto\ \textrm{exp}\left({2\pi i \varphi^\a(\pi_3^{{\rm tor},\b})}\right)
\label{dmapuct}
\end{equation}
does only depend on the homology class $[\pi_{3}^{{\rm tor}, \b}]$. In addition, (\ref{dmapuct}) respects the group law of ${\rm Tor\, } H_3(\cam_6, \IZ)$, and so it is indeed an element of ${\rm Hom} \left({\rm Tor\, }H_{3}(\cam_6, \IZ), \mathbb{Q}/ \IZ\right)$. Hence, we can also think of it as an element of ${\rm Tor\, }H^4(\cam_6, \IZ)$, namely the Poincar\'e dual of $[\pi_{2, \a}^{\rm tor}]$.

Given this identification, it is easy to see that (\ref{dmap}) is nothing but the torsion linking number of $[\pi_{2, \a}^{\rm tor}]$ and $[\pi_{3}^{{\rm tor}, \b}]$. Indeed, following the definition of the main text we have that
\begin{eqnarray}\nonumber
L_{\a}{}^\b = L([\pi_{2, \a}^{{\rm tor}}],[\pi_{3}^{{\rm tor}, \b}]) & \stackrel{{\rm mod\, } 1}{\equiv} & \frac{1}{k_\b} \int_{\Sigma_4^\b} \delta_{4}^\a \, =\, \frac{1}{k_\b} \int_{ k_\b\pi_{3}^{{\rm tor}, \b}} d^{-1} (\delta_4^\a ) \\
&  = & \int_{\pi_{3}^{{\rm tor}, \b}} d^{-1} (\delta_4^\a )\, =\,  \int_{\cam_6} \delta_{3, \b}\wedge d^{-1} (\delta_4^\a )
\label{tln}
\end{eqnarray}
where $k_\b$ is the minimal integer such that $k_\b\pi_{3}^{{\rm tor}, \b}$ is trivial in homology, and we have taken a 4-chain $\Sigma_4^\b$ such that $\p \Sigma_4^\b = k_\b\pi_{3}^{{\rm tor}, \b}$. Finally, we have defined a bump form $\d_{3, \b} = \d_{3, \b}(\pi_{3}^{{\rm tor}, \b})$ for the torsional 3-cycle $\pi_{3}^{{\rm tor}, \b}$. We can also define the torsion linking form $L$ in terms of the latter
\begin{equation}
L^\b{}_\a = L([\pi_{3}^{{\rm tor}, \b}],[\pi_{2, \a}^{{\rm tor}}]) \, \stackrel{{\rm mod\, } 1}{\equiv} \,  \int_{\pi_{2, \a}^{\rm tor}} d^{-1} (\delta_{3, \b} ) \, =\, \int_{\cam_6} \delta_4^\a\wedge d^{-1} (\delta_{3, \b} )
\label{tlf}
\end{equation}
from which is easy to see that for a 6d manifold $L$ is symmetric, and that $kL_\a{}^\b \in \IZ$ for $k = {\rm g.c.d.}(k_\a, k_\b)$.

The torsion linking form is the main topological quantity that one may construct from the finite groups ${\rm Tor\, } H_3(\cam_6, \IZ)$ and ${\rm Tor\, } H_2(\cam_6, \IZ)$ and, by Poincar\'e duality, they express relations that are obeyed by the groups ${\rm Tor\, } H^3(\cam_6, \IZ)$ and ${\rm Tor\, } H^4(\cam_6, \IZ)$. Indeed, from (\ref{tln}) and (\ref{tlf}) we see that we can always construct a set of 2-forms $\{F_{2,\b}\}$ and 3-forms $\{F_3^\a\}$ such that
\begin{equation}
\int_{\cam_6} \d_{3, \b} \wedge F_3^\a \, =\, \int_{\cam_6} \d_4^\a\wedge F_{2,\b} \, =\, \d_\b^\a
\label{normap}
\end{equation}
and
\begin{equation}
dF_{2, \b}\, =\, (L^{-1})_\b{}^\a \d_{3, \a} \quad \quad \quad  dF_3^\a\, =\, -(L^{-1})^\a{}_\b \d_4^\b
\label{basisap}
\end{equation}
where we have used the fact that $L$ is invertible.

As these relations contain topological information of the torsion homology groups, we should impose similar ones to each set of forms with integer coefficients that aim to represent ${\rm Tor\, } H^3(\cam_6, \IZ)$ and ${\rm Tor\, } H^4(\cam_6, \IZ)$. In the main text we have done so for a set of forms that can be thought as smoothed out versions of the bump forms $\d_4^\a$ and $\d_{3, \b}$. More precisely we have the relations
\begin{equation}
\begin{array}{rcl}\vspace*{.5cm}
[\d_{3, \a}] = [\a_\a^{\rm tor}] \in  {\rm Tor}\, H^3 (\cam_6, \IZ) & \quad \quad & [\d_4^\a] = [\tilde{\omega}^{{\rm tor}, \a}] \in  {\rm Tor}\, H^4 (\cam_6, \IZ)\\
F_{2,\a} \sim \omega_\a^{\rm tor} \in \widehat{\textrm{Tor}}\ H^{2} & \quad \quad & F_3^\b \sim \b^{{\rm tor}, \a} \in \widehat{\textrm{Tor}}\ H^{3}
\end{array}
\label{newbasis}
\end{equation}
where the set $\widehat{\textrm{Tor}}\ H^{p}$ is closed under the action of the Laplacian, see eqs.(\ref{eigensp}). That this set of forms exists has been our working assumption in section \ref{mmixing}.

 How can we construct a smoothed out version of our bump functions? One possible way is, following \cite{ckt05}, to consider objects in relative cohomology. Indeed, let us take a set of torsional 2-cycles $\{\pi_{2,\a}^{\rm tor}\}$ and 3-cycles $\{\pi_3^{{\rm tor}, \a}\}$ such that their homology classes generate ${\rm Tor\, } H_2(\cam_6, \IZ)$ and ${\rm Tor\, } H_3(\cam_6, \IZ)$, respectively. We may consider a particular 2-cycle and construct the relative cohomology groups $H^p(\cam_6, \pi_{2,\a}^{\rm tor})$. Those are constructed as in usual de Rham cohomology, but cochains are instead given by pairs of forms $(\sig_p, \tilde{\sig}_{p-1}) \in \Omega^p(\cam_6) \times \Omega^{p-1}( \pi_{2,\a}^{\rm tor})$ and the differential by
 \begin{equation}
d(\sig_p, \tilde{\sig}_{p-1})\,=\, (d\sig_p,\, \sig_p|_{\pi_{2,\a}^{\rm tor}} - d\tilde{\sig}_{p-1})
\end{equation}
Thus, let us take the pair $(\delta_{3,\b}, 0)$, defining a non-trivial class $[(\delta_{3,\b}, 0)] \in H^3(\cam_6, \pi_{2,\a}^{\rm tor})$. Any other 3-form $\a^{\rm tor}_\b$ such that $(\a^{\rm tor}_\b, 0)$ is in the same relative cohomology class $[(\delta_{3,\b}, 0)]$ satisfies that $\a^{\rm tor}_\b - \delta_{3,\b} = d \sig_{2, \b}$ with $\sig_{2, \b}$ such that $\sig_{2, \b}|_{\pi_{2,\a}^{\rm tor}} = d\tilde{\sig}_1$ for some 1-form $\tilde{\sig}_1$ of ${\pi_{2,\a}^{\rm tor}}$. This implies that in (\ref{tlf}) we can replace $\delta_{3,\b}$ with $\a^{\rm tor}_\b$, since
\begin{equation}
\int_{\pi_{2, \a}^{\rm tor}} d^{-1} (\a^{\rm tor}_{\b} )\, =\,  \int_{\pi_{2, \a}^{\rm tor}} d^{-1} (\delta_{3, \b} ) + \int_{\pi_{2, \a}^{\rm tor}}  \sig_2\, =\, \int_{\pi_{2, \a}^{\rm tor}} d^{-1} (\delta_{3, \b} ) + \int_{\pi_{2, \a}^{\rm tor}}  d\tilde{\sig_1}\, = \, L^\b{}_\a
\end{equation}
We can repeat the same construction for $[(\delta_{4}^\a, 0)] \in H^4(\cam_6, \pi_{3}^{{\rm tor}, \b})$. There we have that for any 4-form $\tilde{\omega}^{{\rm tor},\a}$ such that $(\tilde{\omega}^{{\rm tor},\a}, 0) \sim (\d_4^\a, 0)$ in $H^4(\cam_6, \pi_{3}^{{\rm tor}, \b})$, we can replace $\d_4^\a \raw \tilde{\omega}^{{\rm tor},\a}$ in (\ref{tln}) and obtain the same result. It then follows that the set of forms $\{\a^{\rm tor}_\b\}$ and $\{\tilde{\omega}^{{\rm tor},\a}\}$ constructed in this way  satisfy relations equivalent to (\ref{normap}) and (\ref{basisap}), namely \cite{ckt05}
\begin{equation}
\int_{\cam_6} \a^{\rm tor}_\b\wedge \b^{{\rm tor},\a}  \, =\, \int_{\cam_6} \omega^{\rm tor}_{\b} \wedge \tilde{\omega}^{{\rm tor},\a}\, =\, \d_\b^\a
\label{normap2}
\end{equation}
and
\begin{equation}
d\omega^{\rm tor}_\b \, =\, (L^{-1})_\b{}^\a \a^{\rm tor}_\a \quad \quad \quad  d\b^{{\rm tor},\a}\, =\, -(L^{-1})^\a{}_\b\, \tilde{\omega}^{{\rm tor},\b}
\label{basisap2}
\end{equation}
More importantly, this means that the phases (\ref{mapuct}) that these forms associate to each torsional 2 and 3-cycle of our construction are exactly the same as the bump forms $\d_4^\a$ and $\d_{3,\b}$ and, in this sense, they can be thought as the same elements of ${\rm Tor\, } H^4(\cam_6, \IZ)$ and ${\rm Tor\, } H^3(\cam_6, \IZ)$.

In order to complete the construction (\ref{newbasis}) we need to find a set of representatives $\{\a^{\rm tor}_\a\}$ and $\{\tilde{\omega}^{{\rm tor},\b}\}$ of the above relative cohomology classes which form a closed set under the action of the Laplacian, in the sense of eq.(\ref{eigensp2}). That such kind of basis exists has been shown to be the case for simple examples of torsional manifolds like twisted tori, as well as for other manifolds obtained by twists of the form (\ref{twist}) and (\ref{torflat}), see \cite{twist,Cvetic:2007ju}. For those constructions we have that
\begin{equation}
-\int_{\pi_{3}^{{\rm tor}, \b}} \b^{{\rm tor},\a} \, =\, \d_\b^\a
\end{equation}
and so expanding the RR potential $C_5$ as in (\ref{torexpC5}) and dimensionally reducing it over a D6-brane wrapping a torsional 3-cycle we obtain the couplings (\ref{D6coupling}).

The results of this paper, however, do not rely on the above construction and can be derived using the more abstract language of gerbes (see e.g. \cite{Hitchinrw}), which is the precise way to describe RR field strengths and potentials. From such viewpoint we should think of $\a_\a^{\rm tor}$ as the curvature of a 1-gerbe, and $\tilde{\omega}^{{\rm tor}, \a}$ as the curvature of a 2-gerbe. Taking an appropriate covering $\{U_a\}$ of $\cam_6$ we can characterize a 1-gerbe with curvature 3-form $\a$ by a set of forms that satisfy
\begin{equation}
\begin{array}{rcl}
 \a|_{U_a}  & =  & dF_a  \\
F_b - F_a  & =  &  dA_{ab} \\
i\left(A_{ab} + A_{bc} + A_{ca}\right)   & =  &  g_{abc}^{-1}dg_{abc}
\end{array}
\end{equation}
with $g_{abc}:U_a \cap U_b \cap U_c \raw S^1$ a cocycle that defines the gerbe, and that is analogous to a set of transitions functions $g_{ab}: U_a \cap U_b \raw S^1$ for a line bundle.

As discussed in \cite{Hitchinrw}, if the gerbe curvature $\a$ vanishes identically then we can write $F_a = dB_a$ on $U_a$, and we say that we have a gerbe with a flat connection. Similarly to the case of line bundles, where a flat connection defines a homomorphism $\pi_1(\cam_6) \raw S^1$, a 1-gerbe with a flat connection defines a homomorphism $H_2(\cam_6,\IZ) \raw S^1$ , and we dub the phase associated to each 2-cycle of $\cam_6$ as the holonomy induced by the gerbe. If we restrict this homomorphism to ${\rm Tor\, }H_2(\cam_6,\IZ) \raw S^1$, then we see that this holonomy is nothing but the phases of the map (\ref{mapuct}) for $r=3$, and so a 1-gerbe with flat connection can be related to an element of ${\rm Tor\, } H^3(\cam_6, \IZ)$.
If the curvature $\a$ does not vanish then we can still define a holonomy for each 2-cycle $\pi_2$, but now it varies within the homology class $[\pi_2]$. Indeed, let us consider two homologous 2-cycles $\pi_2$ and $\pi_2^\prime$, and a 3-chain $\Sigma_3$ such that $\p \Sigma_3 = \pi_2^\prime - \pi_2$. Then we have that
\begin{equation}
{\rm hol\, }(\pi_2^\prime) \, =\, {\rm hol\, }(\pi_2) \cdot {\rm exp\, } \left( 2\pi i \int_{\Sigma_3} \a \right)
\label{varyinghol}
\end{equation}
which is a well-defined quantity because $\int_{\Pi_3} \a \in \IZ$ for each 3-cycle $\Pi_3 \subset \cam_6$.

Let us now consider a 1-gerbe whose curvature $\a$ does not vanish but it is trivial in $H^3(\cam_6, \IR)$, as it is the case for the torsional 3-forms $\a_\a^{\rm tor}$ considered in this work. In that case we have that on the patch $U_a$, $F_a = F + dB_a$ with $F$ a globally well-defined 2-form such that $dF = \a$. From (\ref{varyinghol}) and the fact that $\int_{\Sigma_3} \a = \int_{\pi_2^\prime} F - \int_{\pi_2} F$ it follows that
\begin{equation}
\widetilde{\rm hol\, }(\pi_2)\, =\,  {\rm hol\, } (\pi_2) \cdot {\rm exp} \left( - 2\pi i \int_{\pi_2} F \right)
\end{equation}
only depends on the homology class of $\pi_2$, and therefore it defines a homomorphism ${\rm Tor\, }H_2(\cam_6,\IZ) \raw S^1$ that allows to identify $\a$ with an element of ${\rm Tor\, } H^3(\cam_6, \IZ)$.

Clearly, we can define $\widetilde{\rm hol\, }$ for a gerbe of any degree. There is however a particularly elegant way to define it for torsional $r$-cycles, based on the topological invariants built on \cite{freed} (see also \cite{Wen:1985qj}). Indeed, let us consider a torsional $r$-cycle $\pi_r^{\rm tor}$ and the holonomy induced on it by a $(r-1)$-gerbe curvature $\a_{r+1}$. Since $\pi_r^{\rm tor}$ is torsional, we have that $k \pi_r^{\rm tor} = \p \Sigma_{r+1}$ for some 4-chain $\Sigma_4$ and $k \in \IZ$. Then we can write
\begin{equation}
\widetilde{\rm hol\, }(\pi_r)\, =\,  {\rm hol\, } (\pi_r) \cdot {\rm exp} \left( - \frac{2\pi i}{k} \int_{\Sigma_{r+1}} \a_{r+1} \right)
\label{freedhol}
\end{equation}

Remarkably, (\ref{freedhol}) is precisely what we obtain when we compute the couplings (\ref{D6coupling}) between D6-brane U(1) gauge bosons and RR massive axions. Indeed, in this case the gerbe curvature is given by $\tilde{\omega}^{{\rm tor}, \b}$, and the torsional cycle by the sum of 3-cycles $\pi_b^-$ that we associate to the open string U(1)$_b$. Naively, the coefficients $c_b^\b$ are obtained from the D6-brane dimensional reduction as
\begin{equation}
\int_{\pi_b^-} \b^{{\rm tor}, \b}\, = \, - k^\b{}_\a \int_{\pi_b^-} d^{-1} (\tilde{\omega}^{{\rm tor}, \a})\, \raw \, \frac{i k^\b{}_\a}{2\pi} {\rm ln\, }{\rm hol}^\a (\pi_b^-)
\label{coeffbare}
\end{equation}
where ${\rm hol}^\a$ is the holonomy induced by $\tilde{\omega}^{{\rm tor}, \a}$. However, to this quantity we need to substract the one that appears in the kinetic mixing of U(1)$_b$ and the torsional RR U(1)'s.
\begin{equation}
 \int_{\IR^{1,3}} (dV^\a - k^\a{}_\b C_2^\b) \wedge F_2^b \ \frac{1}{k_b}\int_{\Sigma_4^b}  \tilde{\omega}^{{\rm tor}, \a}
 \label{kmixRRtor}
\end{equation}
where $\p \Sigma_4^b = k_b \pi_b^-$. Using that the matrix $k$ is symmetric, it is possible to see that subtracting the kinetic mixing coefficient amounts to replace ${\rm hol}^\a (\pi_b^-) \raw \widetilde{\rm hol}{}^\a (\pi_b^-)$ in (\ref{coeffbare}). Therefore, since by definition
\begin{equation}
\frac{1}{2\pi i} {\rm ln\, }\widetilde{\rm hol}{}^\a (\pi_b^-)\, =\, L_\a{}^b\ ,
\end{equation}
we recover via (\ref{decomlink}) the result of the main text.


\end{document}